\documentclass[article]{elsarticle}
\makeatletter
\def\ps@pprintTitle{%
	\let\@oddhead\@empty
	\let\@evenhead\@empty
	\def\@oddfoot{\centerline{\thepage}}%
	\let\@evenfoot\@oddfoot}
\makeatother

\usepackage{lineno,hyperref}
\usepackage{threeparttable}
\usepackage{float}
\usepackage{booktabs}
\usepackage{amsmath}
\usepackage{comment}
\usepackage{color}

\usepackage{amsfonts}
\usepackage{amssymb}
\usepackage{amsmath}
\usepackage{amsthm}
\usepackage{amssymb}

\usepackage{epstopdf}
\usepackage{natbib}

\modulolinenumbers[5]

\begin{document}

\begin{frontmatter}

\title{Atomistic and mean-field estimates of effective stiffness tensor of nanocrystalline materials of cubic symmetry}

\author[]{Katarzyna Kowalczyk-Gajewska\corref{author}}
\author[]{Marcin Ma\'zdziarz}

\cortext[author] {Corresponding author.\\\textit{E-mail address:} kkowalcz@ippt.pan.pl}
\address{Institute of Fundamental Technological Research Polish Academy of Sciences, Warsaw, Poland}

\begin{abstract}
Anisotropic core-shell model of a nano-grained polycrystal, proposed recently for nanocrystalline copper, is applied to estimate elastic effective properties for a set of crystals of cubic symmetry. Materials selected for analysis differ in the lattice geometry  (face-centered cubic vs. body-centered cubic) as well as the value of a Zener factor: a ratio of two shear moduli defining elastic anisotropy of a cubic crystal.  The predictions are verified by means of the atomistic simulations. The dependence of the overall bulk and shear moduli on the average grain diameter is analysed. In the mean-field approach the thickness of the shell is specified by the \emph{cutoff radius} of a corresponding atomistic potential, while the grain shell is isotropic and its properties are identified by molecular simulations performed for very small grains with approximately all atoms belonging to the grain boundary zone. It is shown that the core-shell model provides predictions of satisfactory qualitative and quantitative agreement with atomistic simulations. Performed study indicates that
the variation of the bulk and shear moduli with the grain size changes qualitatively when the Zener anisotropy factor is smaller or greater than one. 
\end{abstract}


\end{frontmatter}

\section{Introduction}
\label{sec:Int}

Nanostructured materials (NsM) considered in the present research are
bulk solids with a nanometre-scale microstructure and a characteristic dimension smaller than 100\,nm \cite{Gleiter00,Gao13}. In particular, a class of such materials is analysed that is composed of equiaxed nanometre-sized building blocks - crystallites of the same chemical composition, which differ only by the crystallographic orientation \cite{Gleiter00}. In the microstructure of nanocrystalline metals two main phases are distinguished: the grain core and the grain boundaries. The impact of grain boundary zone on the effective properties of a bulk polycrystal is the more significant the smaller is a grain size \cite{Sanders97,Gao13}. Usually at the macroscale level the stress-strain response of nanocrystalline metals is governed by the continuum mechanics theory. 

Because for a considered category of nanocrystalline materials continuum mechanics description is applicable at the macro-level, the mean-field estimates are still in use for assessing the bulk properties of nanocrystalline metals. In \cite{Kowalczyk18} an extensive overview of such estimates available in the literature has been presented. In general, in variance with the coarse-grained polycrystals for which one-phase models can be used, for nano-grained polycrystals two-phase or multi-phase frameworks are formulated. In a nutshell they can be categorized into the following three groups:
\begin{itemize}
	\item simplified mixture rule-based models \cite{Carsley95,Kim99,Benson01,Qing06,Zhou07},
	\item inclusion-matrix models \cite{Sharma03},
	\item composite sphere / generalized self-consistent-type models \cite{Jiang04,Chen06,Capolungo07,Mercier07, Ramtani09,Kowalczyk18}.
\end{itemize} 

The main difficulty encountered when applying those multi-phase concepts is to properly identify the properties of a grain boundary zone (or zones) and its volume fraction and morphology. Identification of those parameters and validation of the models are usually performed employing the molecular dynamics/statics simulations \cite{Schiotz99,Chang03,Choi12,Gao13,Mortazavi2014,Fang16,Kowalczyk18}. Occasionally, full-field FEM analyses are also conducted for this purpose, however, in such a case the refined constitutive description with scale effects is used \cite{Kim20123942}. The most often outcome of atomistic simulations is that elastic stiffness decreases with a  grain size \cite{Zhao06,Gao13,Xu17}. The two- or multi-phase mean-field approaches reproduce the observed tendency when the stiffness of the grain boundary is smaller than the stiffness of the grain core \cite{Jiang04,Ramtani09,Gao13}. 

Reported research is continuation of the work undertaken in \cite{Kowalczyk18}.
The aim of the present paper is to verify applicability of a core-shell model proposed therein for estimating the effective elastic stiffness of nanocrystalline metals of cubic symmetry. In particular, the additional assumptions taken in the approach and concerning the properties of a grain boundary zone are checked. Moreover, the possible correlation between the Zener factor - the basic anisotropy parameter in the case of elastic cubic symmetry - and the character of size-dependence of effective properties is analysed.

The next section recalls the construction of the core-shell model starting with the discussion on the spectral form of the elasticity tensor of cubic symmetry and the meaning of a Zener parameter.  Section \ref{sec:Cm} presents the basic elements of atomistic simulations.  In Section~\ref{sec:Res} the results of the atomistic simulations are compared with the predictions of a core-shell model. Dependence of the isotropized overall bulk and shear moduli on the averaged grain diameter is studied. Detailed results of molecular simulations are collected in the Appendix. The paper is closed with conclusions. 

\section{Core-shell model for nanocrystalline materials of cubic symmetry}
\label{sec:Cont}

\begin{figure}
	\centering
	\includegraphics[width=0.85\textwidth]{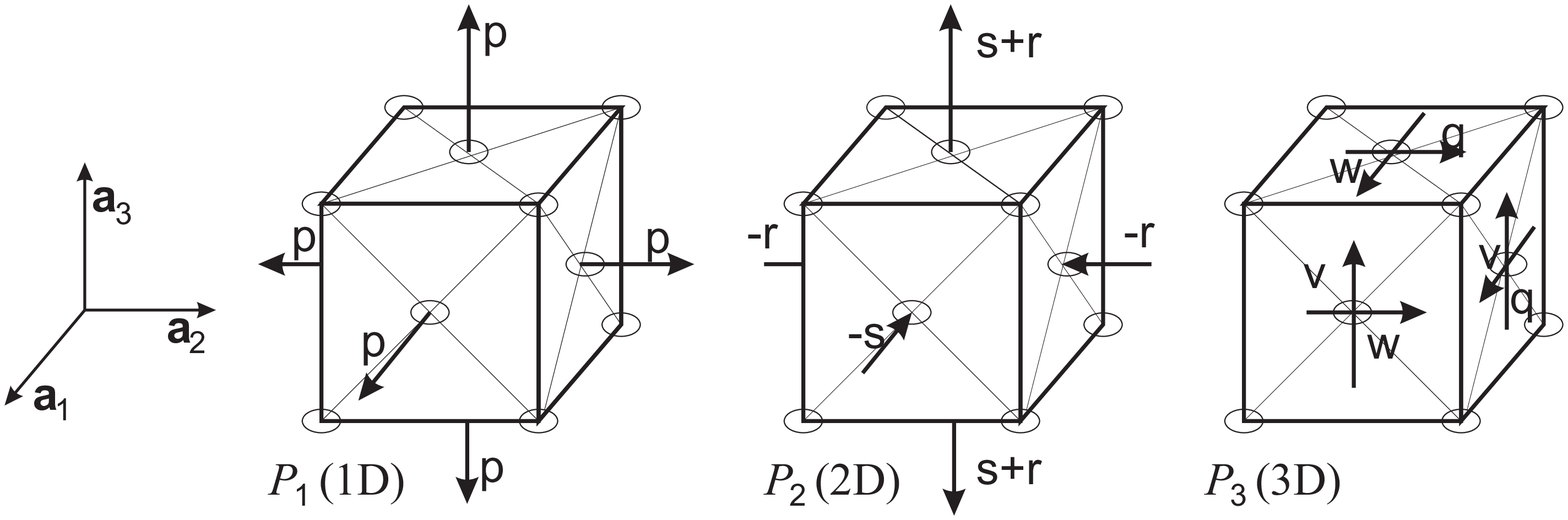}
	\caption{Illustration of eigen-subspaces of the elasticity tensor of cubic symmetry}
	\label{fig:CubicSub}
\end{figure}

The anisotropic constitutive relation between the stress 
$\boldsymbol{\sigma}$ and strain 
$\boldsymbol{\varepsilon}$ in the grain is
\begin{equation}\label{locconst}
\boldsymbol{\sigma}=\mathbb{C}(\phi_c)\cdot\boldsymbol{\varepsilon},\quad
\boldsymbol{\varepsilon}=\mathbb{S}(\phi_c)\cdot\boldsymbol{\sigma},\quad\mathbb{S}(\phi_c)\mathbb{C}(\phi_c)=\mathbb{I}\,,
\end{equation}
where $\mathbb{C}(\phi_c)$, $\mathbb{S}(\phi_c)$ and $\mathbb{I}$ are the fourth order elastic stiffness, compliance and symmetrized identity
tensors, respectively.  Argument $\phi_c$ denotes symbolically an orientation of local axes
$\{\mathbf{a}_k\}$ with respect to some macroscopic frame
$\{\mathbf{e}_k\}$. For crystals of cubic symmetry the local stiffness tensor  $\mathbb{C}(\phi_c)$  can be written in the following spectral form \cite{Walpole81,Rychlewski95,Kowalczyk09}
\begin{equation}\label{Eq:SpekCubic}
\mathbb{C}(\phi^c)=3K\mathbb{I}^{\rm{P}}+2G_1(\mathbb{K}(\phi^c)-\mathbb{I}^{\rm{P}})+2
G_2(\mathbb{I}-\mathbb{K}(\phi^c))\,,
\end{equation}
with
\begin{equation}
\mathbb{I}^{\rm{P}}=\frac{1}{3}\mathbf{1}\otimes\mathbf{1}\quad\textrm{and}\quad\mathbb{K}(\phi_k)=\sum_{k=1}^3\mathbf{a}_k\otimes\mathbf{a}_k\otimes\mathbf{a}_k\otimes\mathbf{a}_k\,.
\end{equation}
$\mathbf{1}$ denotes the second order identity tensor. The material parameters $3K$, $2G_1$ and $2G_2$ are Kelvin moduli which can be expressed by means of the components of the stiffness tensor in the basis of local anisotropy axes $\{\mathbf{a}_k\}$ (see Fig. \ref{fig:CubicSub}), namely
\begin{equation}\label{Eq:moduli}
3K=C_{1111}+2C_{1122}\,,\quad 2G_1=C_{1111}-C_{1122}\,,\quad 2G_2=2C_{2323}\,.
\end{equation}

Three Kelvin moduli $3K$, $2G_1$ and $2G_2$ correspond to three eigen-subspaces of strain or stress states, which are respectively:
\begin{itemize}
	\item the one-dimensional space of hydrostatic states,
	\item the two-dimensional space of deviatoric states with eigen-vectors coaxial with anisotropy axes $\mathbf{a}_k$,
	\item the three-dimensional space of deviatoric states with only shear components in the basis $\mathbf{a}_k$.
\end{itemize}
This subspaces are schematically illustrated in Fig. \ref{fig:CubicSub}. For the states belonging to the respective subspaces the proportionality is observed between stress and strain tensors. 

The  parameter, proposed as an anisotropy measure for cubic crystals by \cite{zener1948elasticity}, is defined as
\begin{equation}\label{Eq:zeta1}
\zeta_1=\frac{C_{1111}-C_{1122}}{2C_{2323}}=\frac{{G}_1}{G_2}\,.
\end{equation}
Note that the condition $\zeta_1=1$ is sufficient to ensure elastic isotropy of cubic crystal. The qualitative difference in elastic response is observed for materials with $\zeta_1>1$ or $\zeta_1<1$.  In particular, if $\zeta_1<1$ the maximal and minimal directional Young moduli are found for directions $\{001\}$ and $\{111\}$, respectively. A reverse result is obtained if $\zeta_1>1$ \cite{Ostrowska01}.

In view of the classical micromechanical theories coarse-grained polycrystals are treated as one-phase heterogeneous materials. Their heterogeneous response results from the varying orientation of crystal axes in the polycrystalline representative volume element. Effective properties of the grain aggregate are estimated assuming that local elastic properties are known and then performing a micro-macro transition. The formulas for the standard estimates, such as the Voigt, Reuss, Hashin-Shtrikman or self-consistent one, can be found in \cite{Kowalczyk18}. These estimates are not sensitive to the grain size.

\begin{figure}[h!]
	\centering
	\includegraphics[width=.7\textwidth]{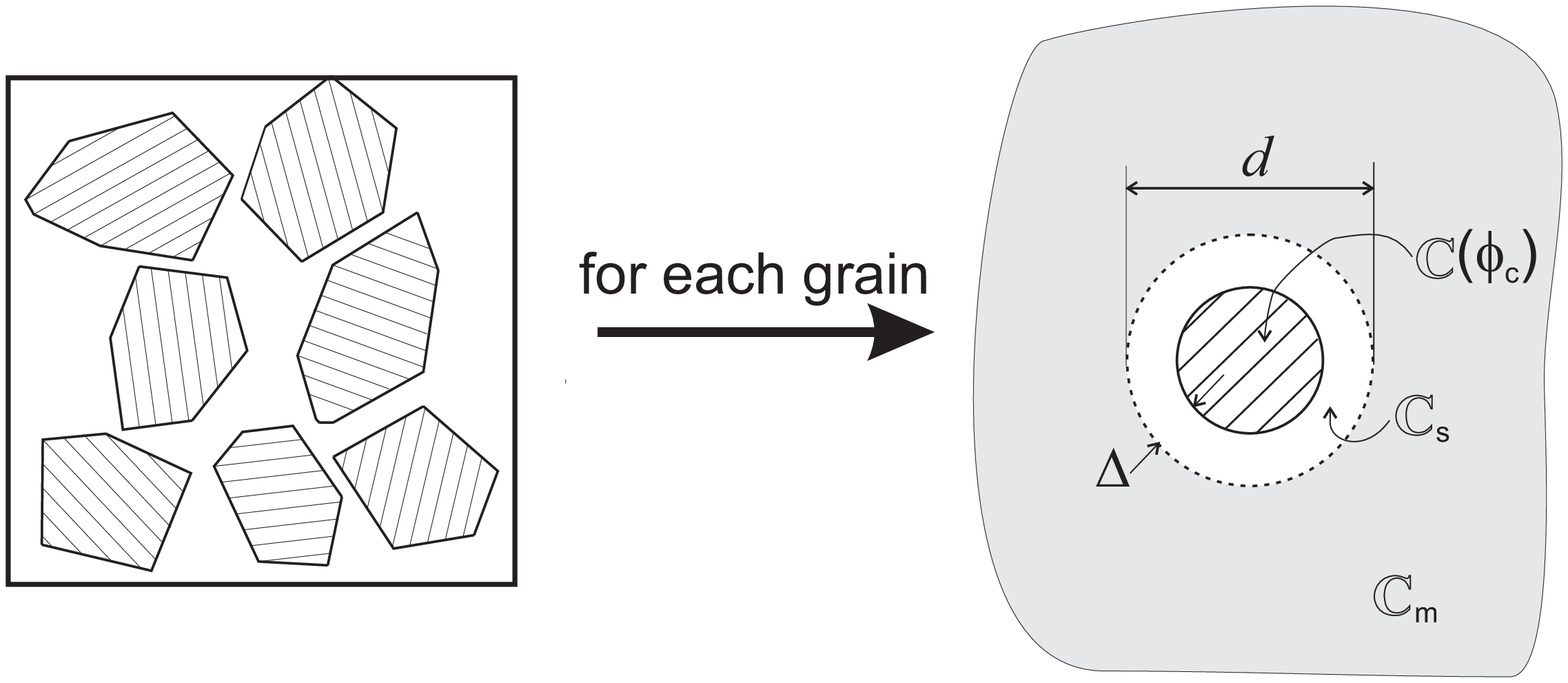}
	\caption{Schematic of the core-shell model of the nanograin polycrystal.}
	\label{fig:CoreShell}
\end{figure}

For nanograined polycrystals the concept of one-phase material has to be abandoned in favour of a two- or multi-phase medium. As listed in Section \ref{sec:Int} a family of approaches using such concept can be found in the literature. Specifically, in \cite{Kowalczyk18} the authors formulated a two-phase model in two variants called \emph{the Mori-Tanaka (MT) and self-consistent (SC) core-shell model}, respectively. As inspired by \cite{Jiang04} and \cite{Capolungo07} an additional phase that forms an \emph{isotropic} coating around the \emph{anisotropic} grain core is introduced. The smaller is the grain this transient zone influences more on the effective properties of polycrystal. 
{Yet another proposal for dealing with interphase layer can be found in \cite{Sevostianov2006,SEVOSTIANOV20071304},  formulated in a different context of metal-matrix composite reinforced by nanosized inclusions, in which both inclusion and the matrix have {isotropic} properties. Extending the proposal by \cite{Shen03}, authors assumed the existence of the {interphase layer} between the inclusion and the matrix. This layer is assumed to have {isotropic} properties which {vary smoothly} with "upward convexity". 
	Note that the schematic configuration considered in the present paper is different: the {nanocrystalline material} is assumed with {anisotropic} properties of a {grain core} and {uniform isotropic shell} (coating). Such anisotropic composite inclusions constitute the polycrystalline material, which, in general, is also {effectively anisotropic}. Therefore, the approach presented in the above mentioned papers is not directly applicable in the present context. Nevertheless the core-shell model can be formulated in yet another variant which follows this idea, so that inhomogeneous shell properties can be assumed. However, such variant leads to more complicated formulation, while in the same time establishment of actual variation of shell stiffness using atomistic simulations is not an easy task and the results are debatable \cite{Kluge1990}. Therefore a respective alternative proposal is only shortly described in Appendix B, demonstrating differences with respect to the original formulation using the example of nanocrystalline copper.}

Below basic relations {for a core-shell model proposed in \cite{Kowalczyk18}} are shortly recapitulated.
By linearity of the local constitutive law overall relations between the averaged strain  $\mathbf{E}=\langle \boldsymbol{\varepsilon}\rangle $ and stress $\boldsymbol{\Sigma}=\langle \boldsymbol{\sigma}\rangle $ in the polycrystalline RVE are: 
\begin{equation}
\boldsymbol{\Sigma}=\bar{\mathbb{C}}\cdot\mathbf{E},\quad
\mathbf{E}=\bar{\mathbb{S}}\cdot\boldsymbol{\Sigma},\quad
\bar{\mathbb{S}}\,{\bar{\mathbb{C}}}=\mathbb{I}\,,
\end{equation}
where $\bar{\mathbb{C}}$ (resp. $\bar{\mathbb{S}}$) is the effective stiffness (resp. compliance) tensor of the polycrystal to be found by the mean-field model. Averaging is defined by the formula $\langle . \rangle =\frac{1}{V}\int_V(.) dV $ and is performed over the representative material volume.
The estimates of two-phase model presented in the next section are obtained under the assumption that RVE contains infinite set of orientations of random uniform distribution. In such case the averaging operation $\langle . \rangle$ is replaced by the averaging over the orientation space $\langle . \rangle_{\mathcal{O}}$ 
and when performed over the fourth order tensor $\mathbb{T}(\phi^c)$ gives \cite{Forte96,Rychlewski01,Kowalczyk09}
\begin{equation}
\langle \mathbb{T}(\phi_c) \rangle_{\mathcal{O}} = \frac{1}{3}(\mathbf{1}\cdot\mathbb{T}(\phi_c)\cdot\mathbf{1})\mathbb{I}^{\rm{P}}+\frac{1}{5}\left(\mathbb{I}^{\rm{D}}\cdot\mathbb{T}(\phi_c)\right)\mathbb{I}^{\rm{D}}\,,
\end{equation}
where $\mathbb{I}^{\rm{P}}$ and $\mathbb{I}^{\rm{D}}=\mathbb{I}-\mathbb{I}^{\rm{D}}$ play a role of the fourth order orthogonal projectors to the hydrostatic and deviatoric subspaces of the second order tensor space, correspondingly. Under such assumption the overall stiffness tensor $\bar{\mathbb{C}}$ is isotropic and specified by 
\begin{displaymath}
\bar{\mathbb{C}}=3\bar{K}\mathbb{I}^{\rm{P}}+2\bar{G}\mathbb{I}^{\rm{D}}\,,
\end{displaymath}
where $\bar{K}$ and $\bar{G}$ are the overall bulk and shear moduli, respectively.

According to the core-shell model formulated in \cite{Kowalczyk18} the effective stiffness $\bar{\mathbb{C}}$ is calculated by embedding the coated grain in the infinite medium of the stiffness $\mathbb{C}_{\rm{m}}$ and next using the procedure of the double-inclusion model \cite{HoriNematNasser93}, namely
\begin{equation}\label{eq:core-shell}
\bar{\mathbb{C}}_{\rm{CS}}=\left[f_0\mathbb{C}_{\rm{s}}\mathbb{A}_{\rm{s}}+(1-f_0)\left<\mathbb{C}(\phi^c)\mathbb{A}(\phi^c)\right>_{\mathcal{O}}\right]\left[f_0\mathbb{A}_{\rm{s}}+(1-f_0)\left<\mathbb{A}(\phi^c)\right>_{\mathcal{O}}\right]^{-1}
\end{equation}						
where
\begin{equation}
\mathbb{A}(\phi^c)=(\mathbb{C}(\phi^c)+\mathbb{C}_*(\mathbb{C}_{\rm{m}}))^{-1}(\mathbb{C}_{\rm{m}}+\mathbb{C}_*(\mathbb{C}_{\rm{m}}))\,,
\end{equation}	
\begin{equation}
\mathbb{A}_{\rm{s}}=(\mathbb{C}_{\rm{s}}+\mathbb{C}_*(\mathbb{C}_{\rm{m}}))^{-1}(\mathbb{C}_{\rm{m}}+\mathbb{C}_*(\mathbb{C}_{\rm{m}}))
\end{equation}
and  $\mathbb{C}_*(\mathbb{C}_{\rm{m}})$ is the Hill tensor \cite{Hill65} depending on the stiffness $\mathbb{C}_{\rm{m}}$ of infinite matrix material and the shape of the coated grain, which is assumed as spherical in the present study. For the MT variant of the core-shell model $\mathbb{C}_{\rm{m}}=\mathbb{C}_{\rm{s}}$ is assumed (the infinite medium has shell properties), while for the SC variant $\mathbb{C}_{\rm{m}}=\bar{\mathbb{C}}_{\rm{CS}}$ (the infinite medium has the effective properties to be found). In formula (\ref{eq:core-shell}) $f_0$ is the volume fraction of the transient zone. Assuming the spherical shape of coated grains and denoting by $\Delta$ the coating thickness, $f_0$ is calculated by the formula
\begin{equation}\label{def:fsa}
f_0=1-\left(1-\frac{2\Delta}{d}\right)^3\,,
\end{equation} 
where $d$ is an averaged grain diameter. The parameter $\Delta$ introduces the size effect to the model. In \cite{Kowalczyk18} it was demonstrated that $\Delta$ can be taken as equal to the \emph{cutoff radius} of the atomistic potential valid for the considered metal.  

In the same paper, in which nanocrystalline copper was analysed, it was additionally assumed that the shell properties are equal to the zeroth order lower bound for the one-phase polycrystal \cite{Nadeau01}, so that the bulk modulus $K_{\rm{s}}=K$ and the shear modulus $G_{\rm{s}}=\min\{G_1,G_2\}$. In the present work this constraint is released, so that the isotropic shell properties need to be identified separately.  

Note that for a coarse-grained polycrystal ($f_0\rightarrow 0$) the effective properties 
$\bar{\mathbb{C}}_{\rm{CS/SC}}$ approach
the self-consistent estimate for a one-phase polycrystal. 
An estimate of shear modulus related to $\bar{\mathbb{C}}_{\rm{CS/MT}}$ with arbitrary shell properties $\mathbb{C}_{\rm{s}}(K_{\rm{s}},G_{\rm{s}})$ and perfectly random orientation distribution, when $f_0\rightarrow 0$, approaches the following value
\begin{equation}
\bar{G}_{\rm{CS/MT}}^{\infty}=\frac{30 G_1 G_2 (2 G_{\rm{s}} + K_{\rm{s}}) + G_{\rm{s}}(2 G_1+3 G_2)(8 G_{\rm{s}} + 9 K_{\rm{s}})}{
	6(3 G_1+2 G_2)(2 G_{\rm{s}} + K_{\rm{s}}) + 5 G_{\rm{s}} (8 G_{\rm{s}} + 9 K_{\rm{s}})}\,.
\end{equation}
This value is some lower (resp. upper) bound estimate of $\bar{\mathbb{C}}$ if the difference $\mathbb{C}_{\rm{s}}-\mathbb{C}(\phi_c)$ is negative (resp. positive)  definite for any $\phi_c$.    
For very small grains ($f_0\rightarrow 1$) the estimates $\bar{\mathbb{C}}_{\rm{CS/MT}}$ and 
$\bar{\mathbb{C}}_{\rm{CS/SC}}$ approach each other and coincide with $\mathbb{C}_{\rm{s}}$. 

Since the bulk modulus of a coating $K_{\rm{s}}$ can be different from the local bulk modulus of a cubic crystal $K$ also the overall bulk modulus $\bar{K}$, next to the overall shear modulus $\bar{G}$, can be affected by a grain size, contrary to the result obtained in \cite{Kowalczyk18} where the bulk modulus of a coating was equal to the local one. Note that the latter situation (i.e. $\bar{K}=K$) is always obtained for a one-phase polycrystals of cubic symmetry \cite{Walpole85}.

\section{Computational methods}
\label{sec:Cm}
{All atomistic calculations were carried out with the use of the Large-scale Atomic/Molecular Massively Parallel Simulator (LAMMPS) \cite{Plimpton1995}, the Embedded Atom Model (EAM) \cite{Tadmor2011,Becker2013277}, and the molecular statics (MS) method (i.e. at 0K temperature) \cite{Tadmor2011,Maz2010,Maz2011}.  The results obtained were analysed and visualized with the Open Visualization Tool OVITO \cite{Ovito2010}. The methodology for generating polycrystal samples by the Voronoi tessellation method implemented in the Atomsk program  \cite{Hirel2015212}, their pre-relaxation and numerical simulations was taken directly from \cite{Kowalczyk18}. The components of stiffness tensor, of all pre-relaxed structures, $\bar{C}_{ijkl}$ were computed by the stress-strain method with the maximum strain amplitude set to be 10$^{-4}$ \cite{Plimpton1995,Mazdziarz2015}. 
	
	In order to study the effect of the cubic crystal system (face-centered cubic (FCC) vs. body-centered cubic (BCC)), the value of a Zener factor as well as the number and size of grains on mechanical properties of polycrystalline material, eight metals with nine grain sizes each were considered in this work, see the following enumeration \ref{i:Cu}--\ref{i:V}, Tabs. \ref{tab:SamplesCu}--\ref{tab:SamplesV} and Fig. \ref{fig:Zener}.
	
	\begin{figure}[h!]
		\centering
		\includegraphics[width=.6\textwidth]{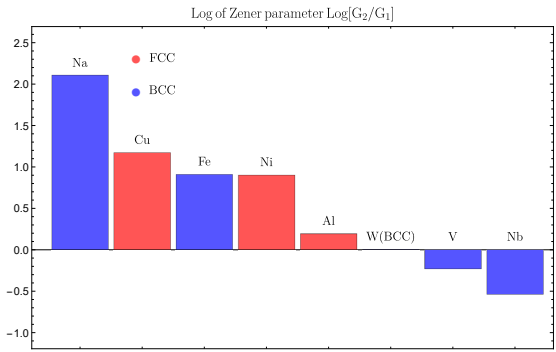}
		\caption{Zener parameter and the lattice geometry for the analysed metals.}
		\label{fig:Zener}
	\end{figure}
	
	{There is a generally known problem at the nano level with how to assume local effective material parameters for the interfaces between atomistic layers in the local continuum approach. In such situations the gradation of parameters is sometimes proposed, either step-wise \cite{Jurczak_2019} or continuous \cite{Sevostianov2006,SEVOSTIANOV20071304}. The actual properties and their variation along the interphase layer are usually established performing atomistic simulations in a bicrystal configuration \cite{Kluge1990}. They then depend on the disorientation angle between two grains. In the present study the set of parameters that are representative for an averaged stiffness of an interphase layers between any pair of grain orientations is of interest. Therefore, alternative way of identification of the shell elastic parameters has been adopted.}
	In order to obtain elastic properties for the shells themselves, the size of some generated polycrystal samples was reduced so that the fraction of transient shell atoms {approaches unity}, $f_0\rightarrow 1$ (see Eq.(\ref{def:fsa})). For each material there are two samples with the smallest grain.

	\begin{enumerate}[I.]
		\item \textbf{FCC copper (Cu)} \label{i:Cu}
		
		The copper EAM potential parametrized by \cite{Mishin2001} was utilized. This potential reproduces the copper monocrystal FCC lattice constant $a_{FCC}$=3.615\,\AA, the cohesive energy $E_{c}$=-3.54\,eV, Zener factor $\zeta_1$=3.22, and the elastic constants in crystallographic axes coinciding with Cartesian coordinate system axes: $C_{1111}$=169.88\,GPa, $C_{1122}$=122.60\,GPa and  $C_{2323}$=76.19\,GPa. The characteristics of computational copper samples are listed in the Tab.\ref{tab:SamplesCu}.
		
		\item  \textbf{FCC aluminum (Al)} \label{i:Al}
		
		The aluminum EAM potential parametrized by \cite{Mishin1999} was utilized. This potential reproduces the aluminum monocrystal FCC lattice constant $a_{FCC}$=4.05\,\AA, the cohesive energy $E_{c}$=-3.36\,eV, Zener factor $\zeta_1$=1.21, and the elastic constants in crystallographic axes coinciding with Cartesian coordinate system axes: $C_{1111}$=113.80\,GPa, $C_{1122}$=61.56\,GPa and  $C_{2323}$=31.60\,GPa. The characteristics of computational aluminium samples are listed in the Tab.\ref{tab:SamplesAl}.

		\item \textbf{FCC nickel (Ni)} \label{i:Ni}
		
		The nickel EAM potential parametrized by \cite{Mendelev2012} was utilized. This potential reproduces the nickel monocrystal FCC lattice constant $a_{FCC}$=3.518\,\AA, the cohesive energy $E_{c}$=-4.39\,eV, Zener factor $\zeta_1$=2.46, and the elastic constants in crystallographic axes coinciding with Cartesian coordinate system axes: $C_{1111}$=247.0\,GPa, $C_{1122}$=147.29\,GPa and  $C_{2323}$=122.77\,GPa. The characteristics of computational nickel samples are listed in the Tab.\ref{tab:SamplesNi}.
		
		\item \textbf{BCC tungsten (W)} \label{i:W}
		
		The tungsten EAM potential parametrized by \cite{Marinica2013} was utilized. This potential reproduces the tungsten monocrystal BCC lattice constant $a_{BCC}$=3.14\,\AA, the cohesive energy $E_{c}$=-8.9\,eV, Zener factor $\zeta_1$=1.00, and the elastic constants in crystallographic axes coinciding with Cartesian coordinate system axes: $C_{1111}$=523.04\,GPa, $C_{1122}$=202.19\,GPa and  $C_{2323}$=160.88\,GPa. The characteristics of computational tungsten samples are listed in the Tab.\ref{tab:SamplesT}.
		
		\item \textbf{BCC iron (Fe)} \label{i:Fe}
		
		The iron EAM potential parametrized by \cite{Johnson2004} was utilized. This potential reproduces the iron monocrystal BCC lattice constant $a_{BCC}$=2.87\,\AA, the cohesive energy $E_{c}$=-4.29\,eV, Zener factor $\zeta_1$=2.48, and the elastic constants in crystallographic axes coinciding with Cartesian coordinate system axes: $C_{1111}$=229.65\,GPa, $C_{1122}$=135.50\,GPa and  $C_{2323}$=116.76 \,GPa. The characteristics of computational iron samples are listed in the Tab.\ref{tab:SamplesFe}.
		
		\item \textbf{BCC sodium (Na)} \label{i:Na}
		
		The sodium EAM potential parametrized by \cite{Wilson2015} was utilized. This potential reproduces the sodium monocrystal BCC lattice constant $a_{BCC}$=4.23\,\AA, the cohesive energy $E_{c}$=-1.11\,eV, Zener factor $\zeta_1$=8.16, and the elastic constants in crystallographic axes coinciding with Cartesian coordinate system axes: $C_{1111}$=8.26\,GPa, $C_{1122}$=6.83\,GPa and  $C_{2323}$=5.84\,GPa. The characteristics of computational sodium samples are listed in the Tab.\ref{tab:SamplesNa}.
		
		\item \textbf{BCC niobium (Nb)} \label{i:Nb}
		
		The niobium EAM potential parametrized by \cite{Fellinger2010} was utilized. This potential reproduces the niobium monocrystal BCC lattice constant $a_{BCC}$=3.31\,\AA, the cohesive energy $E_{c}$=-7.09\,eV, Zener factor $\zeta_1$=0.59, and the elastic constants in crystallographic axes coinciding with Cartesian coordinate system axes: $C_{1111}$=233.08\,GPa, $C_{1122}$=123.89\,GPa and  $C_{2323}$=32.13\,GPa. The characteristics of computational niobium samples are listed in the Tab.\ref{tab:SamplesNb}.
		
		\item \textbf{BCC vanadium (V)} \label{i:V}
		
		The vanadium EAM potential parametrized by \cite{Han2003} was utilized. This potential reproduces the vanadium monocrystal BCC lattice constant $a_{BCC}$=3.03\,\AA, the cohesive energy $E_{c}$=-5.3\,eV, Zener factor $\zeta_1$=0.79, and the elastic constants in crystallographic axes coinciding with Cartesian coordinate system axes: $C_{1111}$=227.57\,GPa, $C_{1122}$=119.10\,GPa and  $C_{2323}$=43.16\,GPa. The characteristics of computational vanadium samples are listed in the Tab.\ref{tab:SamplesV}.
	\end{enumerate}
}
\section{Results}
\label{sec:Res}

\subsection{Results of atomistic simulations}
\label{ssec:ResAS}

{All computational samples of nanocrystalline material subjected to the atomistic simulations are denoted as $$N_{\rm{UC}}-N_g-\rm{SYS}$$ where $N_{\rm{UC}}$ is a number of unit cells, $N_g$ - a number of grain orientations and $\rm{SYS}$ denotes the system of grain distribution, i.e.: BCC or random, see Tables \ref{tab:SamplesCu}--\ref{tab:SamplesV} included in the Appendix.
	Similarly to \cite{Kowalczyk18} the finite set of $N_g$ orientations has been considered, namely polycrystalline representative volumes composed of grains with 16, 54, 125, 128 or 250 randomly selected orientations, defined in terms of Euler angles, have been analysed. 
	Detailed results, in the form of the full elasticity tensors $\bar{\mathbb{C}}$\,, derived from molecular simulations of analysed samples for eight cubic metals are listed in the Tables \ref{tab:Cij-c}, \ref{tab:Cij-cAl}, \ref{tab:Cij-cFe}, \ref{tab:Cij-cNa}, \ref{tab:Cij-cNb}, \ref{tab:Cij-cNi}, \ref{tab:Cij-cW} and \ref{tab:Cij-cV}, respectively. As it was concluded in the previous study devoted to the nanocrystalline copper the system of grain distribution and the number of orientations in the sample have secondary effect on the magnitude of elastic stiffness, while the primary effect is related to the grain size equivalent to a number of atoms per grain \cite{Kowalczyk18}.}
\begin{figure}[!h]
	\centering
	\begin{tabular}{ccc}
		\textbf{Aluminum (FCC)}& & \\
		\includegraphics[width=0.34\linewidth]{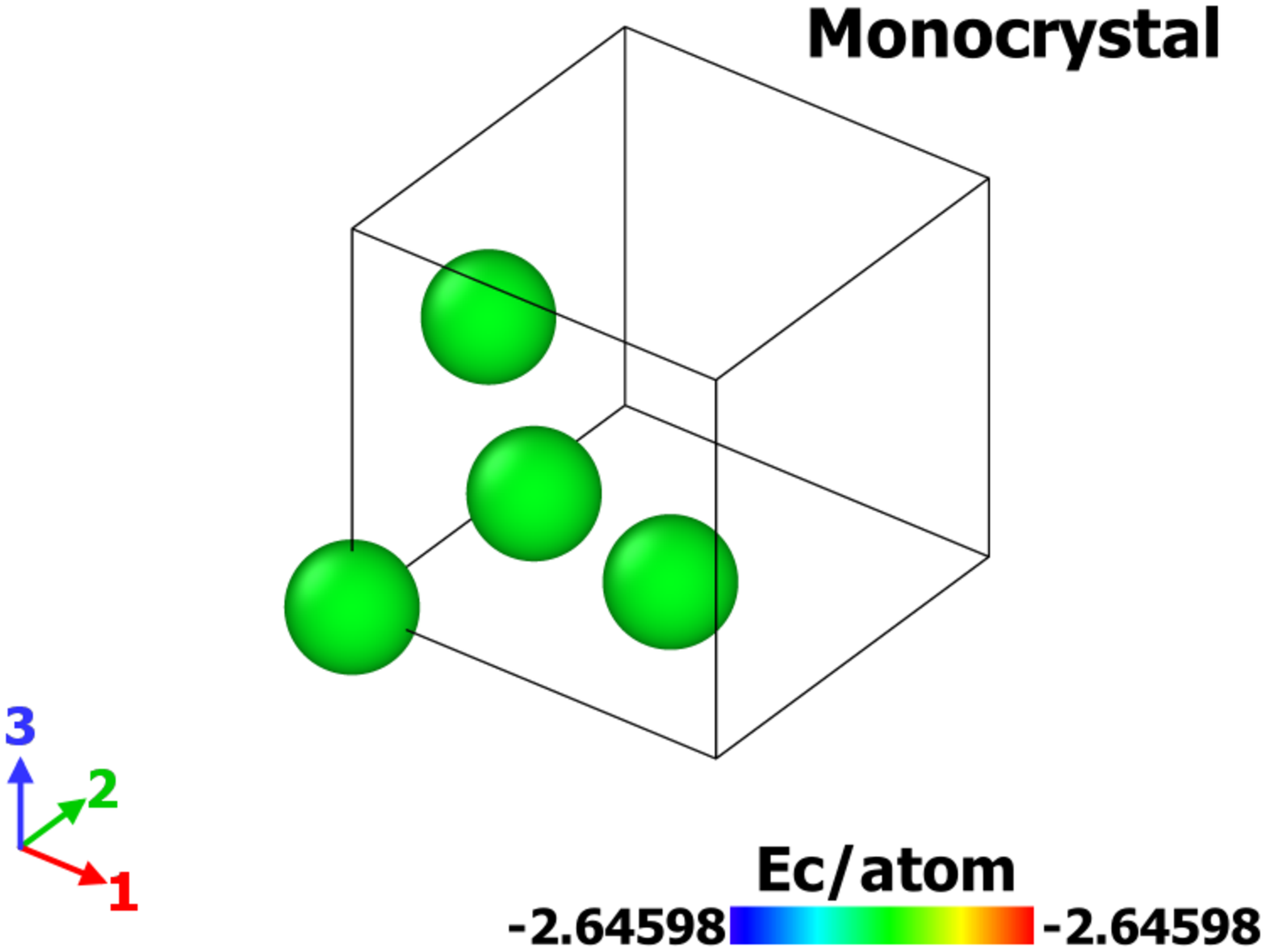} &
		\includegraphics[width=0.34\linewidth]{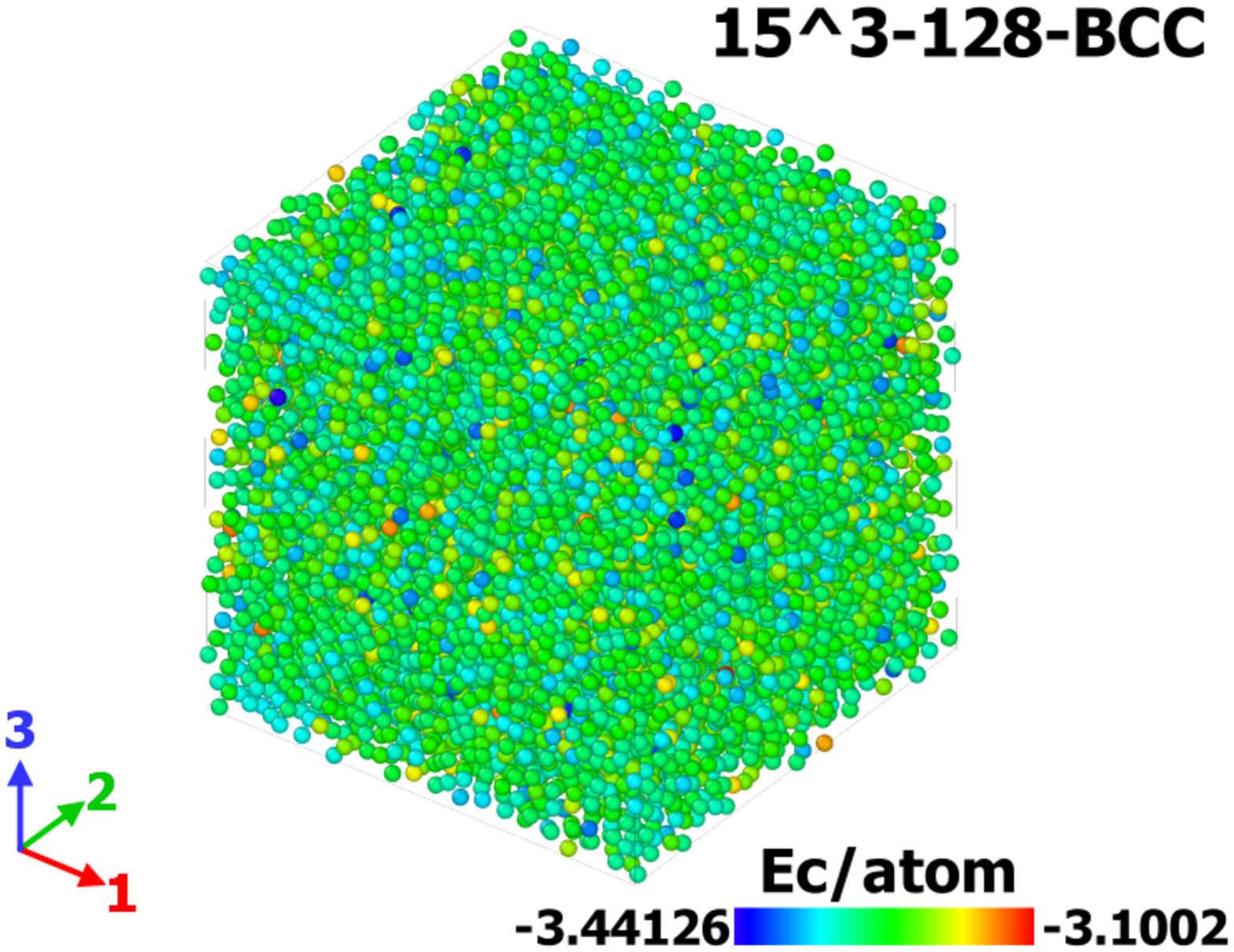} &
		\includegraphics[width=0.34\linewidth]{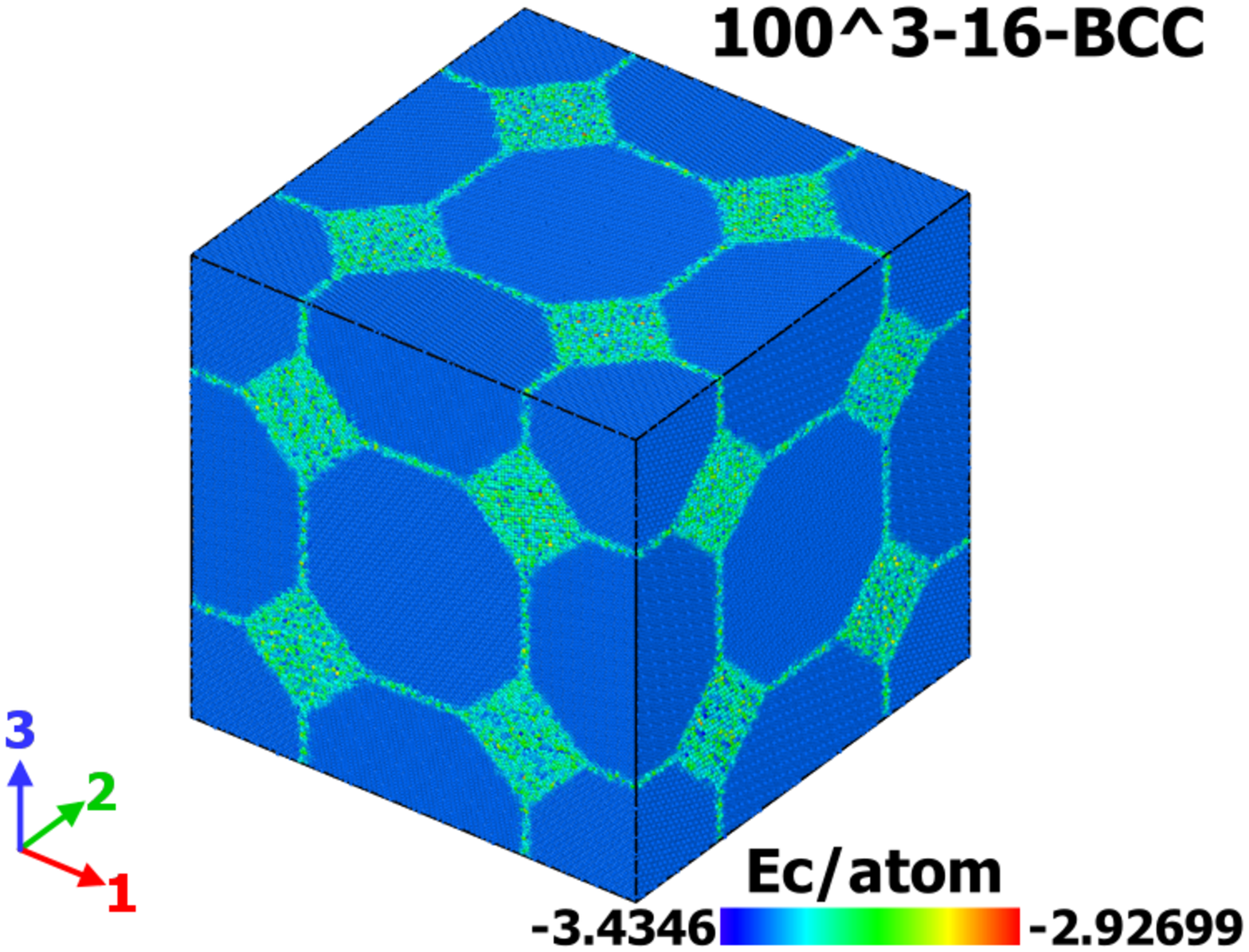} \\
		\textbf{Niobium (BCC)}& &\\
		\includegraphics[width=0.34\linewidth]{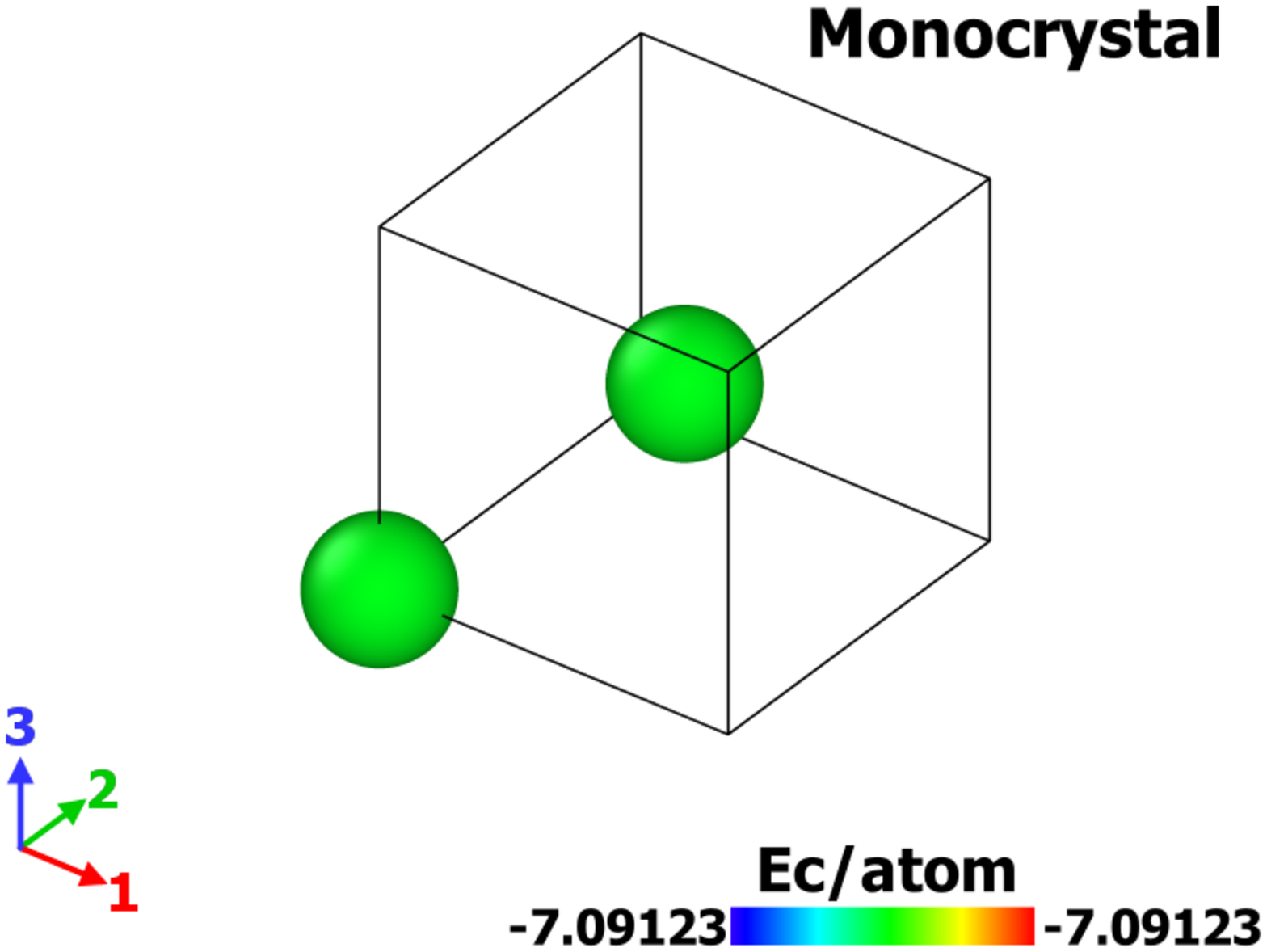} &
		\includegraphics[width=0.34\linewidth]{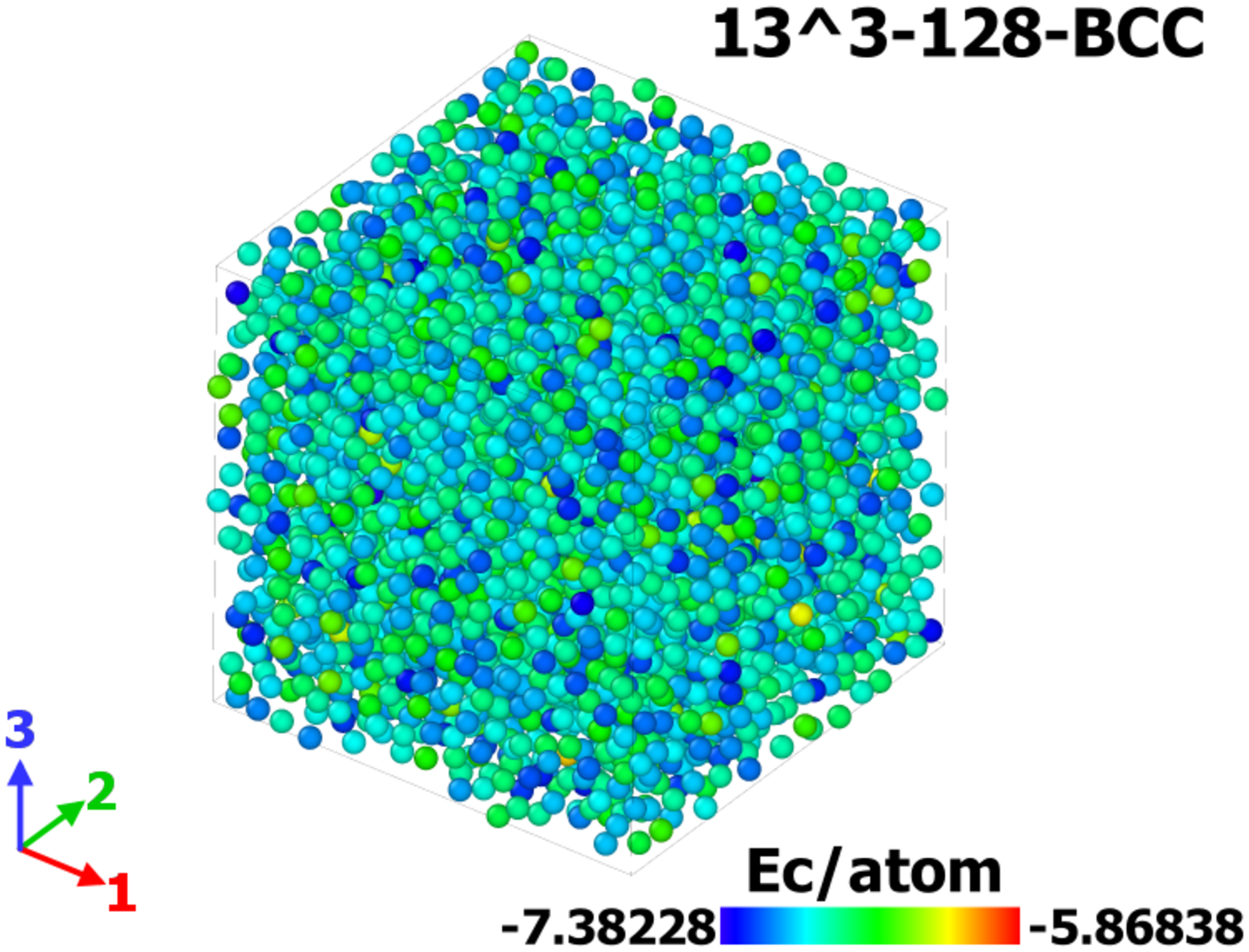} &
		\includegraphics[width=0.34\linewidth]{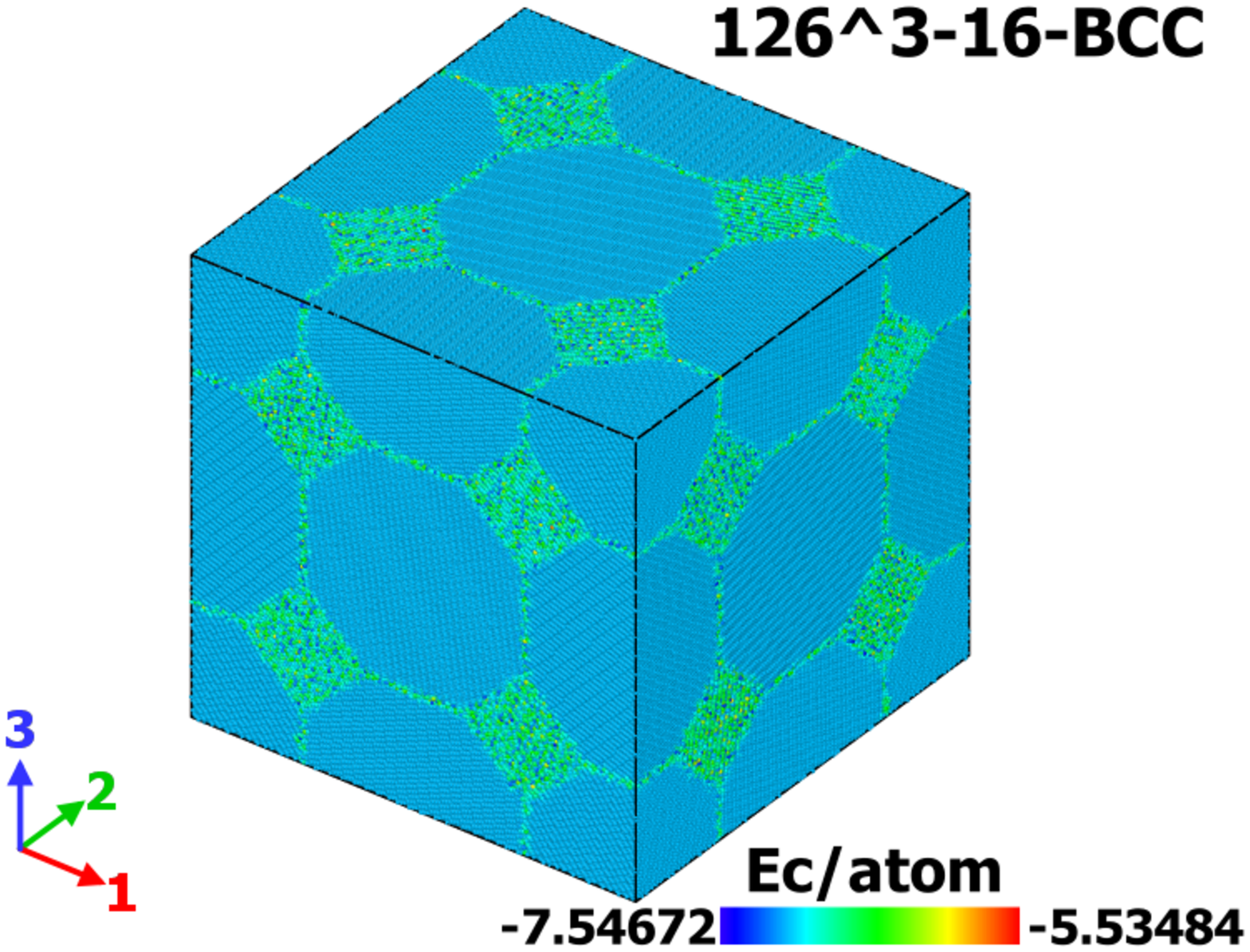}
	\end{tabular}
	\caption{Visualization of selected atomistic computational samples and cohesive energy $E_{c}$\,(eV/atom) for aluminium and niobium.  }
	\label{fig:Samples}
\end{figure}

Selected atomistic computational samples and cohesive energy $E_{c}$\,(eV/atom) are visualized in Fig. \ref{fig:Samples}. As expected, with a decreasing average grain size, the fraction of transient shell atoms in the sample rises. It can be seen that for the samples with a smallest ratio $N_{\rm{UC}}/N_g$ almost all atoms belong to this zone. As $N_{\rm{UC}}/N_g$ increases samples can be described as composed of crystalline cores of monocrystal pattern surrounded by amorphous sheaths. It is consistent with the idea of a core-shell model recalled in Sec.\ref{sec:Cont} (see also \cite{Palosz2002}). As already mentioned, in conjunction with atomistic simulations, the shell thickness $\Delta$ can be approximately assessed as equal to \textit{the cutoff radius} of the used potential. The corresponding values of $d$ and $f_0$ calculated by formula (\ref{def:fsa}) for analysed samples are collected in Tables \ref{tab:SamplesCu}, \ref{tab:SamplesAl}, \ref{tab:SamplesNi}, \ref{tab:SamplesT}, \ref{tab:SamplesFe}, \ref{tab:SamplesNa}, \ref{tab:SamplesNb} and \ref{tab:SamplesV}  in \ref{App}.

By the assumption that the orientation distribution in the considered samples is random, so it is justified to find the closest isotropic approximation of the calculated elasticity tensors collected in Tables \ref{tab:Cij-c}-\ref{tab:Cij-cV} in \ref{App}.

{The closest isotropic approximation $\bar{\mathbb{C}}^{\mathcal{L}}_{\rm{iso}}$ of anisotropic $\bar{\mathbb{C}}$ is calculated on the basis of the Log-Euclidean metric \cite{Moakher06}, namely 
	\begin{equation}\label{Eq:IsoBulkShear}
	\bar{\mathbb{C}}^{\mathcal{L}}_{\rm{iso}}=\underbrace{\exp\left(\frac{1}{3}\mathbf{1}\cdot\mathbb{Z}\cdot\mathbf{1}\right)}_{=3\bar{K}^{\mathcal{L}}_{\rm{iso}}}\mathbb{I}^{\rm{P}}+\underbrace{\exp\left(\frac{1}{5}\mathbb{I}^{\rm{D}}\cdot\mathbb{Z}\right)}_{=2\bar{G}^{\mathcal{L}}_{\rm{iso}}}\mathbb{I}^{\rm{D}}\,\quad\textrm{where}\quad
	\mathbb{Z}=\mathrm{Log}\bar{\mathbb{C}}\,.
	\end{equation}
	The two	scalars $\bar{K}^{\mathcal{L}}_{\rm{iso}}$ and $\bar{G}^{\mathcal{L}}_{\rm{iso}}$ will be next compared with the estimates of isotropic bulk and shear moduli obtained by two variants of the core-shell model.} 
The error $\zeta_2$ related to the proposed isotropic approximation  is defined by a normalized difference between $\bar{\mathbb{C}}^{\mathcal{L}}_{\rm{iso}}$ and the actual $\bar{\mathbb{C}}$. It is defined as \cite{Kowalczyk11b}
\begin{equation}\label{Eq:zeta2}
\zeta_2=\frac{||\mathrm{Log}\bar{\mathbb{C}}-\mathrm{Log}\bar{\mathbb{C}}^{\mathcal{L}}_{\rm{iso}}||}{||\mathrm{Log}\bar{\mathbb{C}}||}\times 100\%\geq 0\,,
\end{equation}
where $||\mathbb{A}||=\sqrt{\mathbb{A}\cdot\mathbb{A}}=\sqrt{A_{ijkl}A_{ijkl}}$ and $\mathrm{Log}\mathbb{A}=\sum_{K}\mathrm{log}\lambda_L\mathbb{P}_K$ ($\lambda_K$ - eigenvalues of $\mathbb{A}$, $\mathbb{P}_K$ - eigenprojectors of $\mathbb{A}$ obtained by its spectral decomposition). 
Note that $\zeta_2$ can be also used as an anisotropy measure of $\bar{\mathbb{C}}$.

The isotropized bulk and shear moduli obtained by formula (\ref{Eq:IsoBulkShear}), together with the anisotropy factor (\ref{Eq:zeta2}), are collected in Tables \ref{tab:SamplesCuAlNi} and \ref{tab:SamplesNaFeWVNb} for the analysed FCC and BCC cubic metals, respectively. In the tables, in order to facilitate analysis samples are ordered according to the increasing averaged grain size.  
For each metal the shell elastic parameters: $K_s$ and $G_s$, were identified as an average value for two  samples with $f_0$ approaching unity.  The identified values, together with the Kelvin moduli of monocrystals and the \emph{cutoff radius} $\Delta$ of the applied atomistic potential are collected in Table \ref{tab:Shell-CutOff}. Metals in the table are ordered according to the decreasing Zener parameter. Interestingly, it can be observed that for the Zener parameter higher than one the identified shell bulk modulus $K_{\rm{s}}$ is smaller than the bulk modulus of single crystal $K$, while a reverse situation takes place for the Zener parameter smaller than one. {It is also worth noting that the elastic parameters for the shell for both sample sizes, i.e. averaged over different volumes,  are almost identical, which means that resignation from the gradation of elastic parameters within the shell is justified.} As it will be shown in the next subsection similar quantitative difference is seen for the shear modulus of shell (grain boundary zone) and the coarse-grained polycrystal.  A common trend observed when analysing Tables \ref{tab:SamplesCuAlNi} and \ref{tab:SamplesNaFeWVNb} is that the elastic moduli increase with a grain size for metals with Zener parameter larger than one (i.e. for sodium, copper, iron and nickel) and decrease with increasing grain size in an opposite case (i.e. for vanadium and niobium). A rule is not clear for tungsten with the Zener parameter approximately equal one for which the bulk modulus decreases, while shear modulus increases with an increasing average grain size. Note that the lattice geometry (FCC vs. BCC) seems not to have such a qualitative influence on the present results.  
In conclusion the simplifying assumption accepted for nanocrystalline copper in \cite{Kowalczyk18} that $K_s=K$ and  $G_s=\min\{G_1,G_2\}$ is not valid for other cubic metals. 

\begin{table}[H] 
	\caption{The overall isotropized bulk and shear moduli $\bar{K}^{\mathcal{L}}_{\rm{iso}}$\,[GPa] and $\bar{G}^{\mathcal{L}}_{\rm{iso}}$\,[GPa] and anisotropy factor $\zeta_2$\,[\%] calculated for the effective stiffness tensors resulting from the atomistic simulations for metals of FCC lattice geometry. Samples are ordered according to the increasing average grain size $d$, while metals according to the decreasing Zener parameter}
	\label{tab:SamplesCuAlNi}
	\centering
	\renewcommand{\arraystretch}{1.5}
	\tiny 
	\begin{tabular}{|c c c c c c c c c c|}
		\hline
		Sample	&\multicolumn{3}{c}{Cu}&\multicolumn{3}{c}{Ni}&\multicolumn{3}{c|}{Al}\\  
		& $\bar{K}^{\mathcal{L}}_{\rm{iso}}$ & $\bar{G}^{\mathcal{L}}_{\rm{iso}}$ & $\zeta_2$ &  $\bar{K}^{\mathcal{L}}_{\rm{iso}}$ & $\bar{G}^{\mathcal{L}}_{\rm{iso}}$ & $\zeta_2$  & $\bar{K}^{\mathcal{L}}_{\rm{iso}}$ & $\bar{G}^{\mathcal{L}}_{\rm{iso}}$ & $\zeta_2$\\ 	
		10$^3$-128-BCC & 134.9 & 21.99 & 1.80  & 91.23  & 31.44 & 1.39 & 74.22 & 15.51 & 1.46\\
		15$^3$-128-BCC &  135.2 & 22.14 & 1.44 & 92.73 & 32.04 &  1.10 & 73.31& 15.18& 1.10\\
		50$^3$-250-BCC& 137.2 & 27.69 & 1.85 & 125.7 & 47.00 & 1.65 & 74.84 & 18.38& 3.02\\
		50$^3$-128-BCC & 137.0 & 29.50 & 1.71 & 133.9  & 50.71 & 1.94  & 75.22 & 21.65 & 1.06\\
		50$^3$-125-Random & 136.6 & 29.91 & 2.16 & 131.5 & 49.66 & 2.09 & 75.20 & 19.55& 2.87\\
		50$^3$-54-BCC& 136.4 & 33.65 & 2.38 & 142.5 & 50.70 & 3.33 &76.02& 20.85 & 2.58\\
		50$^3$-16-BCC &	137.5 & 36.43 & 2.86 & 153.4 & 59.65 & 4.23 &77.29 & 24.22 & 1.91\\
		100$^3$-16-BCC& 137.9 & 40.26 & 2.60 & 166.2 & 70.07 & 2.05 & 77.69 & 26.22 & 0.81\\
		\hline 
	\end{tabular}
\end{table}

\begin{table}[H] 
	\caption{The overall isotropized bulk and shear moduli $\bar{K}^{\mathcal{L}}_{\rm{iso}}$\,[GPa] and $\bar{G}^{\mathcal{L}}_{\rm{iso}}$\,[GPa] and anisotropy factor $\zeta_2$\,[\%] calculated for the effective stiffness tensors resulting from the atomistic simulations for metals of BCC lattice geometry. Samples are ordered according to the increasing average grain size $d$, while metals according to the decreasing Zener parameter}
	\label{tab:SamplesNaFeWVNb}
	\centering
	\renewcommand{\arraystretch}{1.5}
	\tiny 
	\begin{tabular}{|c c c c c c c c c c c c c c c c|}
		\hline
		Sample	&\multicolumn{3}{c}{Na}&\multicolumn{3}{c}{Fe}&\multicolumn{3}{c}{W}&\multicolumn{3}{c}{V}&\multicolumn{3}{c|}{Nb}\\  
		& $\bar{K}^{\mathcal{L}}_{\rm{iso}}$ & $\bar{G}^{\mathcal{L}}_{\rm{iso}}$ & $\zeta_2$ &  $\bar{K}^{\mathcal{L}}_{\rm{iso}}$ & $\bar{G}^{\mathcal{L}}_{\rm{iso}}$ & $\zeta_2$  & $\bar{K}^{\mathcal{L}}_{\rm{iso}}$ & $\bar{G}^{\mathcal{L}}_{\rm{iso}}$ & $\zeta_2$& $\bar{K}^{\mathcal{L}}_{\rm{iso}}$ & $\bar{G}^{\mathcal{L}}_{\rm{iso}}$ & $\zeta_2$& $\bar{K}^{\mathcal{L}}_{\rm{iso}}$ & $\bar{G}^{\mathcal{L}}_{\rm{iso}}$ & $\zeta_2$\\ 	
		X$^3$-128-BCC & 6.718 & 1.667 & 3.72 & 134.2 & 42.77& 1.00 &  325.9& 84.65& 2.14 &180.8 &65.65 &3.31 &177.5 &44.58 & 0.52\\
		Y$^3$-128-BCC & 6.727  &1.736  & 2.45 & 134.00& 40.95& 1.23 & 82.85 & 327.7 & 0.84 &185.4 &65.13 &2.60 &177.4 & 45.77& 0.37\\
		63$^3$-250-BCC& 7.093 & 1.952 & 5.39 & 147.8& 54.66& 1.04& 309.8  &109.1 &1.88 & 166.1 &51.17 &1.63 &167.5 &40.86 & 0.77\\
		63$^3$-128-BCC & 7.165 & 2.092 & 3.65 & 150.1& 57.08 & 1.23& 310.0& 117.6& 1.34 &165.6 &51.23 &0.97 &167.2 &40.87 & 0.64\\
		63$^3$-125-Random & 7.157 & 1.948 & 7.88  & 153.1 & 55.21& 2.92& 312.1&117.7 & 0.43 &161.4 &52.25 &2.16 &167.7 &40.80 & 0.74\\
		63$^3$-54-BCC& 7.234 & 2.216 & 8.68 & 154.3 &  64.11&  1.34&  307.2& 115.8& 3.47&162.6 &51.34 &1.54 &166.2 & 41.69 & 0.45\\
		63$^3$-16-BCC & 7.265 & 2.361  & 10.05 & 158.1 & 65.52& 2.50& 309.1& 118.7 &2.84 &162.9 &49.80 &0.96 &164.3 &40.99 & 0.81\\
		126$^3$-16-BCC& 7.293  & 2.531 & 10.44 & 161.8 &73.80 & 2.10& 309.2& 143.8 &0.49 &159.0 &49.39 &0.18 & 162.2&40.21 & 1.04\\
		\hline 
	\end{tabular}
\end{table}

\begin{table}[H] 
	\caption{Three Kelvin moduli $K$, $G_1$ and $G_2$ of monocrystal samples, identified shell elastic moduli $K_{\rm{s}}$ and $G_{\rm{s}}$ and \emph{cutoff radius} $\Delta$ of the applied atomistic potential for analysed metals. Metals are ordered with a decreasing Zener parameter $\zeta_1$ }
	\label{tab:Shell-CutOff}
	\centering
	\renewcommand{\arraystretch}{1.5}
	\footnotesize 
	\begin{tabular}{|c c c c c c c c|}
		\hline
		Metal & $K$ & $G_1$ & $G_2$ & $\zeta_1$ & $K_s$ & $G_s$ & $\Delta$\\ 
		& [GPa] & [GPa] & [GPa] &  & [GPa] & [GPa] & [$\AA$]\\ \hline
		Na & 7.306 & 0.7159 & 5.842 & 8.16 &6.723 & 1.701&9.2 \\
		Cu & 138.4 & 23.64 & 79.19 & 3.22 &135.0   &22.06 &5.5 \\
		Fe & 166.9 & 47.07 & 116.8 & 2.48 &134.1 & 41.86&5.550\\ 
		Ni & 180.5 & 49.85 & 122.8 & 2.46 &91.98 & 31.74&6.0\\
		Al & 78.97 & 26.12 & 31.59 & 1.21 &73.77 & 15.35&6.287\\
		W  & 309.1 & 160.4 & 160.9 & 1.003 &326.8 &83.75&5.5\\	
		V & 155.25 & 54.23 & 43.15 & 0.796 &183.1 & 65.39&3.939\\
		Nb & 160.3 & 54.59 & 32.13 & 0.588 &177.5 &45.18&4.75\\ 
		\hline 
	\end{tabular}
\end{table}

\newpage

\subsection{Comparison of atomistic and mean-field estimates}
\label{ssec:CompACest}

\begin{figure}
	\centering
	\begin{tabular}{ccc}
		&bulk modulus & shear modulus\\
		a) \textbf{Na}&\includegraphics[angle=0,width=0.39\textwidth]{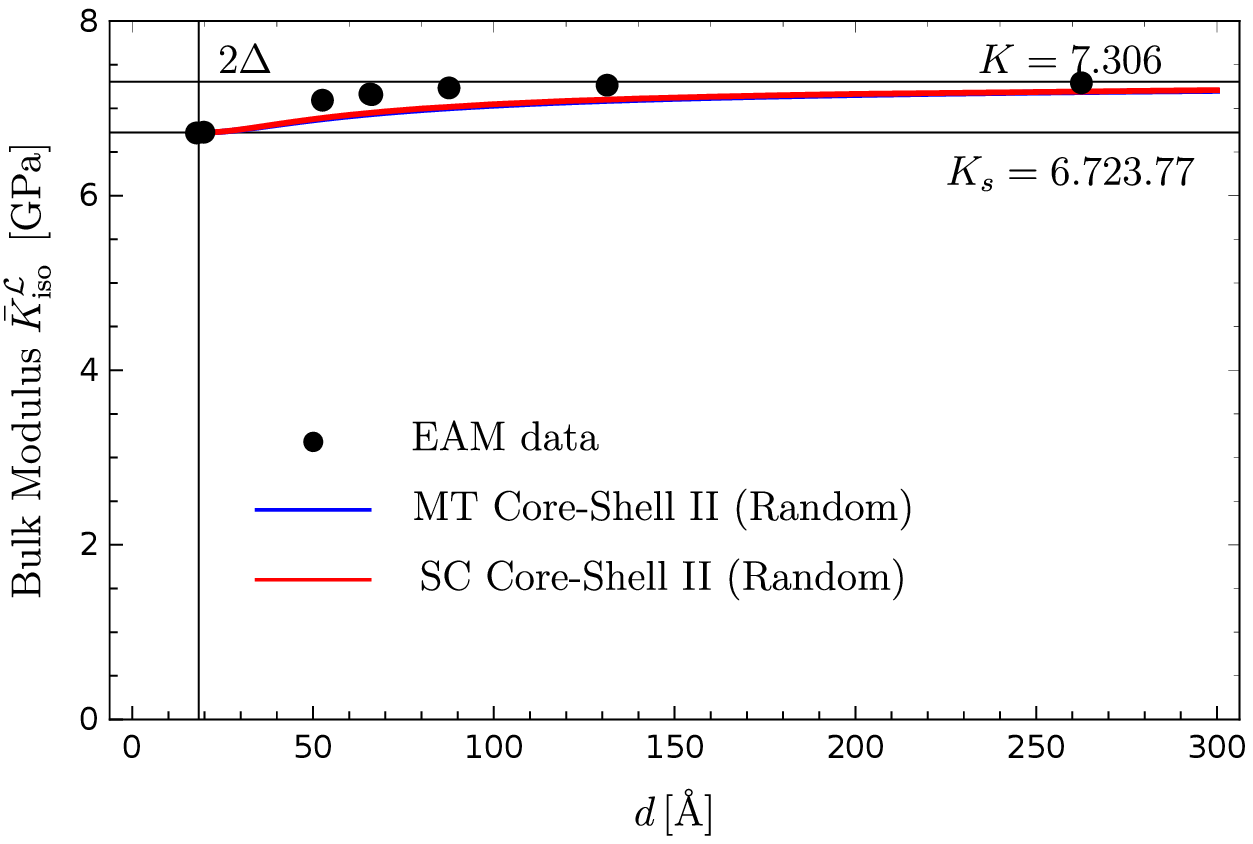}&
		\includegraphics[angle=0,width=0.39\textwidth]{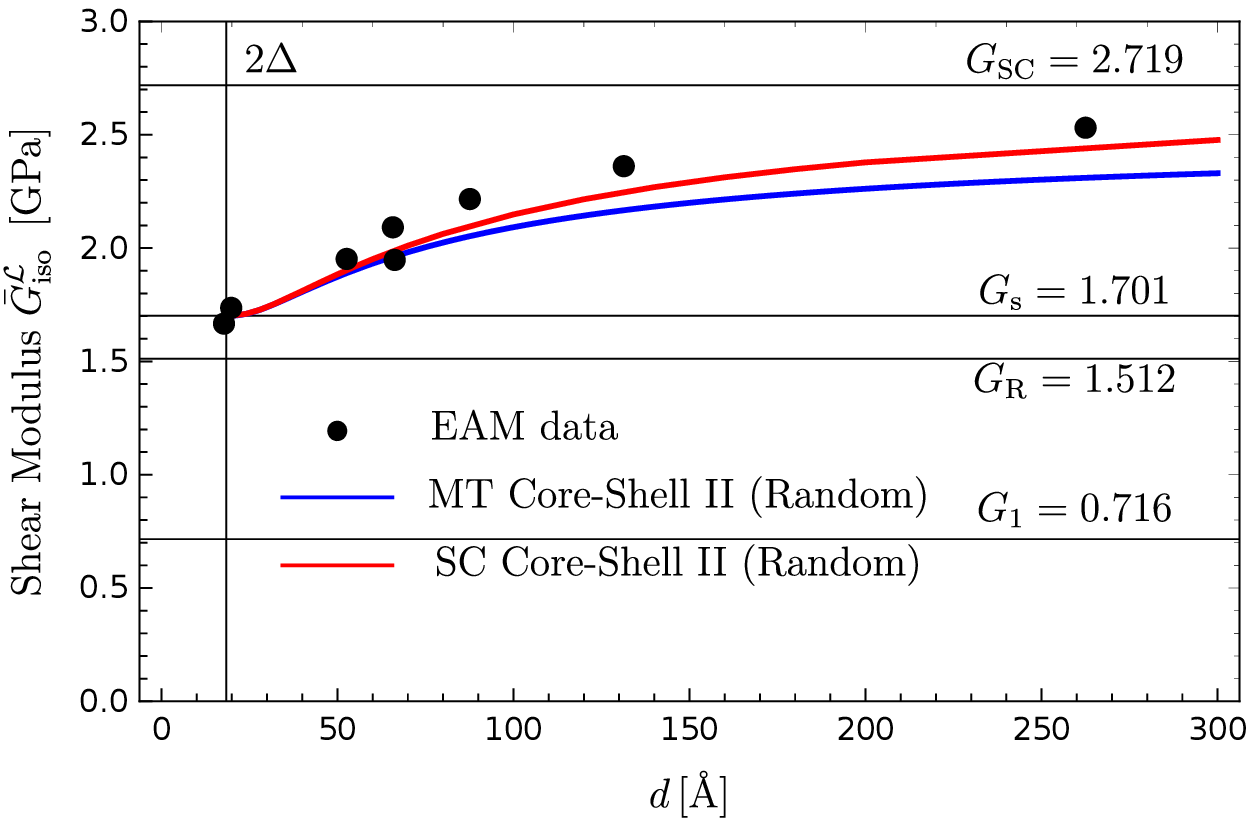}\\
		b) \textbf{Cu}&	\includegraphics[angle=0,width=0.39\textwidth]{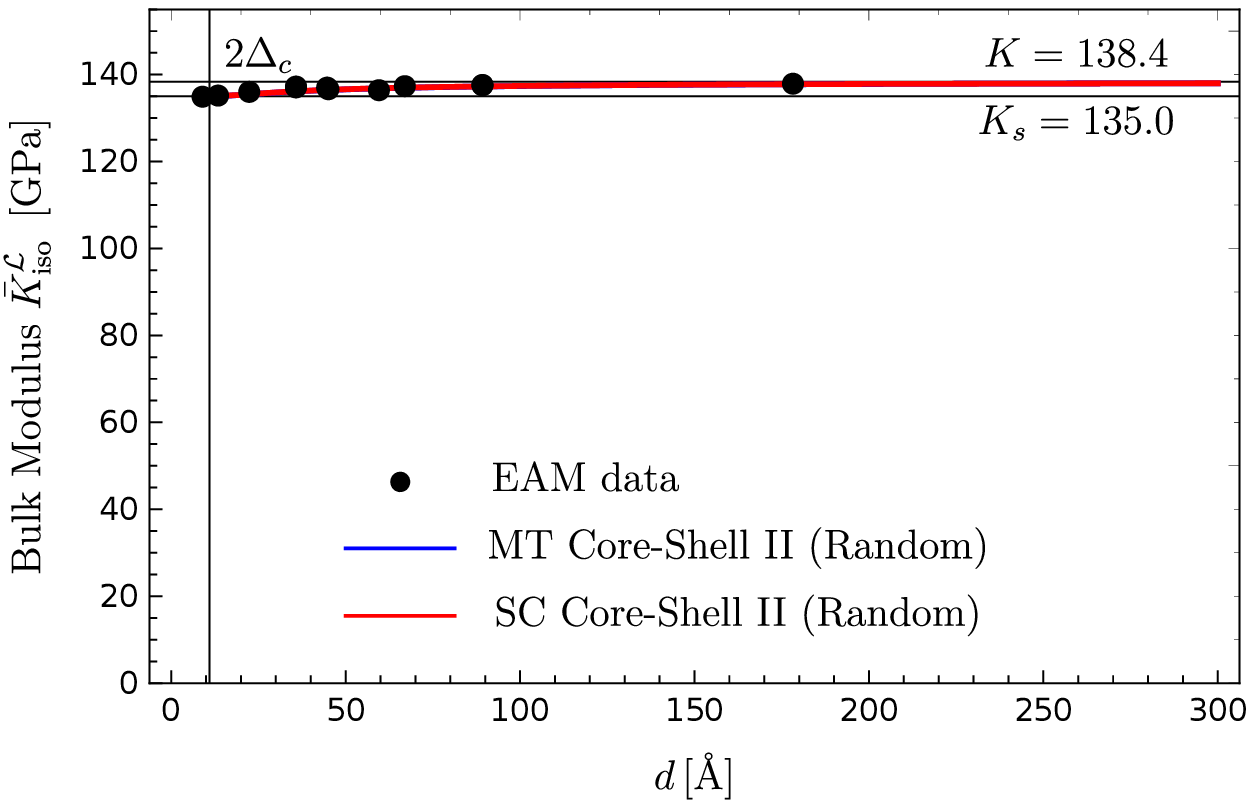}&
		\includegraphics[angle=0,width=0.39\textwidth]{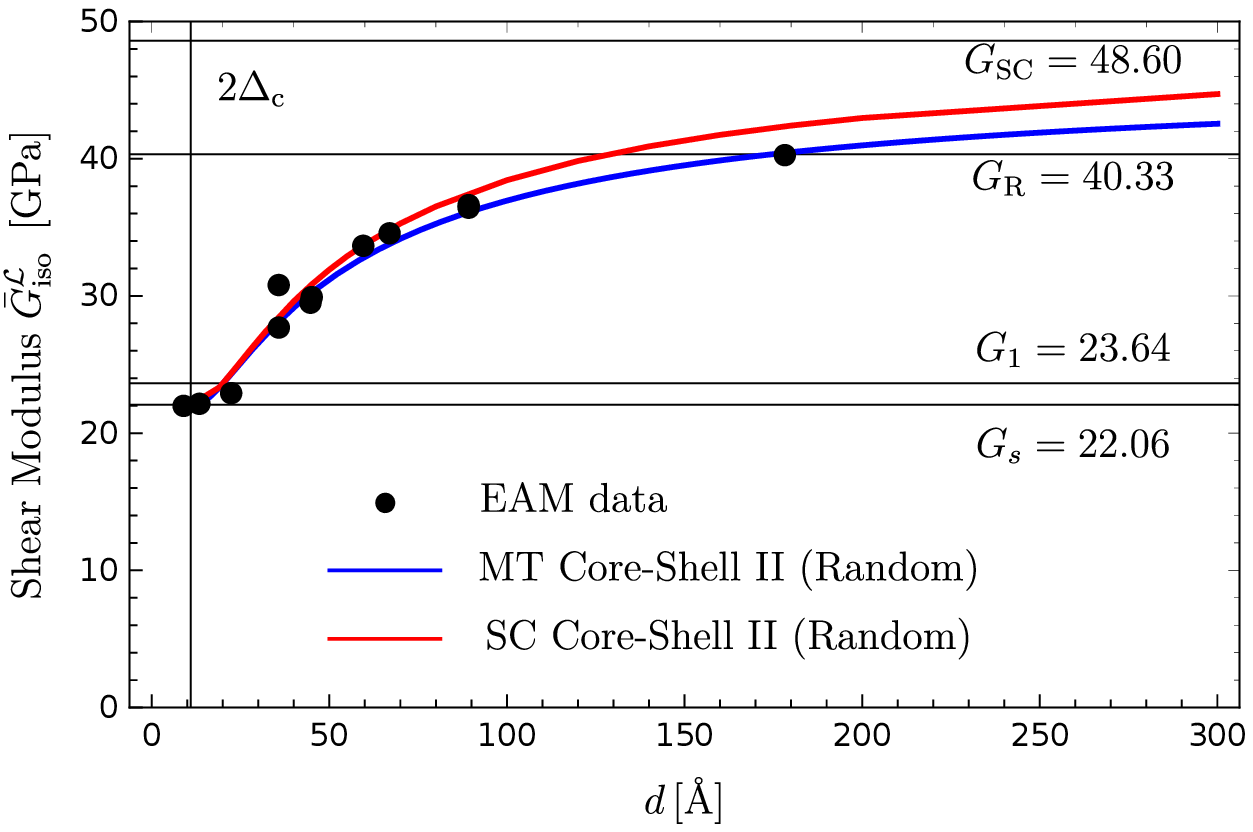}\\
		c) \textbf{Fe}&	\includegraphics[angle=0,width=0.39\textwidth]{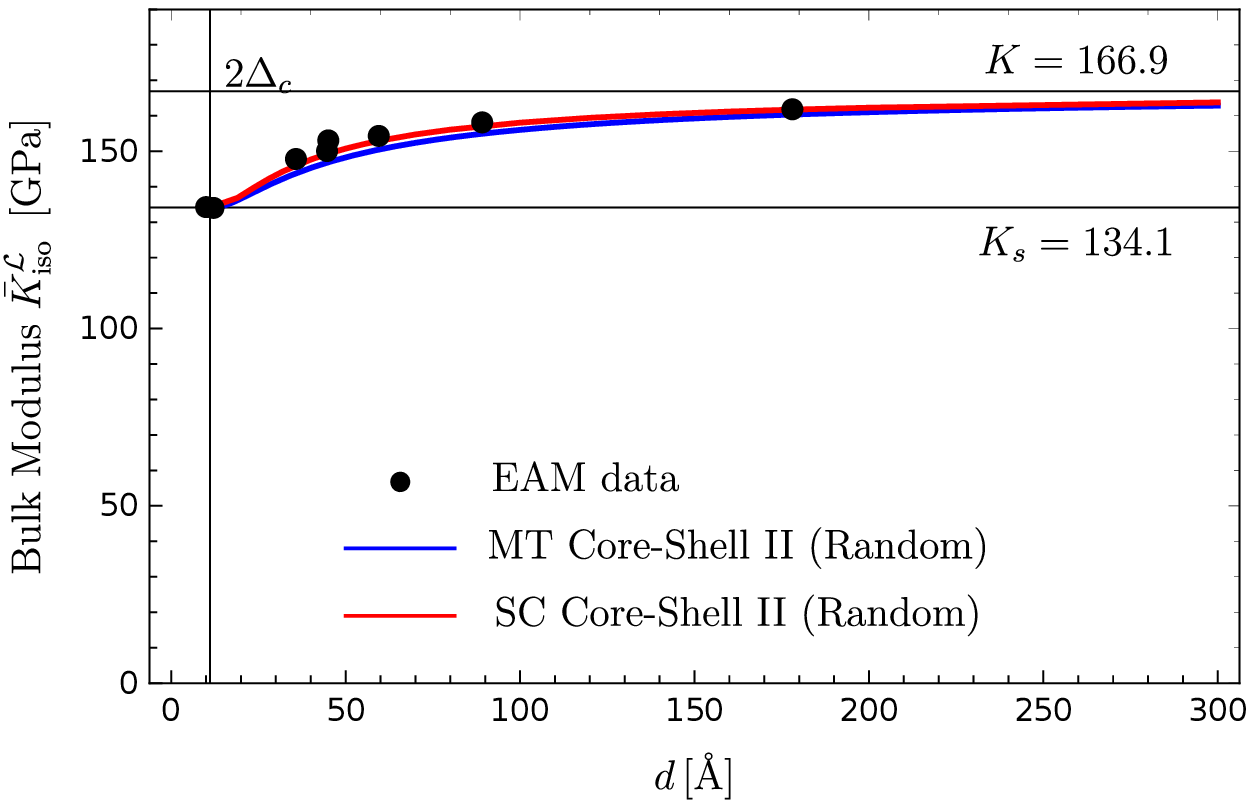}&
		\includegraphics[angle=0,width=0.39\textwidth]{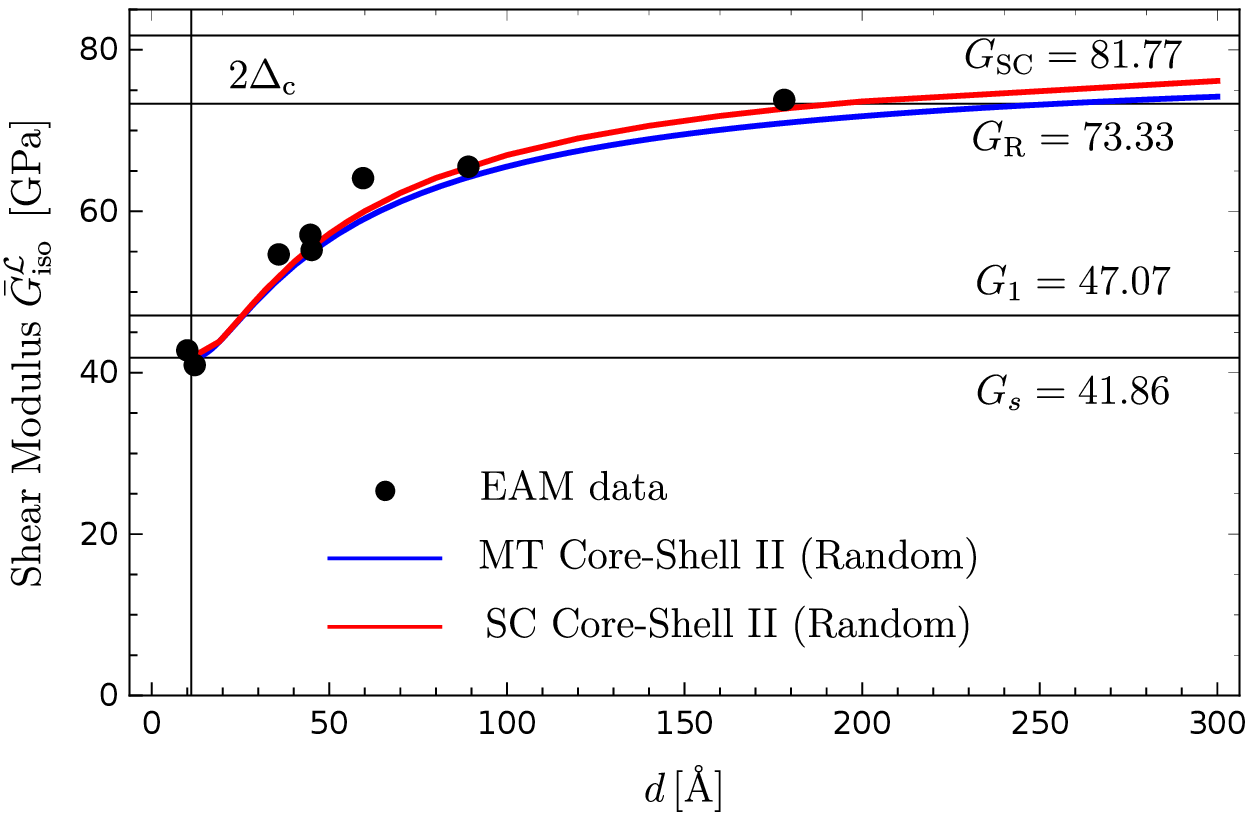}\\
		d) \textbf{Ni}& \includegraphics[angle=0,width=0.39\textwidth]{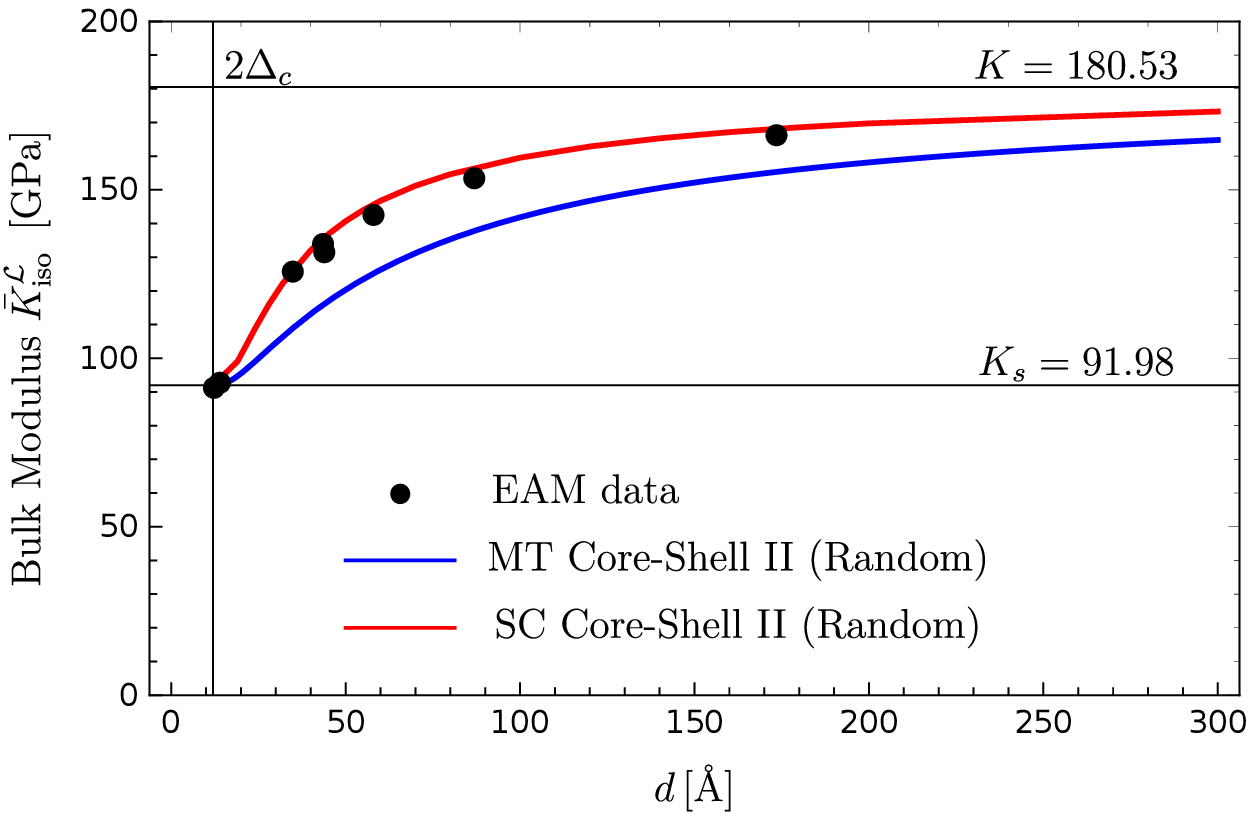}&
		\includegraphics[angle=0,width=0.4\textwidth]{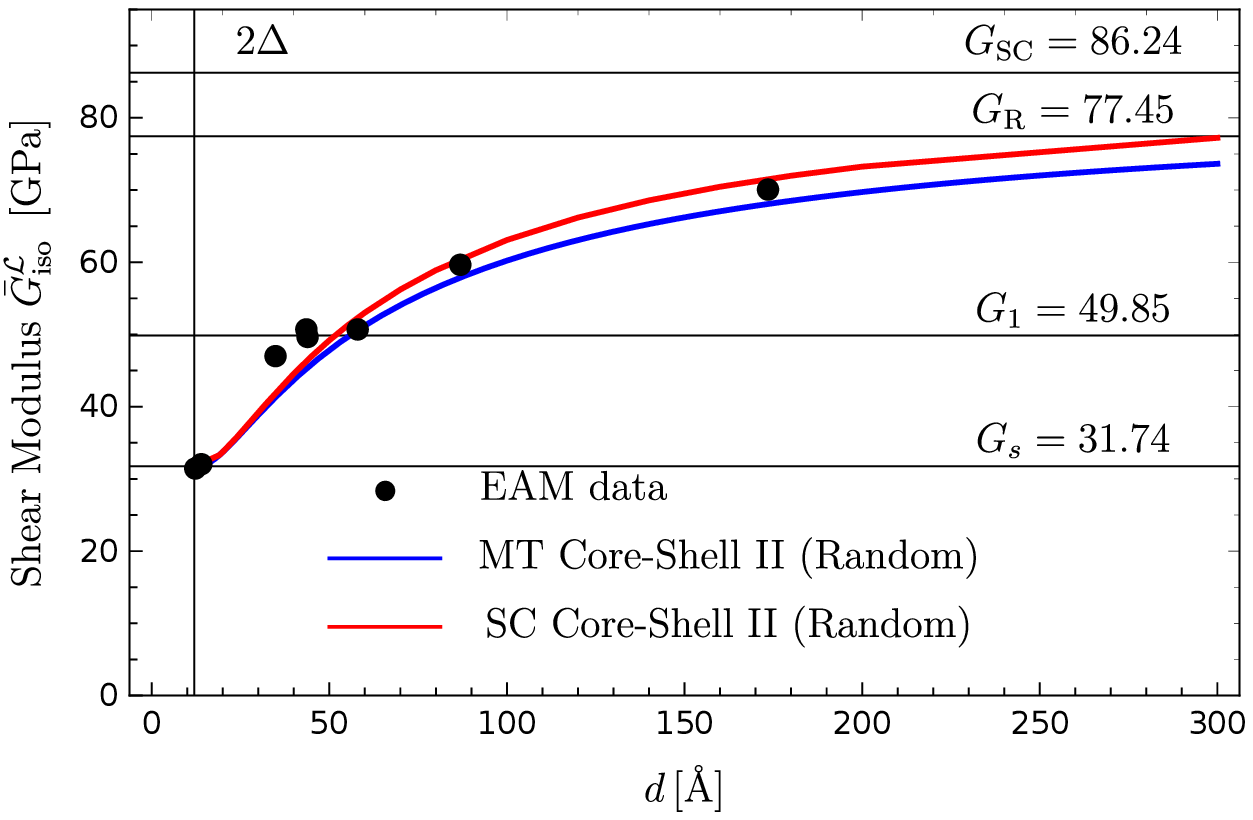}\\
		e) \textbf{Al}& \includegraphics[angle=0,width=0.39\textwidth]{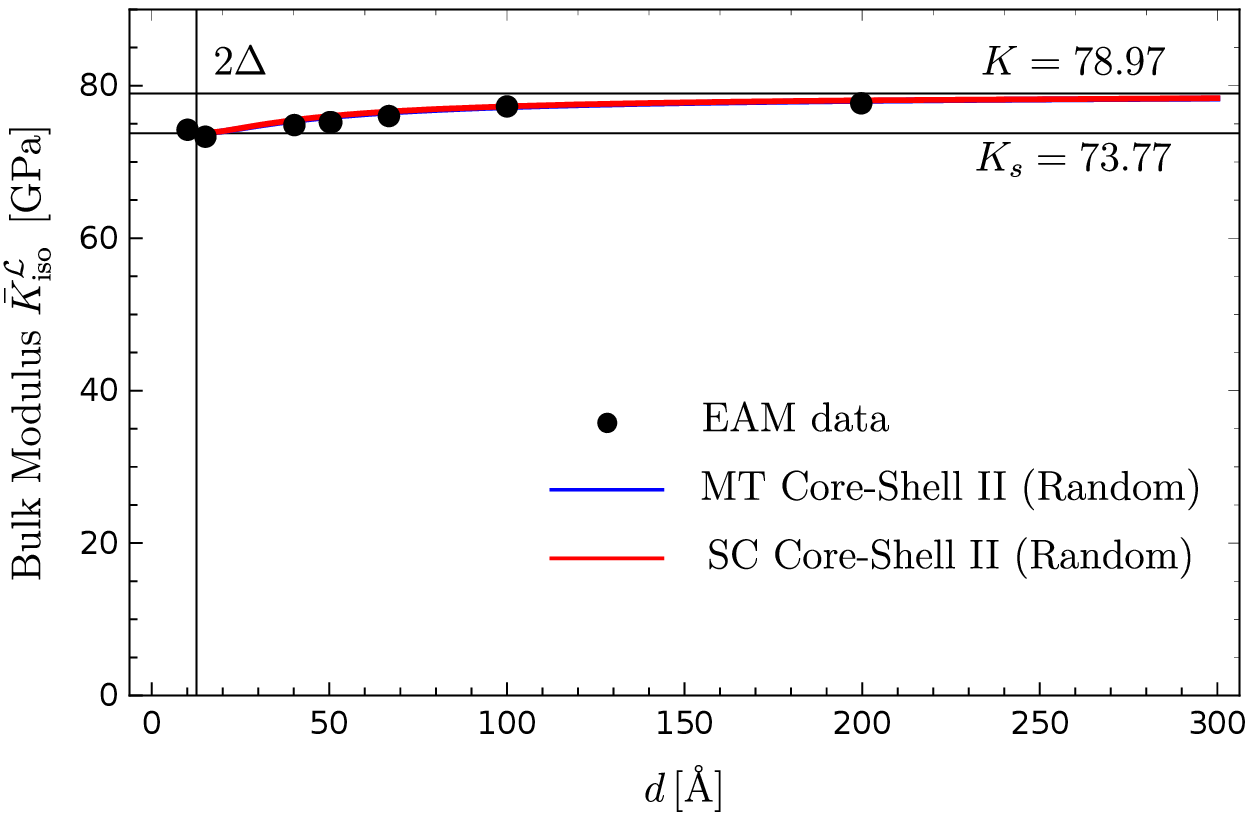}&
		\includegraphics[angle=0,width=0.39\textwidth]{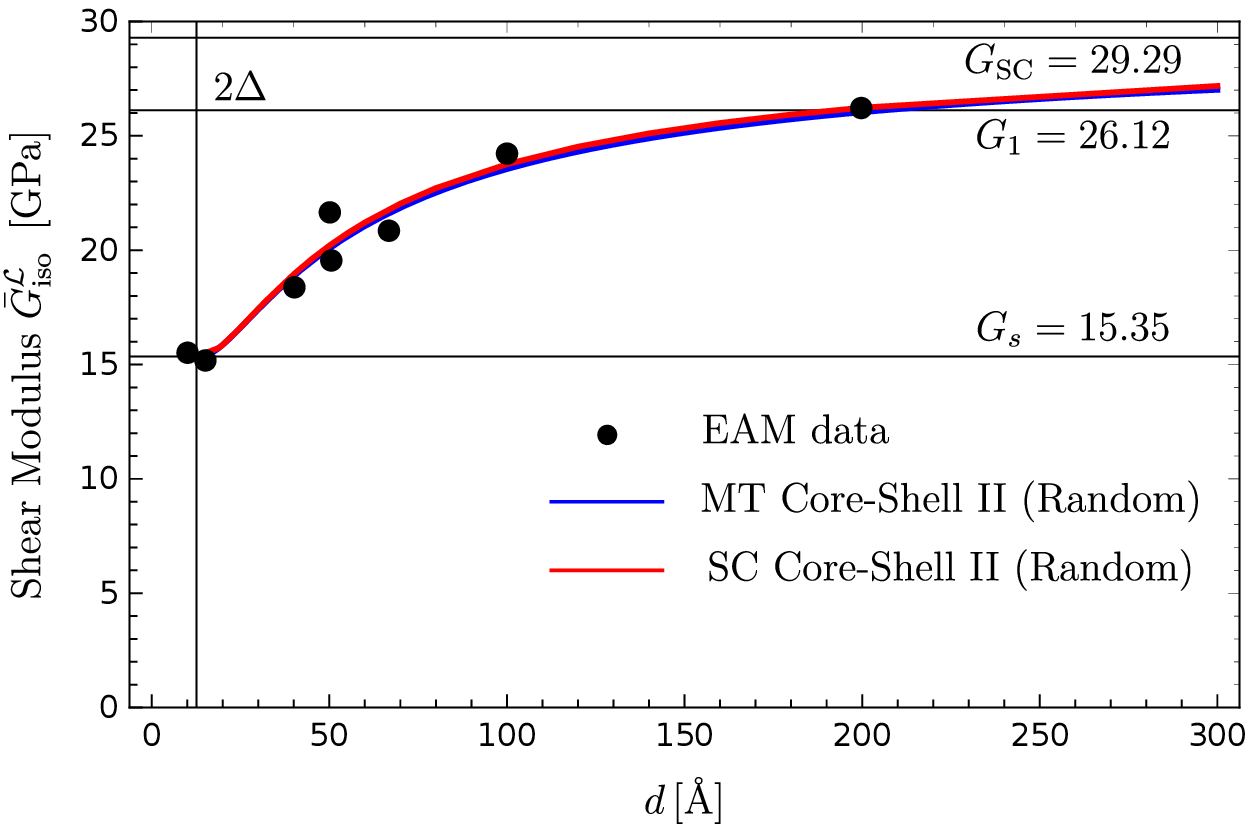}\\
	\end{tabular}
	\caption{The isotropic bulk and shear moduli $\bar{K}^{\mathcal{L}}_{\rm{iso}}$ and $\bar{G}^{\mathcal{L}}_{\rm{iso}}$ as a function of the average grain diameter $d$ by the two variants of the core-shell model - comparison with results of atomistic simulations reported in Tables \ref{tab:Cij-cNa}, \ref{tab:Cij-c}, \ref{tab:Cij-cFe}, \ref{tab:Cij-cNi}, \ref{tab:Cij-cAl} for cubic metals with a Zener parameter $\zeta_1>1$:  (a) Na, (b) Cu (c) Fe (d) Ni (f) Al.} 
	\label{fig:ZetaG1}
\end{figure}

\begin{figure}
	\centering
	\begin{tabular}{ccc}
		&bulk modulus & shear modulus\\
		a)  \textbf{W}&\includegraphics[angle=0,width=0.39\textwidth]{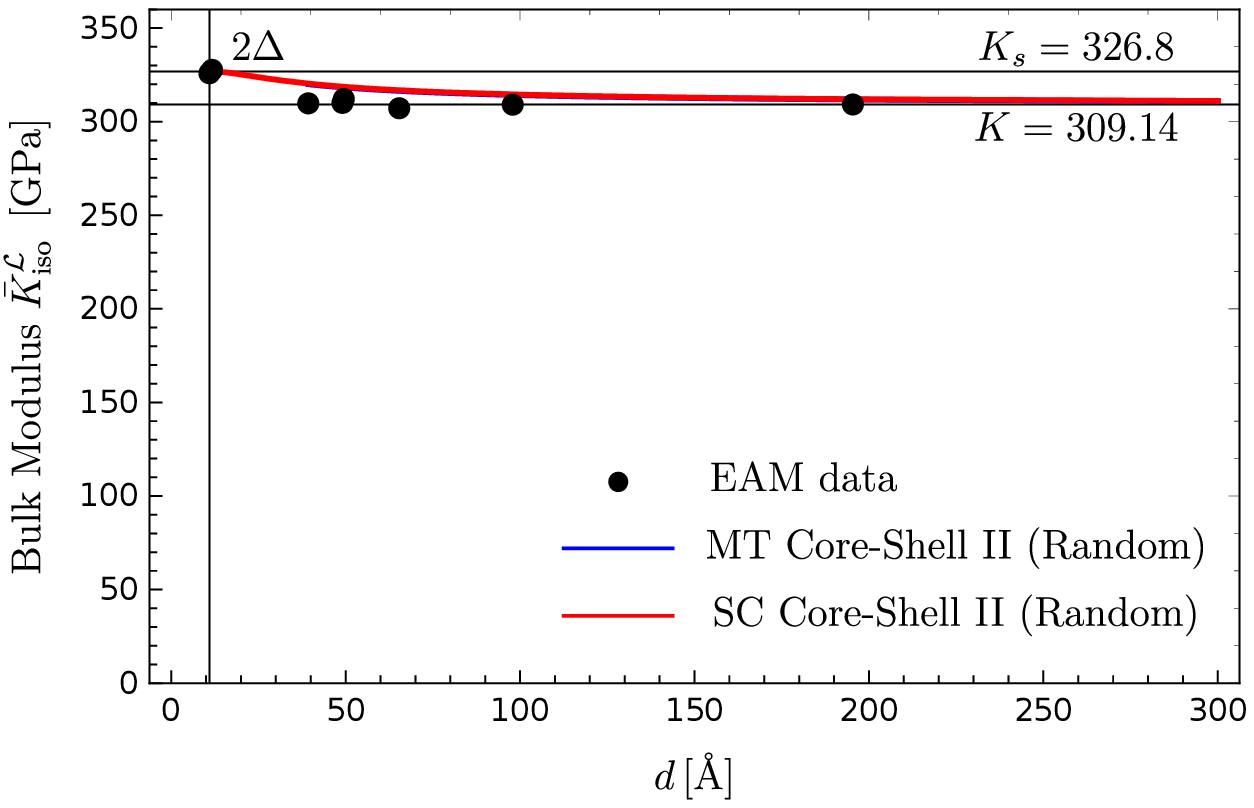}&
		\includegraphics[angle=0,width=0.39\textwidth]{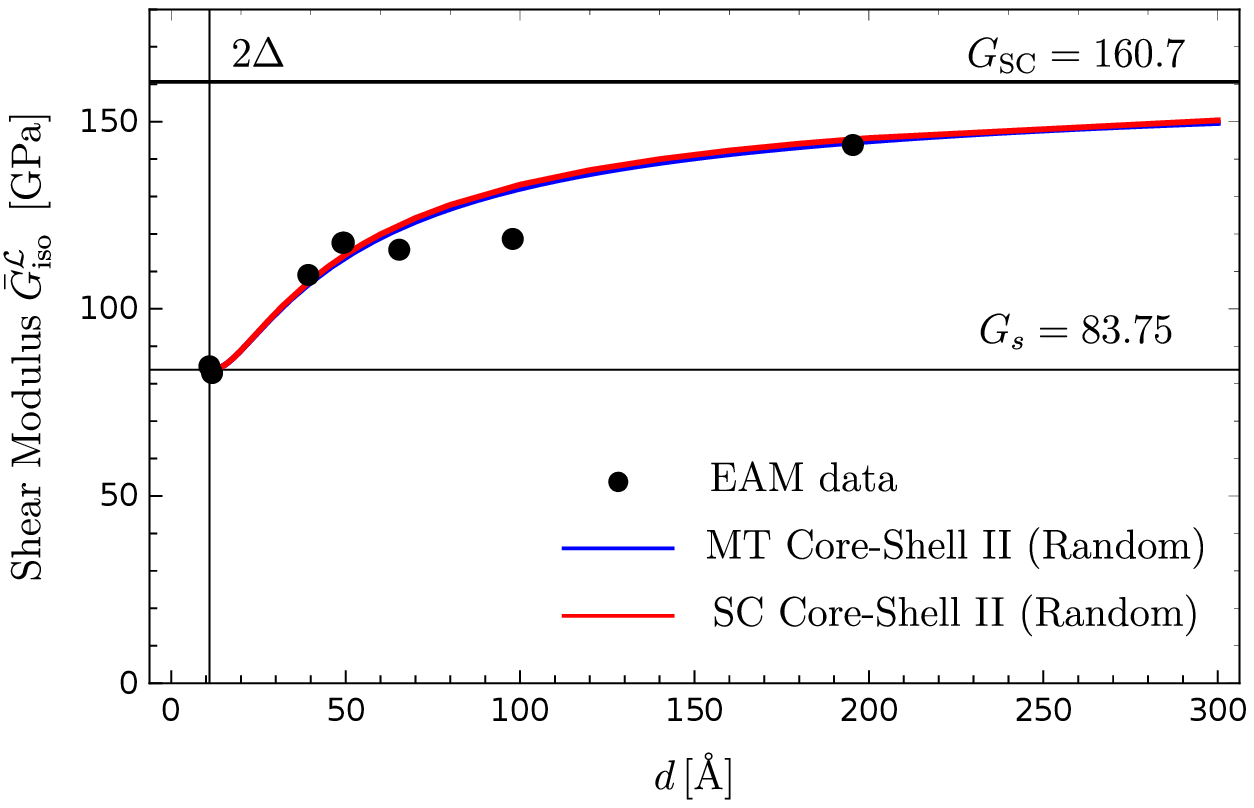}\\
		b) \textbf{V}&	\includegraphics[angle=0,width=0.39\textwidth]{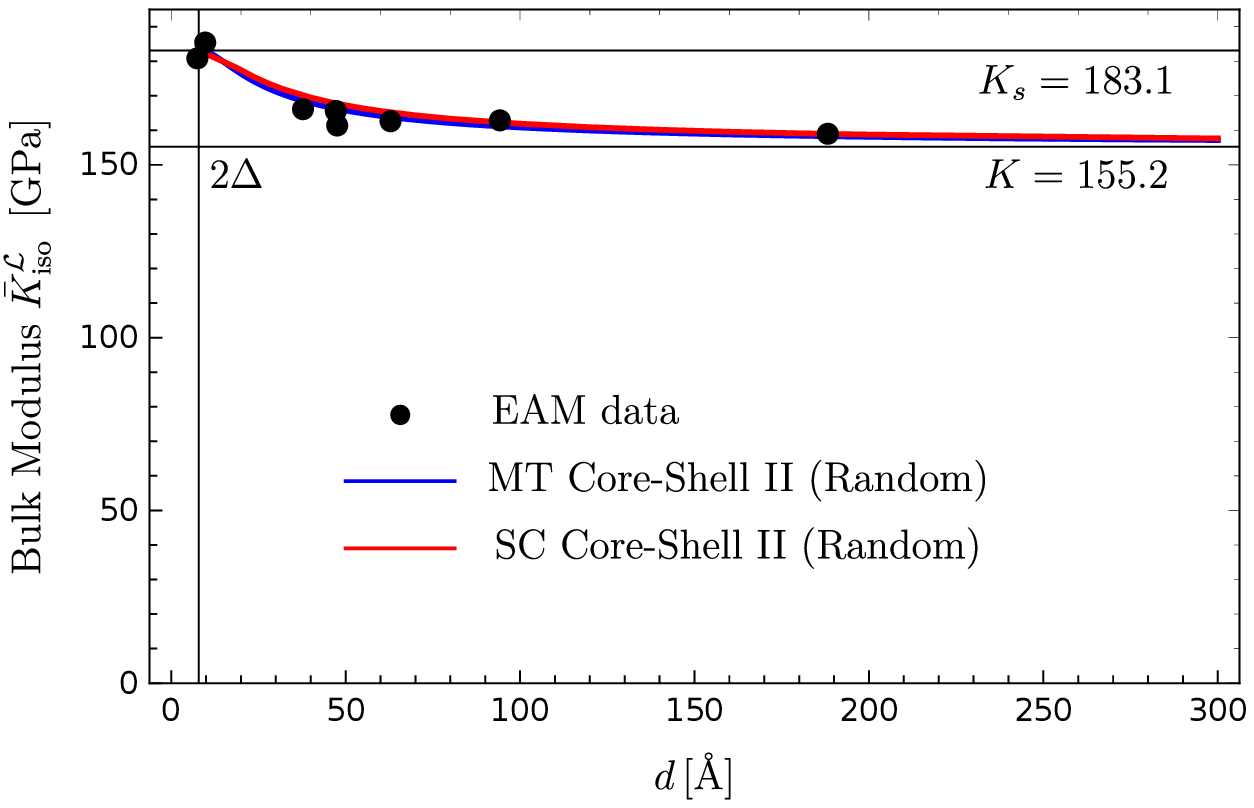}&
		\includegraphics[angle=0,width=0.39\textwidth]{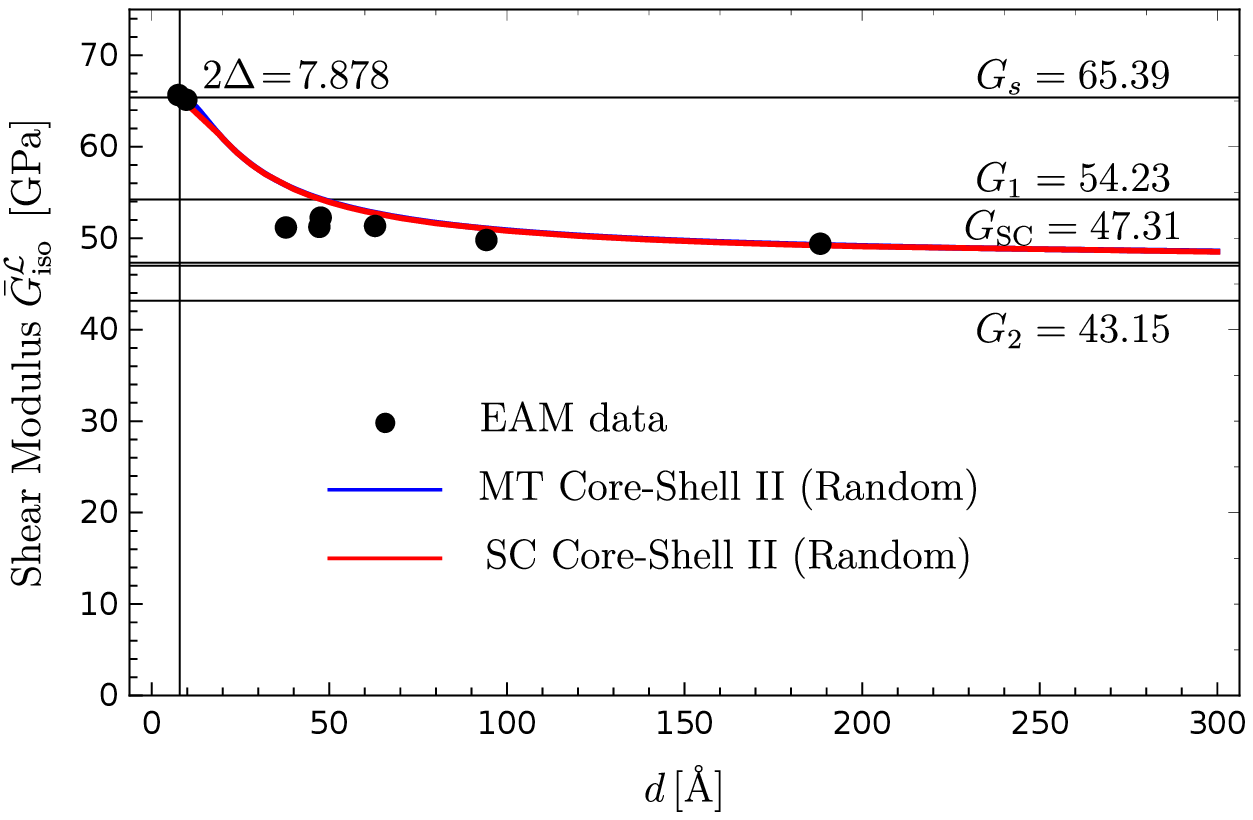}\\
		c) \textbf{Nb}&	\includegraphics[angle=0,width=0.39\textwidth]{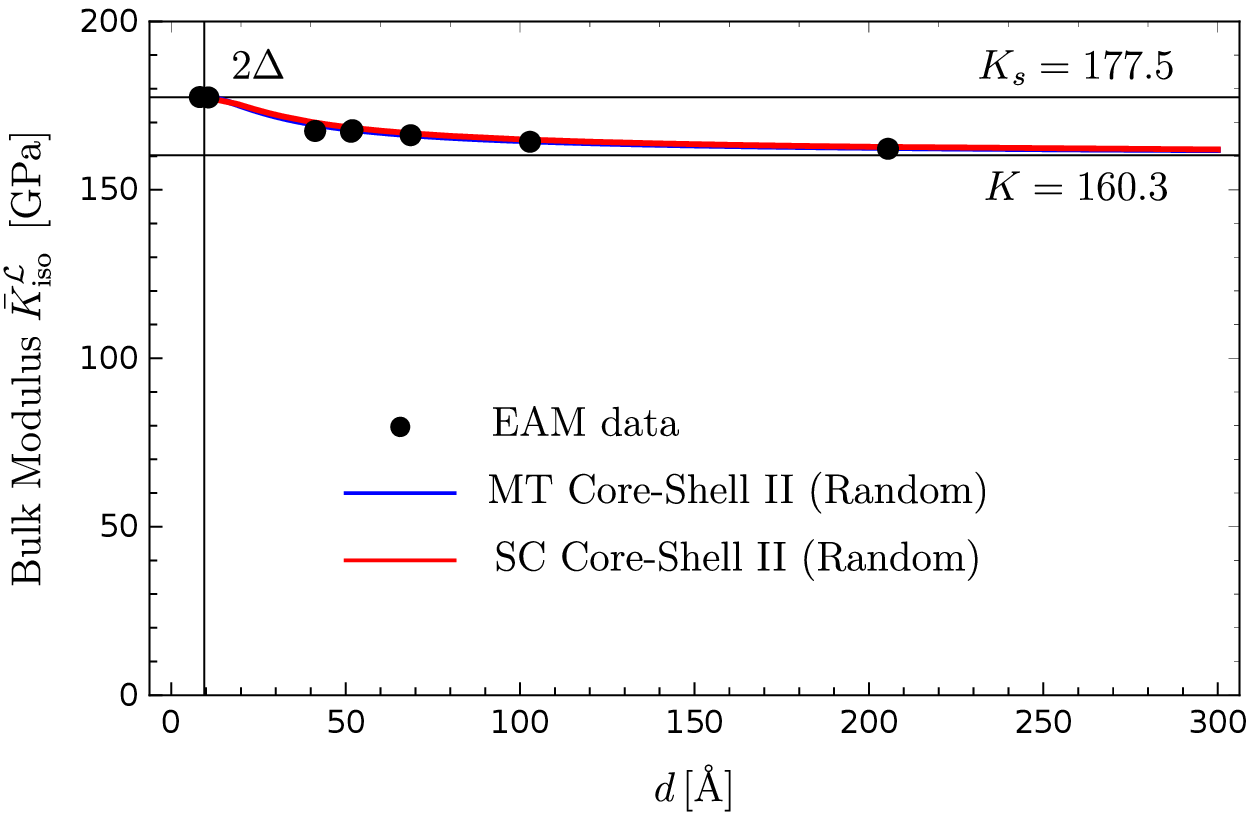}&
		\includegraphics[angle=0,width=0.39\textwidth]{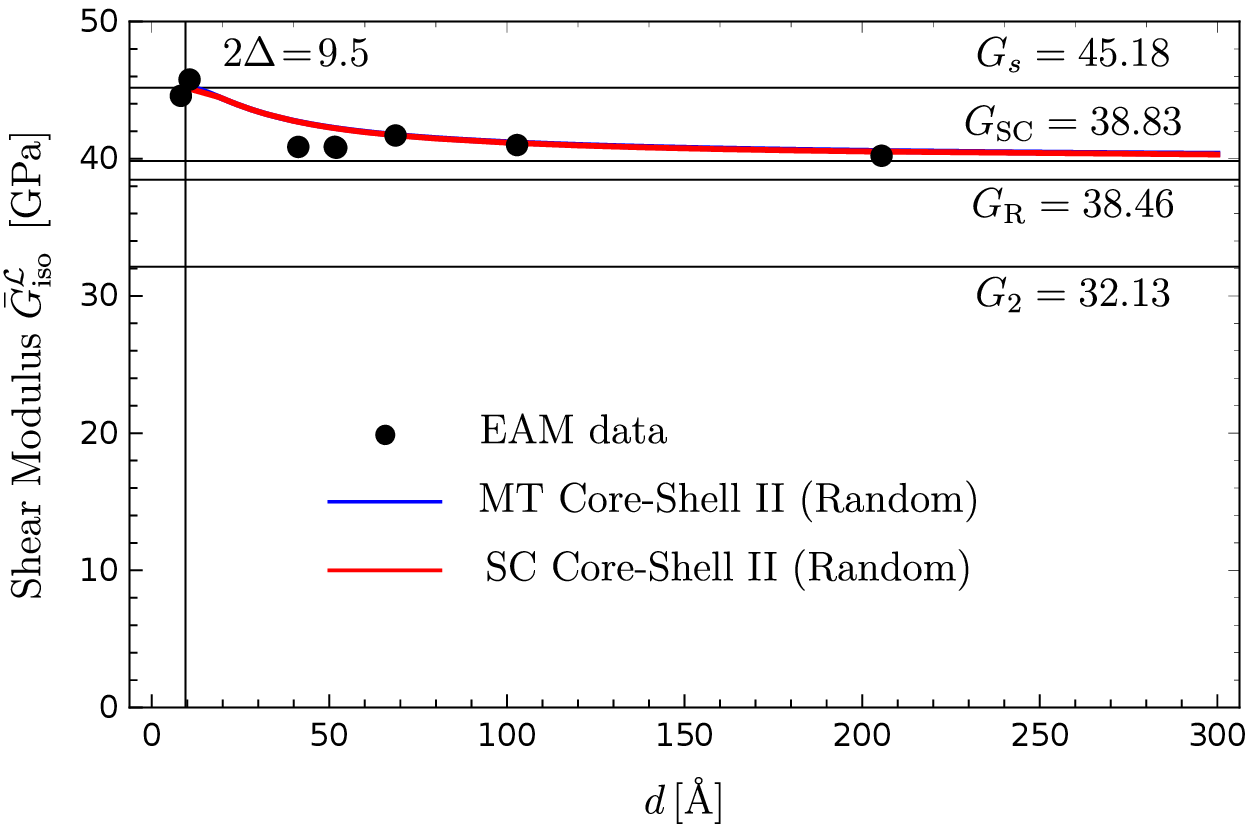}\\
	\end{tabular}
	\caption{The isotropic bulk and shear moduli $\bar{K}^{\mathcal{L}}_{\rm{iso}}$ and $\bar{G}^{\mathcal{L}}_{\rm{iso}}$ as a function of the average grain diameter $d$ by the two variants of the core-shell model - comparison with results of atomistic simulations reported in Tables \ref{tab:Cij-cW}, \ref{tab:Cij-cV}, \ref{tab:Cij-cNb} for cubic metals with a Zener parameter $\zeta_1\leq1$:  (a) W, (b) V (c) Nb.} 
	\label{fig:ZetaG2}
\end{figure}

The effective elastic properties collected in Tables \ref{tab:SamplesCuAlNi} and \ref{tab:SamplesNaFeWVNb} are now compared with predictions of a core-shell model in two variants presented in Section \ref{sec:Cont}. 
In order to calculate estimates infinite number of randomly distributed orientations is assumed, so overall quantities are calculated under the assumption 
of the overall isotropy. Grains are taken as equal in size and spherical. As already mentioned the shell thickness is assumed to be equal to the \emph{cutoff radius} $\Delta$ of the respective EAM atomistic potential for a given metal applied in the molecular simulations, while the shell elastic properties are identified as an average value for two samples with $f_0$ approaching unity.  Figures \ref{fig:ZetaG1} and \ref{fig:ZetaG2} present variation of bulk and shear moduli $\bar{K}$ and $\bar{G}$ with a increasing averaged grain diameter as predicted by a core-shell model. Figures concern metals of $\zeta_1>1$ and $\zeta_1\leq 1$, respectively. The results are compared with the outcomes of molecular simulations collected in Tables \ref{tab:SamplesCuAlNi} and \ref{tab:SamplesNaFeWVNb}.   
In the figures additionally classical estimates of overall shear modulus for coarse-grained polycrystals of cubic constituents obtained by self-consistent ($\bar{G}_{SC}$) and Reuss ($\bar{G}_R$) mean-field schemes are shown as reference values. Table \ref{tab:BoundsCM} includes the collection of all basic classical estimates of overall bulk modulus calculated for the coarse-grained polycrystals of analysed cubic metals with perfectly random distribution of orientations. Let us recall that $\bar{G}_{SC}$ provides the limit of the SC core-shell model for a coarse-grained polycrystal. The corresponding limit value obtained by the MT core-shell model, denoted by $\bar{G}^{\infty}_{CS/MT}$, is also included in the table. Note that this estimate, contrary to $\bar{G}^{\infty}_{CS/SC}=\bar{G}_{SC}$, depends on the assumed shell properties. It should be underlined that the respective estimate of the bulk modulus for all the classical averaging schemes is equal to the local bulk modulus $K$ of single crystal.    

\begin{table}[!htp]
	\caption{The overall shear modulus $\bar{G}^{\mathcal{L}}_{\rm{iso}}$\,[GPa] obtained by the Voigt (V), Reuss (R), Hashin-Shtrikman  bounds (HS-U, HS-L), self-consistent (SC) estimate and the limit (coarse-grained) value obtained by MT core-shell model for polycrystals with perfectly random orientation distributions and eight cubic metals considered in the present study. Local properties are collected in Table \ref{tab:Shell-CutOff}.
	}
	\label{tab:BoundsCM}\vspace{.05in}
	\centering
	\begin{tabular}{ccccccc}
		\hline
		Metal & R & HS-L & SC & HS-U &  V & CS/MT\\ 
		&\multicolumn{6}{c}{[GPa]}\\\hline	
		Na & 1.512 & 2.106 & 2.719 & 3.053 & 3.792 & 2.489\\ 
		Cu & 40.33 & 46.35 & 48.60 & 49.90 & 55.17 & 46.12\\
		Fe & 73.34 & 80.03 & 81.79 & 82.87 & 88.91 & 79.52\\
		Ni & 77.46 & 84.43 & 86.25 & 87.37 & 93.62 & 82.59\\
		Al & 29.15 & 29.28 & 29.29 & 29.29 & 29.40 & 29.25\\
		W & 160.7 & 160.7 & 160.7 & 160.7 & 160.7 & 160.7\\
		V & 46.99 & 47.29 & 47.30 & 47.32 & 47.58 &47.35\\ 
		Nb & 38.46 & 39.71 & 39.83 & 40.00 & 41.11 & 39.91\\\hline
	\end{tabular}\\[.051in]
\end{table}

It is seen that for all analysed nanocrystalline metals the core-shell model is able to appropriately reproduce, both qualitatively and quantitatively, the trends observed in the molecular simulations. When comparing two variants of the core-shell model, the estimates obtained by the SC variant are in better agreement with atomistic simulations.

\begin{figure}
	\centering
	\begin{tabular}{ccc}
		&Young's modulus &  Poisson's ratio\\
		a) \textbf{Na}&\includegraphics[angle=0,width=0.39\textwidth]{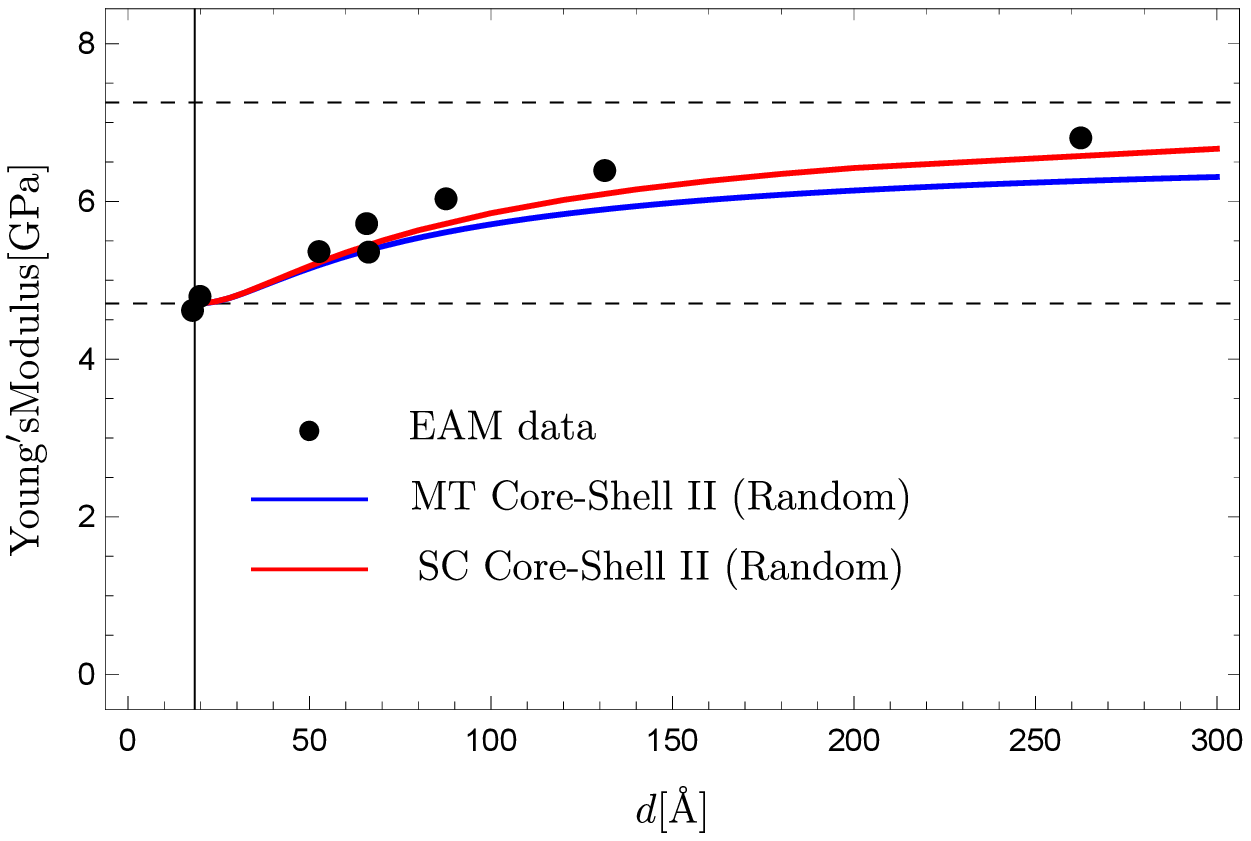}&
		\includegraphics[angle=0,width=0.39\textwidth]{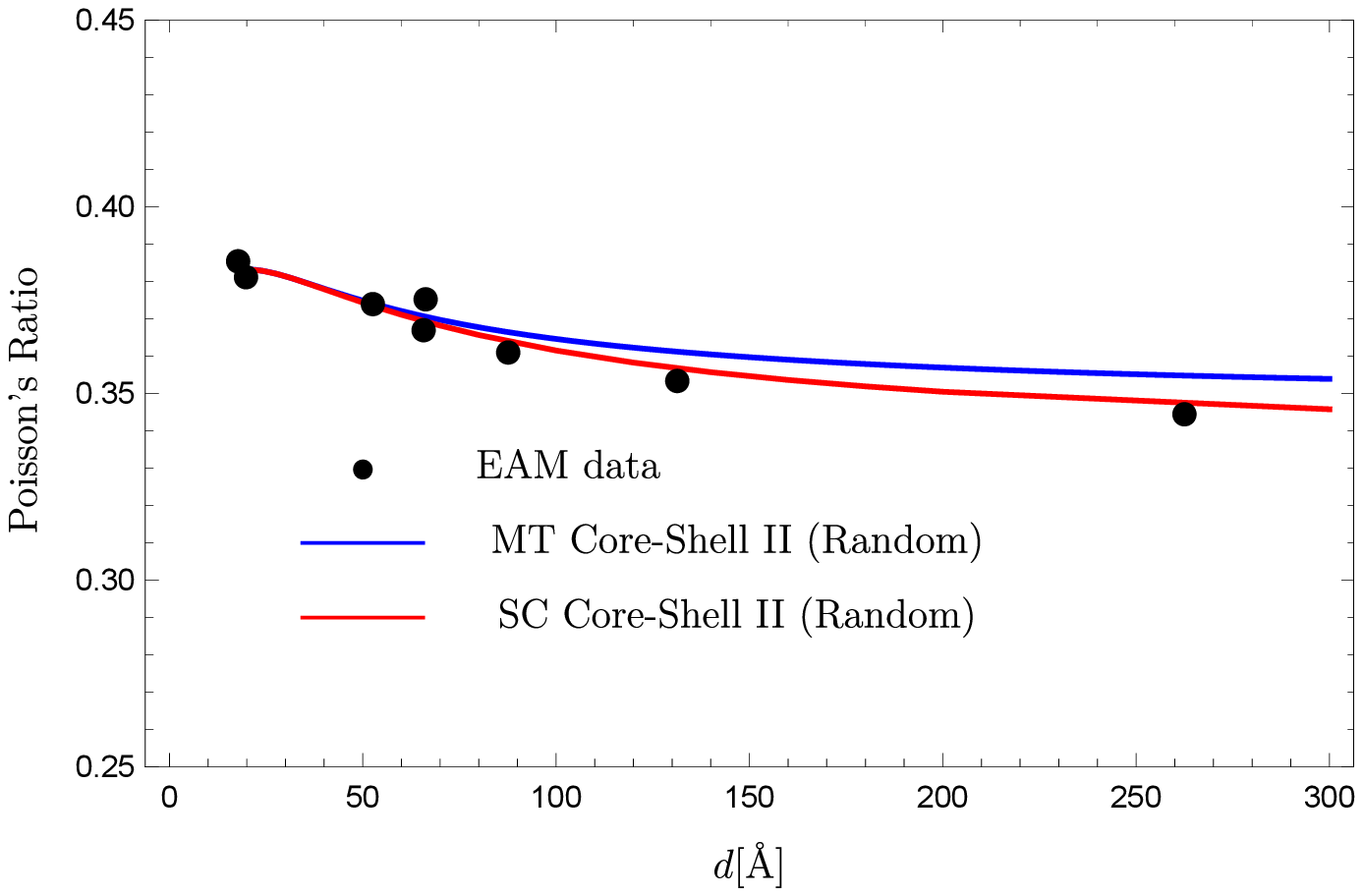}\\
		b) \textbf{Cu}&	\includegraphics[angle=0,width=0.39\textwidth]{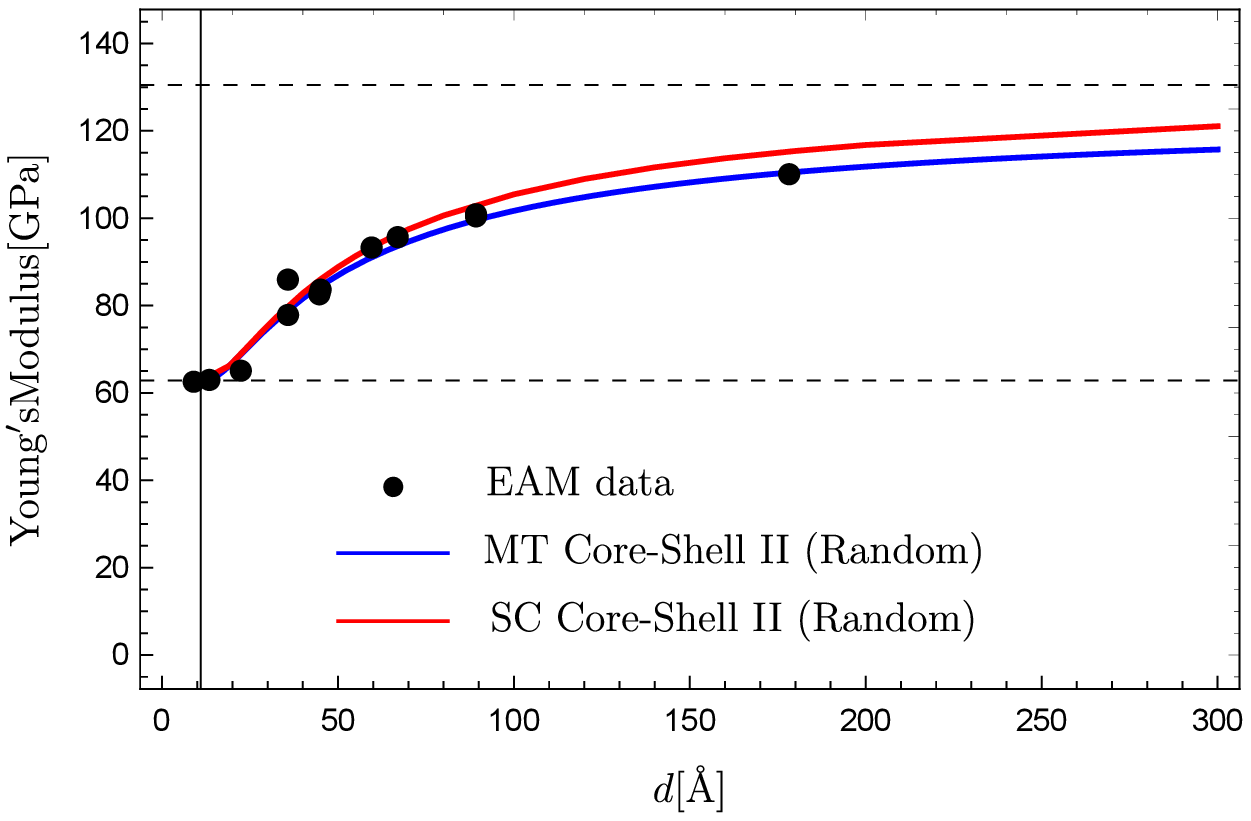}&
		\includegraphics[angle=0,width=0.39\textwidth]{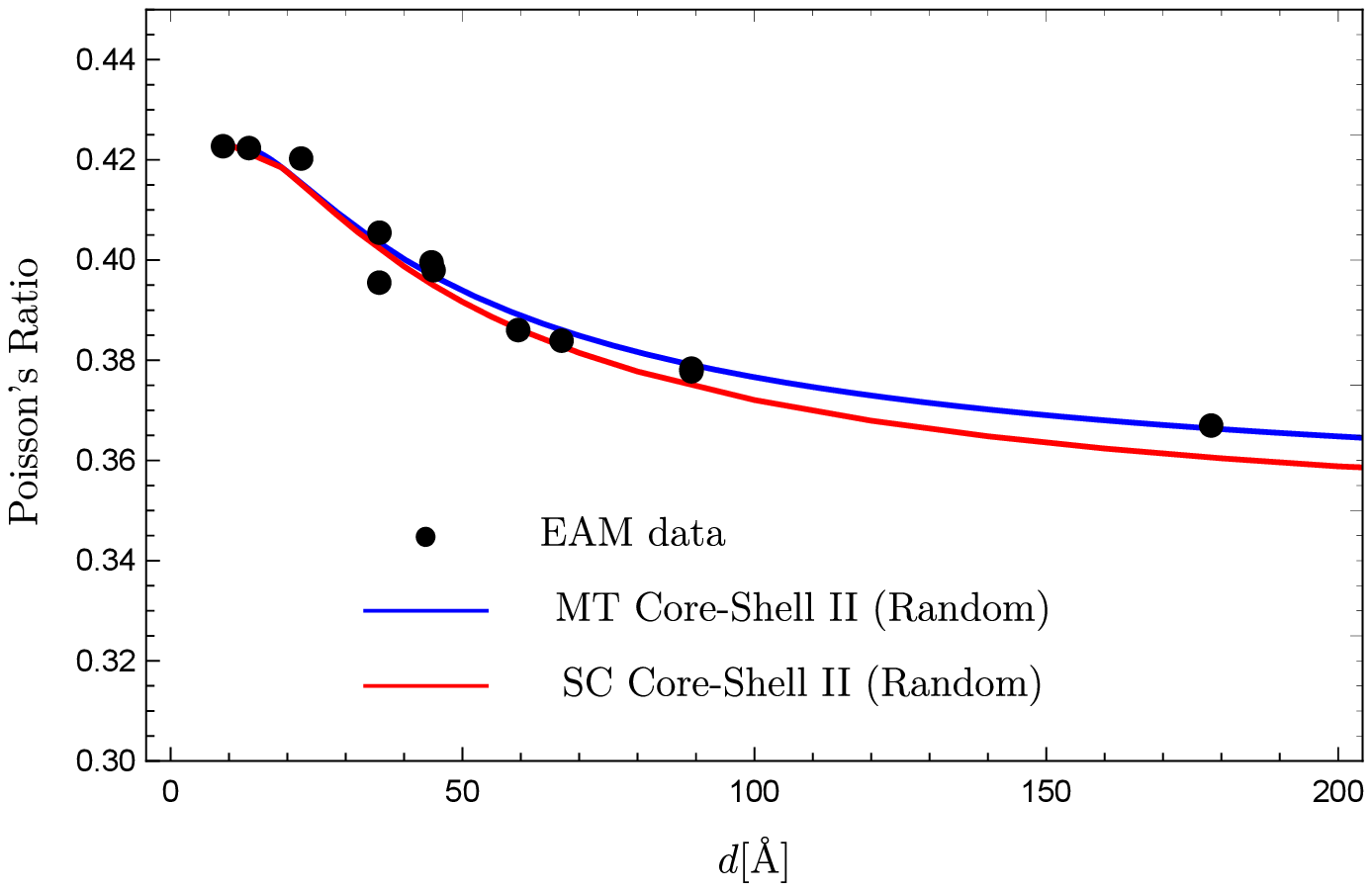}\\
		c) \textbf{Fe}&	\includegraphics[angle=0,width=0.39\textwidth]{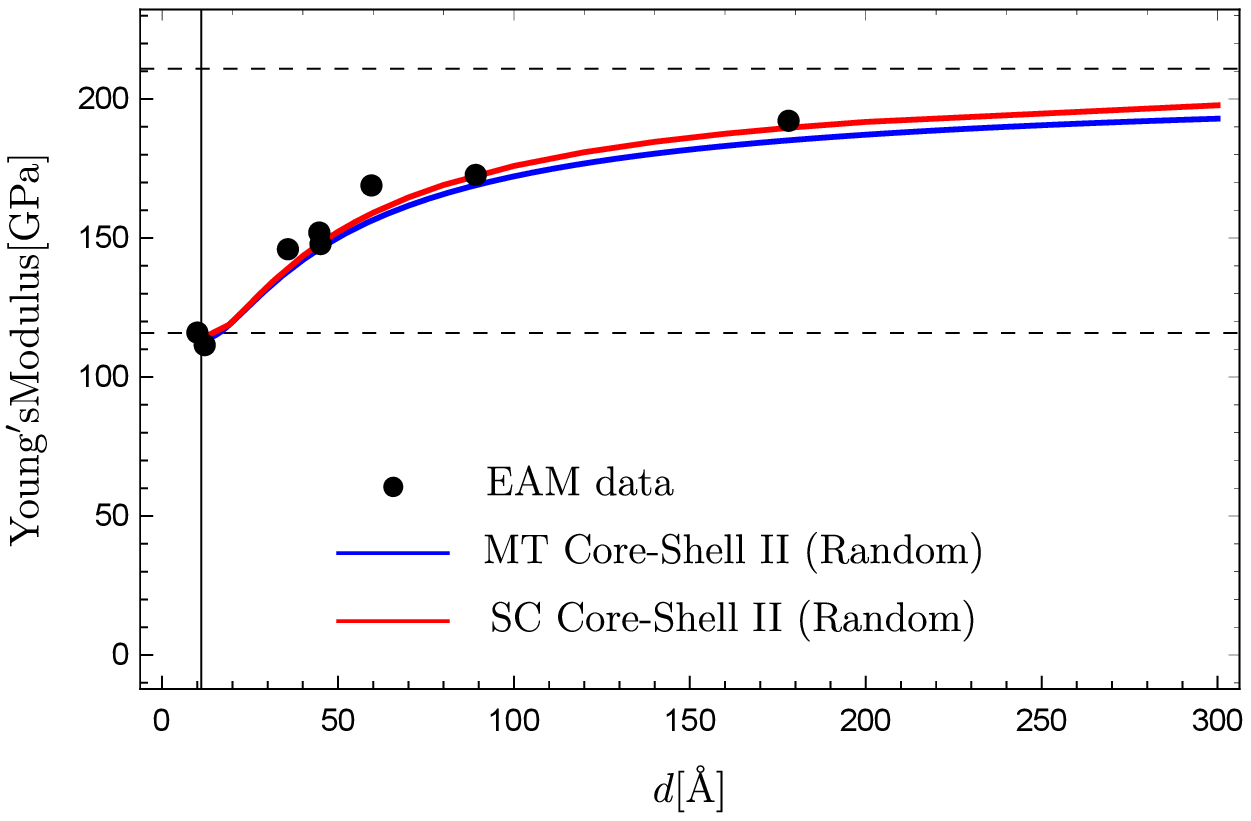}&
		\includegraphics[angle=0,width=0.39\textwidth]{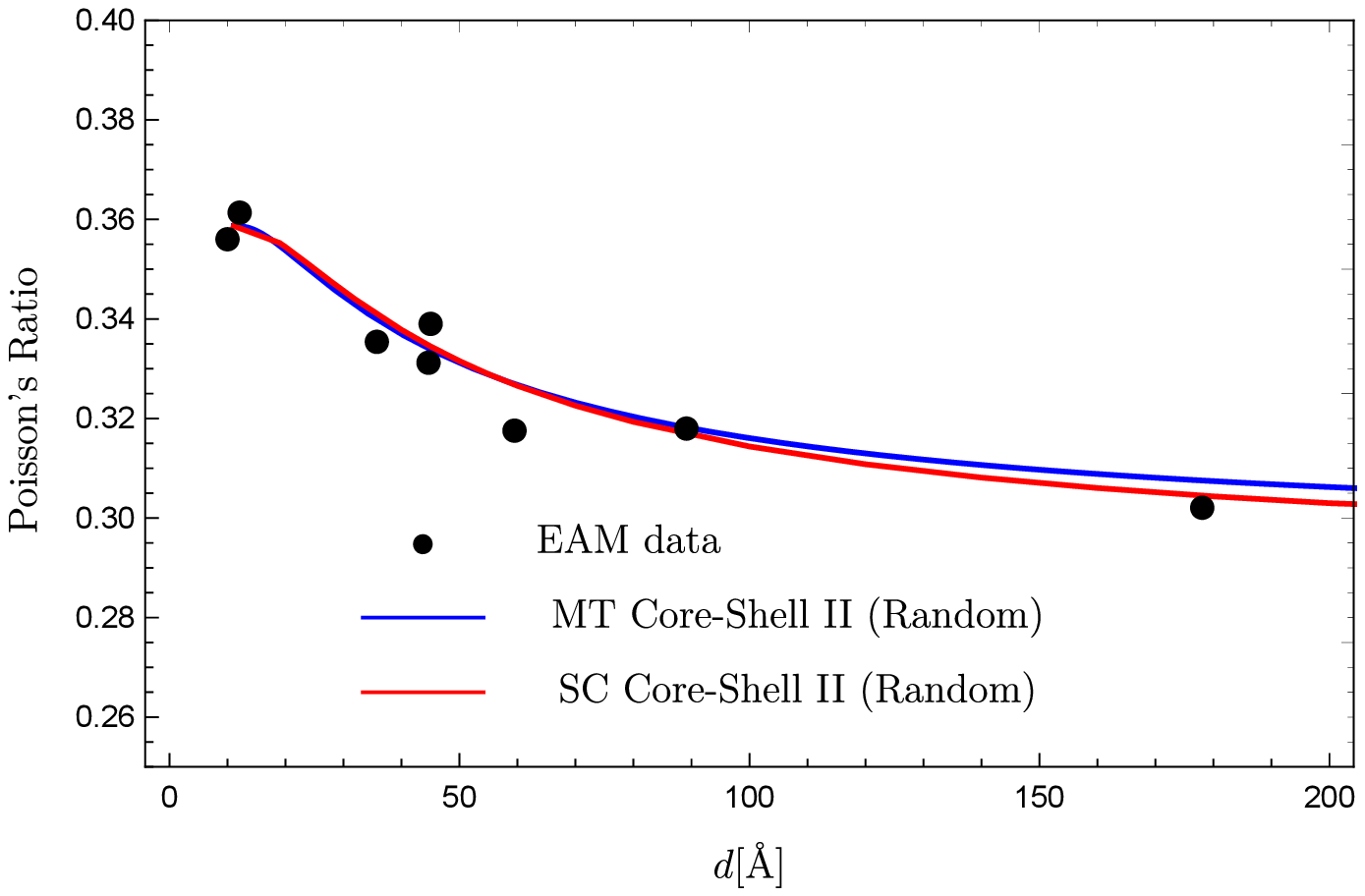}\\
		d) \textbf{Ni}& \includegraphics[angle=0,width=0.39\textwidth]{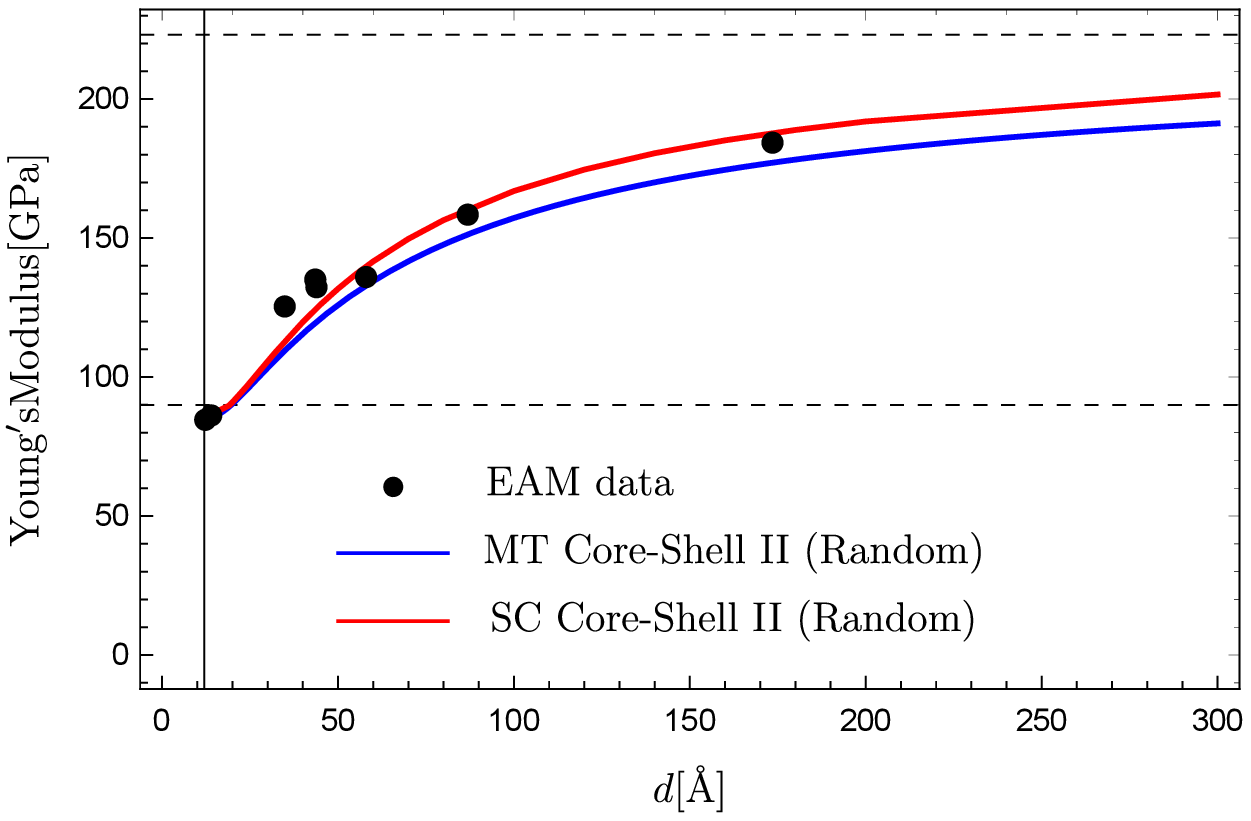}&
		\includegraphics[angle=0,width=0.4\textwidth]{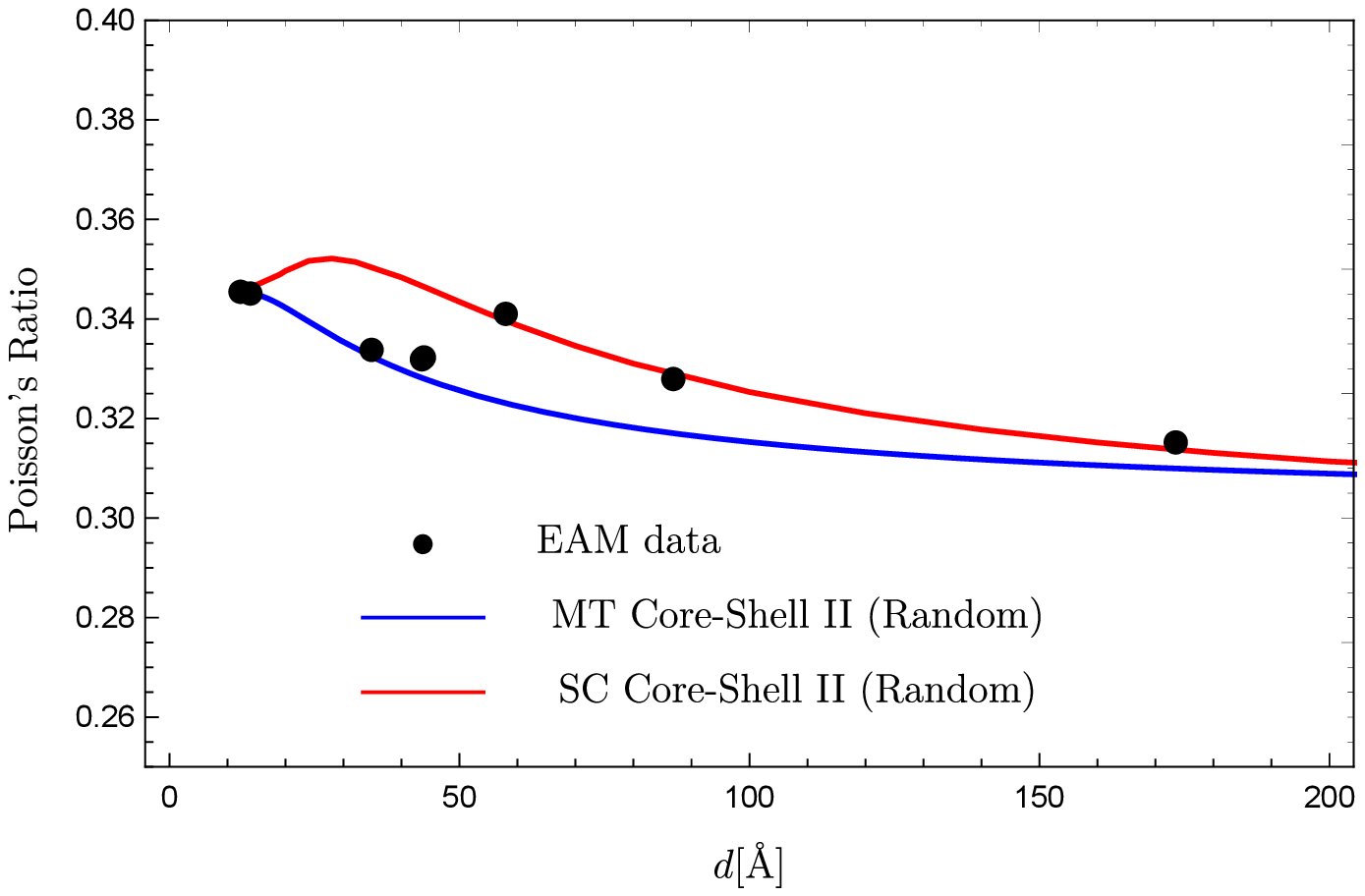}\\
		e) \textbf{Al}& \includegraphics[angle=0,width=0.39\textwidth]{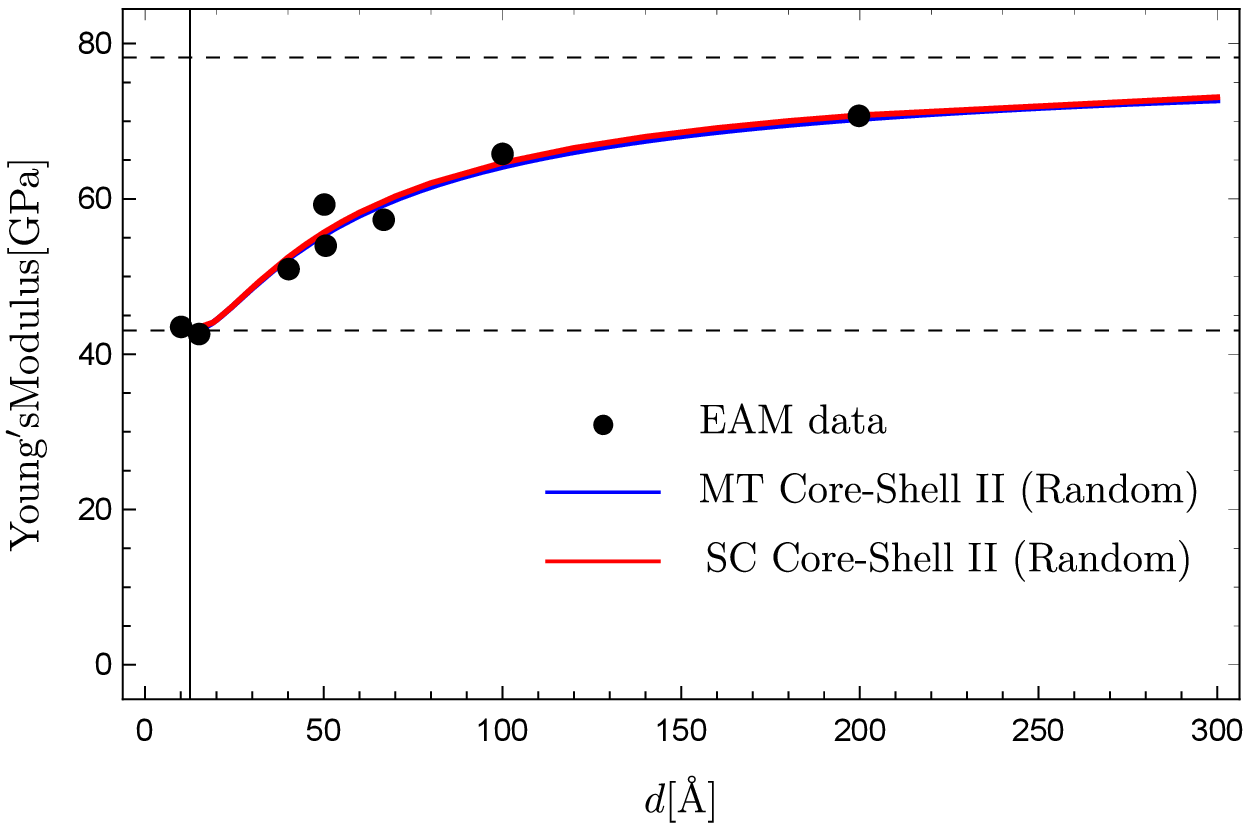}&
		\includegraphics[angle=0,width=0.39\textwidth]{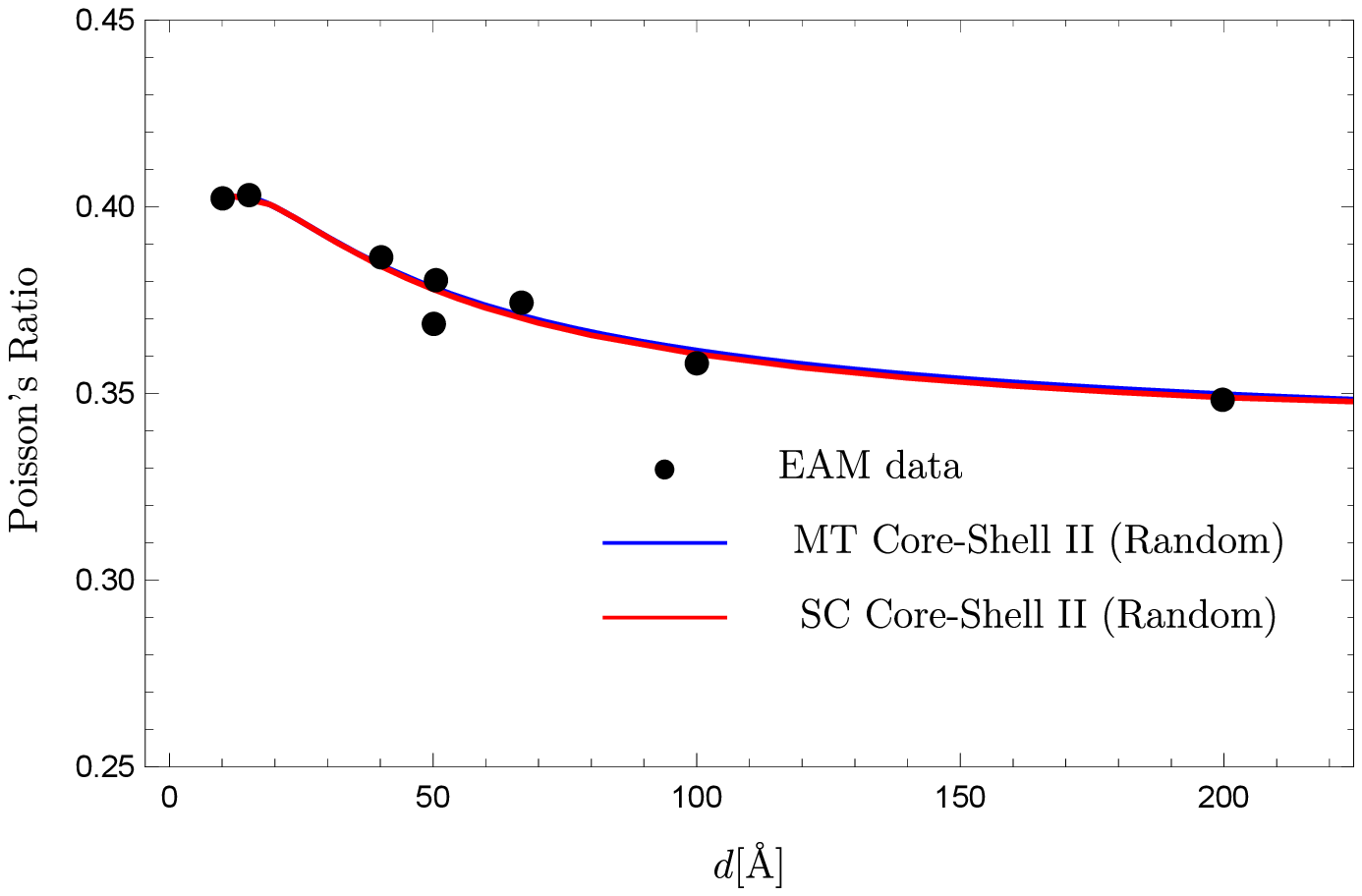}\\
	\end{tabular}
	\caption{The isotropic Young modulus $\bar{E}^{\mathcal{L}}_{\rm{iso}}$ and Poisson's ratio  $\bar{\nu}^{\mathcal{L}}_{\rm{iso}}$ as a function of the average grain diameter $d$ by the two variants of the core-shell model - comparison with results ofatomistic simulations, calculated using Eq. \ref{Eq:E-nu}, cubic metals with a Zener parameter $\zeta_1>1$:  (a) Na, (b) Cu (c) Fe (d) Ni (f) Al.} 
	\label{fig:ZetaE1}
\end{figure}

\begin{figure}
	\centering
	\begin{tabular}{ccc}
		&Young's modulus &  Poisson's ratio\\
		a)  \textbf{W}&\includegraphics[angle=0,width=0.39\textwidth]{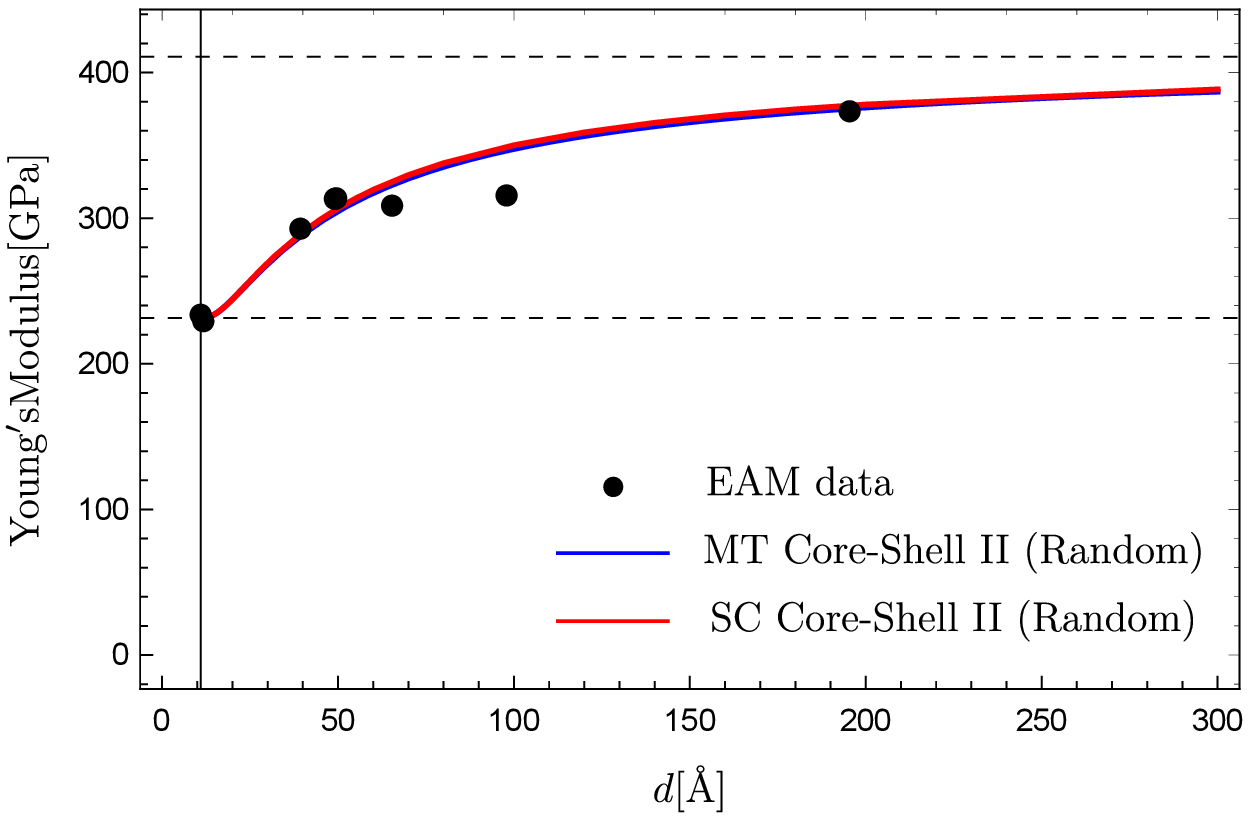}&
		\includegraphics[angle=0,width=0.39\textwidth]{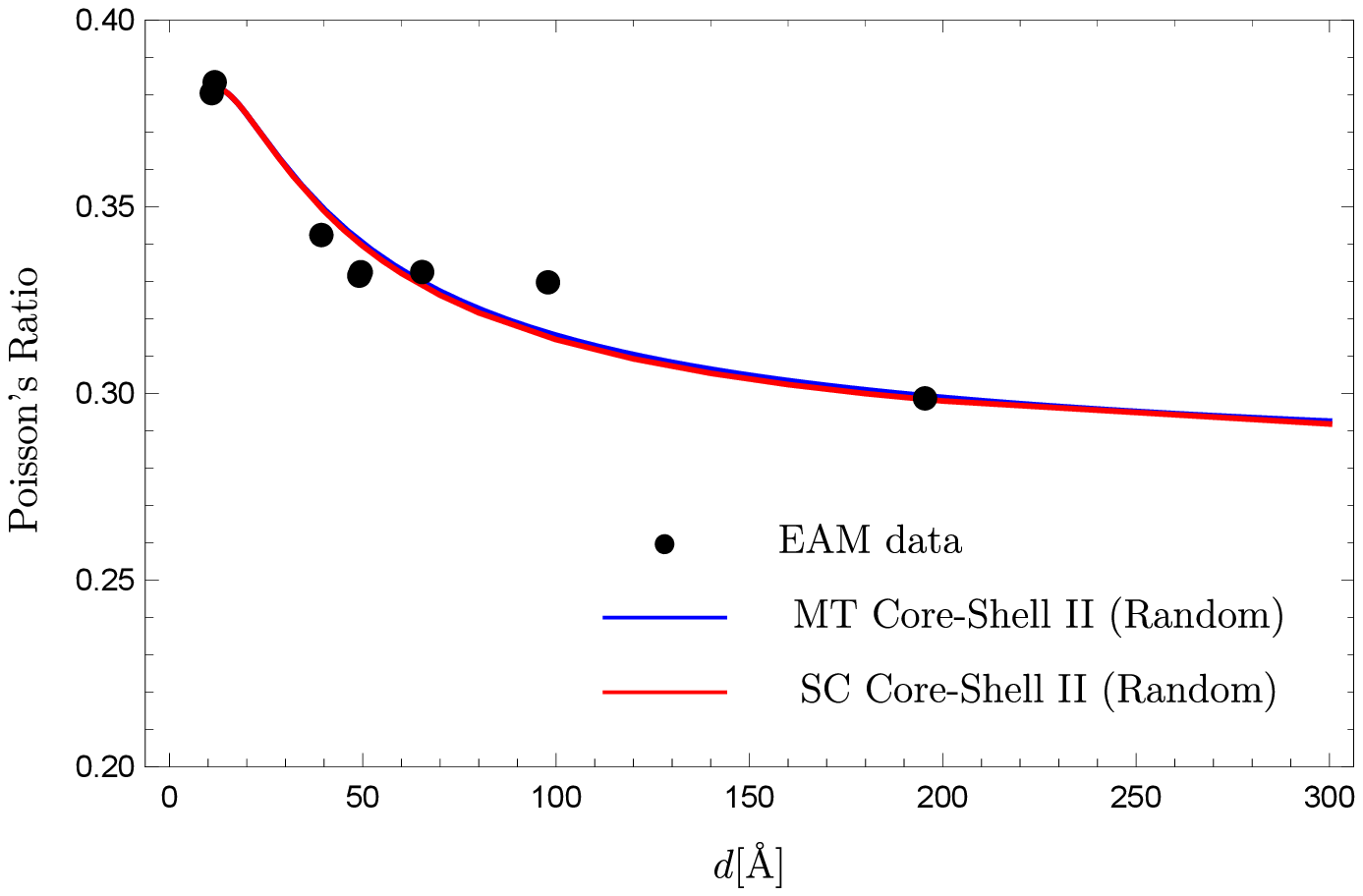}\\
		b) \textbf{V}&	\includegraphics[angle=0,width=0.39\textwidth]{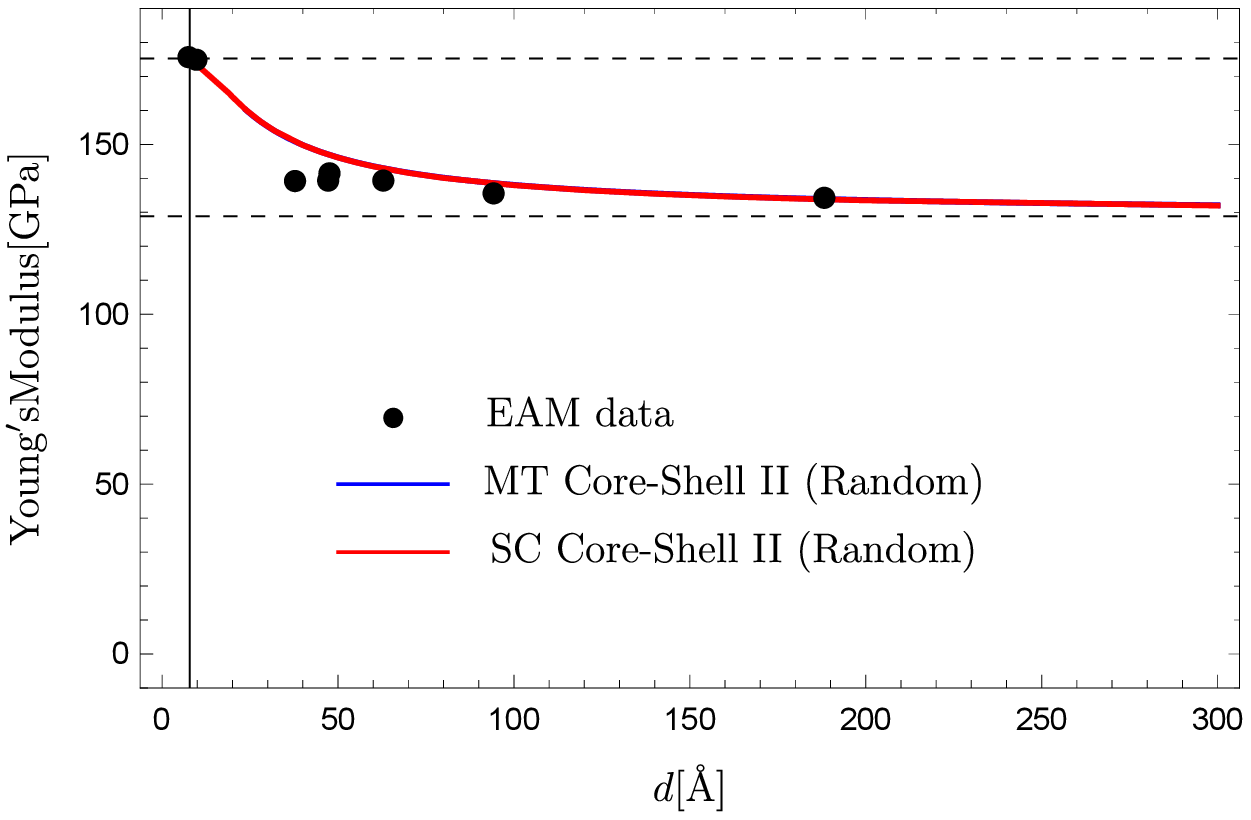}&
		\includegraphics[angle=0,width=0.39\textwidth]{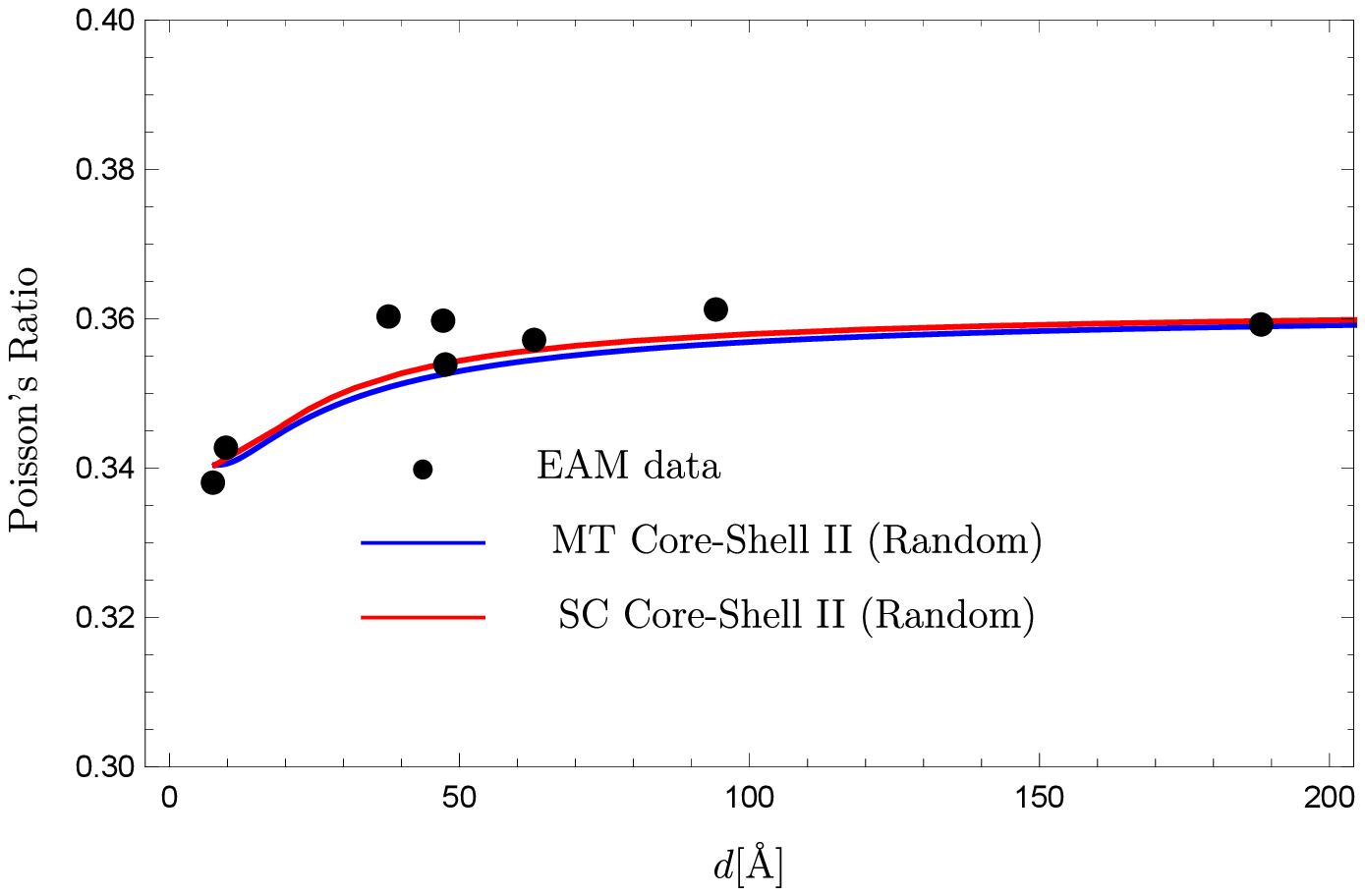}\\
		c) \textbf{Nb}&	\includegraphics[angle=0,width=0.39\textwidth]{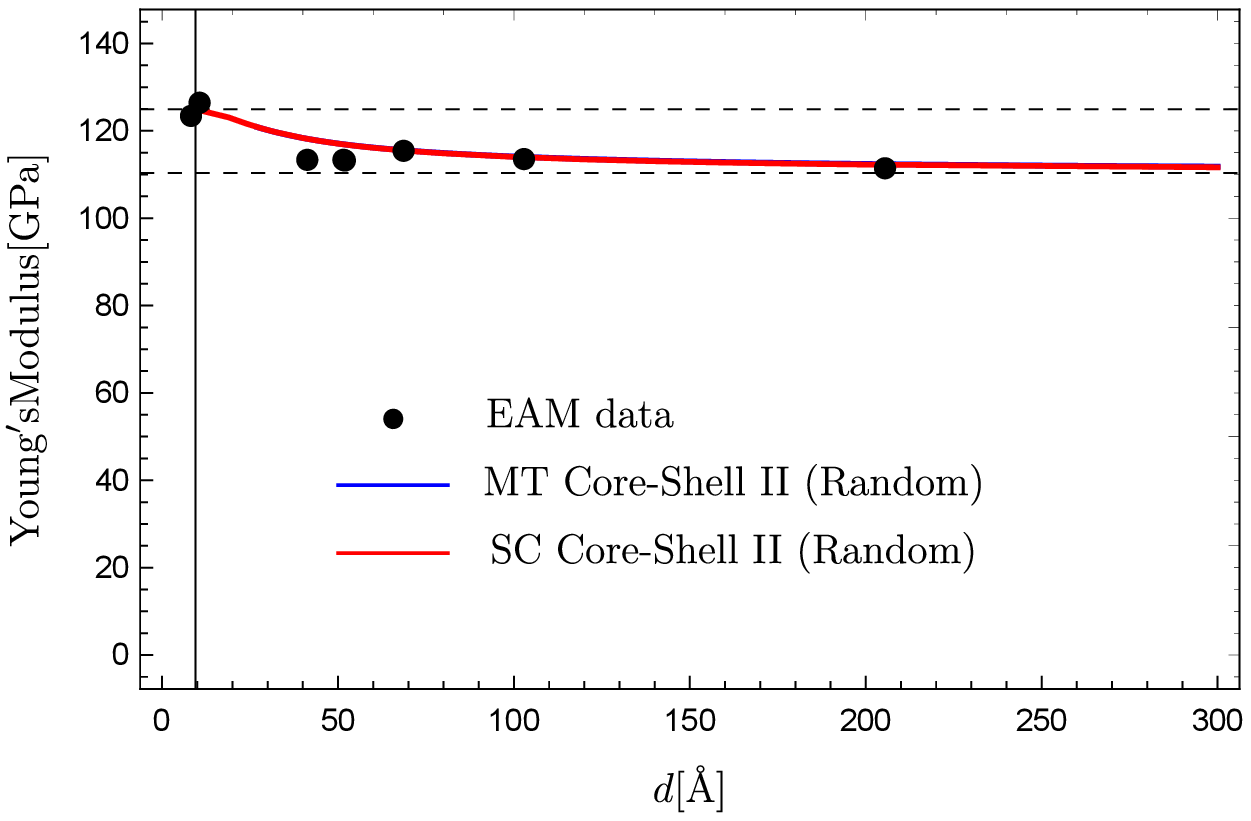}&
		\includegraphics[angle=0,width=0.39\textwidth]{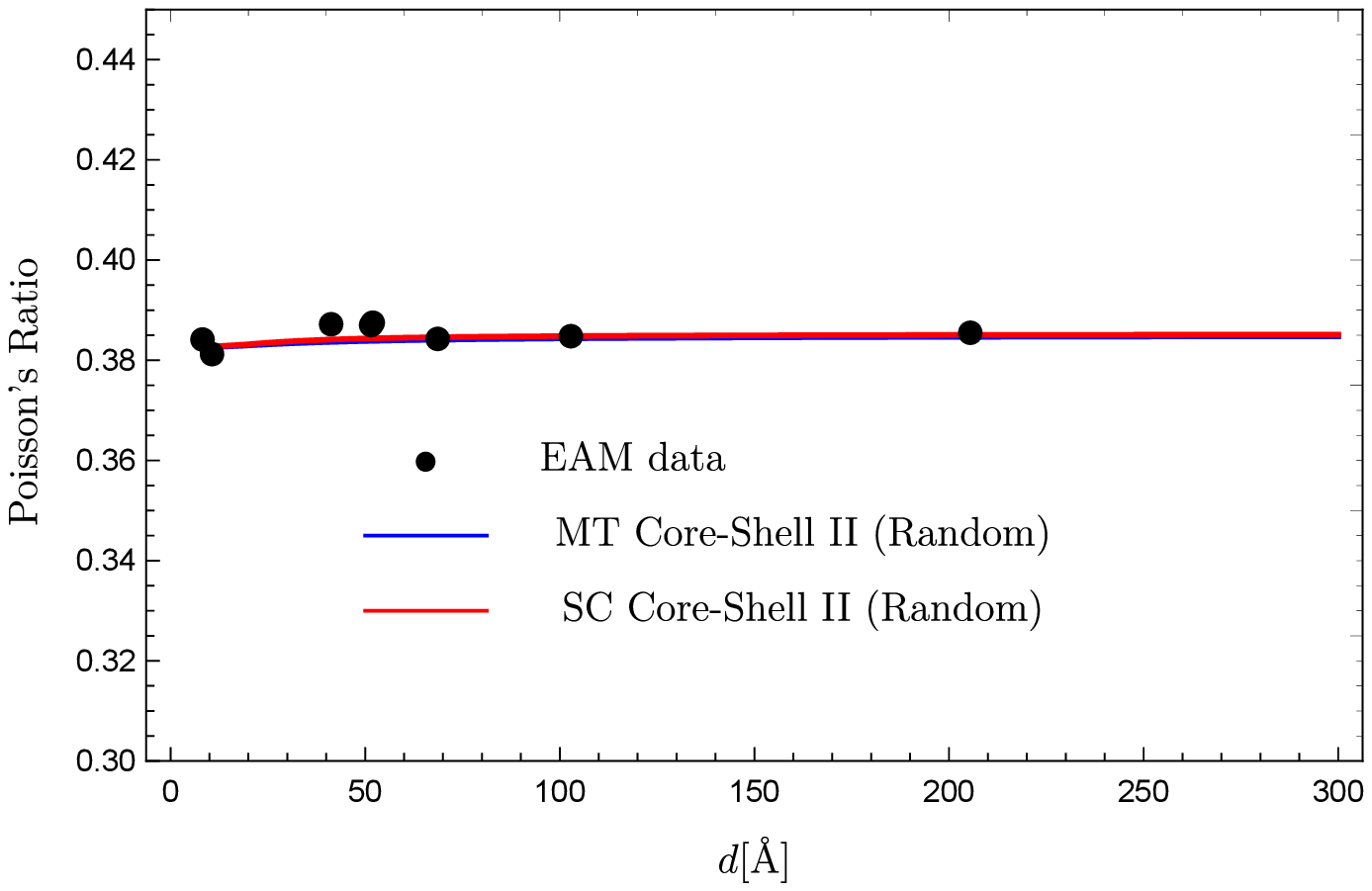}\\
	\end{tabular}
	\caption{The isotropic Young modulus $\bar{E}^{\mathcal{L}}_{\rm{iso}}$ and Poisson's ratio  $\bar{\nu}^{\mathcal{L}}_{\rm{iso}}$ as a function of the average grain diameter $d$ by the two variants of the core-shell model - comparison with results of atomistic simulations, calculated by Eq. \ref{Eq:E-nu}, cubic metals with a Zener parameter: $\zeta_1\leq1$:  (a) W, (b) V (c) Nb.} 
	\label{fig:ZetaE2}
\end{figure}

The comparison of core-shell model and atomistic estimates of isotropized Young's modulus $\bar{E}^{\mathcal{L}}_{\rm{iso}}$ and Poisson's ratio $\bar{\nu}^{\mathcal{L}}_{\rm{iso}}$ are shown in Figs. \ref{fig:ZetaE1} and \ref{fig:ZetaE2}.
{Presented values were obtained by applying well-known relations: 
	\begin{equation}\label{Eq:E-nu}
	\bar{E}^{\mathcal{L}}_{\rm{iso}}=\frac{9\bar{K}^{\mathcal{L}}_{\rm{iso}}\bar{G}^{\mathcal{L}}_{\rm{iso}}}{3\bar{K}^{\mathcal{L}}_{\rm{iso}}+\bar{G}^{\mathcal{L}}_{\rm{iso}}}\,,\quad\bar{\nu}^{\mathcal{L}}_{\rm{iso}}=\frac{3\bar{K}^{\mathcal{L}}_{\rm{iso}}-2\bar{G}^{\mathcal{L}}_{\rm{iso}}}{6\bar{K}^{\mathcal{L}}_{\rm{iso}}+2\bar{G}^{\mathcal{L}}_{\rm{iso}}}\,.
	\end{equation}}	

\section{Conclusions}
\label{sec:Con}

Applicability of a mean-field core-shell model due to \cite{Kowalczyk18} for estimation of elastic properties of bulk nanocrystalline cubic metals has been validated. Validation was performed using the results of atomistic simulations since the experimental data for the considered materials are scarce \cite{Sanders97,Haque02,Zhao06,Xu17}. For an extensive verification of the proposed approach eight metals of cubic symmetry with BCC or FCC lattice geometry and different value of the anisotropy Zener factor (\ref{Eq:zeta1}) were selected (see the collective figure \ref{fig:Zener}).   For each metal atomistic simulations have been conducted on eight generated samples of polycrystalline materials with randomly selected orientations. All components of the elastic stiffness tensor are identified by performing six numerical atomistic tests on the generated samples. They vary mainly with respect to the number of atoms per grain, which parameter corresponds to the averaged grain size. In the present analyses an averaged grain diameter takes values between ca. 1 nm to 20 nm.      

For the obtained anisotropic stiffness tensor the closest isotropic approximation is found using the Log-Euclidean norm \cite{Moakher06}. Next the variation of resulting bulk and shear moduli on a grain size is studied. Interestingly, it is found that the dependence varies quantitatively with the Zener factor: these two stiffness moduli increases (resp. decreases) with the grain size if the Zener factor is higher (resp. lower) than one.    

In \cite{Kowalczyk18} the two-phase core-shell model was proposed in two variants. The concept follows earlier ideas of \cite{Jiang04,Capolungo07}. The model is size dependent and the basic length scale parameter is $\Delta$ - the thickness of a core coating (see Figure \ref{fig:CoreShell}), representing the grain boundary zone.  This thickness is specified by a \emph{cutoff radius} of the corresponding atomistic potential. Outcomes of present molecular simulations summarized above indicated that the assumptions concerning the stiffness of boundary zone, which were made for nanocrystalline copper in \cite{Kowalczyk18}, cannot be transferred on other cubic metals. Specifically, it is observed that the bulk modulus may vary with a grain size 
and the shear modulus of a boundary zone can be equated with none of two modulus of a single crystal. Therefore, in variance with \cite{Kowalczyk18}, the shell properties are identified separately using the atomistic simulations results for samples with very small grains.
When the shell thickness and properties are identified in the described way, obtained mean-field estimates are in satisfactory qualitative and quantitative agreement with the results of atomistic simulations for all considered cubic metals.

In the future it will be interesting to confirm experimentally the observed qualitative difference in stiffness related to the Zener parameter, since the present observations strongly relays on validity of the applied atomistic potential and results may change if different EAM potential function is selected for the considered metals. It should be also mentioned that the applied mean-field two-phase model can be extended to estimate a non-linear response of a nano-grained polycrystal and specifically the yield strength as in \cite{Jiang04,Capolungo07}. Atomistic simulations can be also used to validate such extension of the model.


\appendix

\section{Detailed results of atomistic simulations}\label{App} 

Detailed results of atomistic simulations for nine samples of eight metals are collected in the next subsections. For each metal the first table for a given metal collects quantitative data related to the analysed samples and the second table the calculated 21 components of the anisotropic elasticity tensor for each sample written in the Voigt notation, i.e. in the form of the following $6\times 6$ array: 
\begin{eqnarray}
\centering
\left[C_{KL}\right]=\left[
\begin{array}{cccccc}
{C_{1111}} & {C_{1122}} & {C_{1133}} & {C_{1123}} & {C_{1131}} & {C_{1112}} \\
& {C_{2222}} & {C_{2233}} & {C_{2223}} & {C_{2231}} & {C_{2212}} \\
&  & {C_{3333}} & {C_{3323}} & {C_{3331}} & {C_{3312}} \\
& & & {C_{2323}} & {C_{2331}} & {C_{2312}} \\
\multicolumn{4}{c}{\textbf{Sym.}} & {C_{3131}} & {C_{3112}} \\
& & & & & {C_{1212}} \\
\end{array}
\right]\,.
\label{eqn:CuCij}
\end{eqnarray}

\subsection{Nanocrystalline copper}

\begin{table}[H] 
	\caption{Copper: Volume (\AA$^3$), box lengths (\AA), number of atoms, average grain diameter $d$ (\AA), fraction of transient shell atoms $f_{0}$ (\ref{def:fsa}), average cohesive energy $E_{c}$\,(eV/atom) of analysed computational samples.}
	\label{tab:SamplesCu}
	\centering
	\renewcommand{\arraystretch}{1.5}
	\tiny 
	\begin{tabular}{|c c c c c c c|}
		\hline Sample & V & L & No.of atoms & $d$ & $f_{0}$ &  E$_{c}$\\ 
		\hline Monocrystal & 47.24 & 3.615 & 4 & & &-3.54 \\
		10$^3$-128-BCC & 48634.1 & 36.50 & 4016 & 9.00 & 1.00 & -3.454 \\
		15$^3$-128-BCC & 162305.7 & 54.52 & 13430 & 13.43 & 0.99 & -3.460 \\
		50$^3$-128-BCC & 5985342.9 & 181.56 & 499984 & 44.7 & 0.57 & -3.497 \\
		50$^3$-16-BCC & 5950371.6 & 181.21 & 500020 &  89.2 & 0.33 &  -3.516 \\
		50$^3$-54-BCC & 5969399.2 & 181.40 & 500058 & 59.6 & 0.46 & -3.506 \\
		50$^3$-250-BCC & 5997791.1 & 181.69 & 500008 & 35.8 & 0.67 & -3.489 \\
		50$^3$-125-Random & 5985474.7 & 181.57 & 499836 & 45.1 & 0.57 & -3.495 \\
		100$^3$-16-BCC & 47432793 & 361.99 & 4000010 & 178.2 & 0.17 & -3.528 \\
		\hline 
	\end{tabular}
\end{table}

\begin{table}[H] 
	\caption{{{Copper: Elasticity tensors $\bar{\mathbb{C}}$\,[GPa]  of analysed samples (for notation used see Eq. (\ref{eqn:CuCij}))}}.}
	\label{tab:Cij-c}
	\begin{threeparttable}[b]
		\centering
		\scriptsize 
		\begin{tabular}{ c c }
			\hline
			{Monocrystal} & {10$^3$ unit cells, 128 grains in BCC system, 4016 atoms} \\
			& {(10$^3$-128-BCC)}	  \\ \hline
			$\begin{bmatrix}  {169.88} & {122.60} & {122.60} & 0       &       0 & 0 \\
			& {169.88} & {122.60} & 0       &       0 & 0 \\
			&          & {169.88} & 0       & 0       & 0 \\
			&          &          & {76.19} &       0 & 0 \\
			\multicolumn{4}{c}{\textbf{Sym.}}                          & {76.19} & 0 \\
			&          &          &         &         & {76.19} \end{bmatrix}$ &	
			
			$\begin{bmatrix}  {164.64} & {118.31} & {121.89} & {-0.99} & {-0.26} & {0.76} \\
			& {165.65} & {120.89} & {-0.26} & {1.00}  & {0.12} \\
			&          & {161.70} & {0.98}  & {-0.58} & {-1.13} \\
			&          &          & {23.15} & {-0.07} & {1.06} \\
			\multicolumn{4}{c}{\textbf{Sym.}}                          & {22.50} & {-0.43} \\
			&          &          &         &         & {21.02} \end{bmatrix}$\\  
			
			\hline {15$^3$ unit cells, 128 grains in BCC system, 13430 atoms} & {50$^3$ unit cells, 128 grains in BCC system, 499984 atoms} \\  
			{(15$^3$-128-BCC)} & {(50$^3$-128-BCC)} \\  \hline
			$\begin{bmatrix}  {162.44} & {121.61} & {121.23} & {0.18}  & {-0.40} & {1.24} \\
			& {165.78} & {118.19} & {0.17}  & {0.64}  & {-0.66} \\
			&          & {166.25} & {-0.31} & {0.13}  & {-0.78} \\
			&          &          & {21.14} & {-0.21} & {0.08} \\
			\multicolumn{4}{c}{\textbf{Sym.}}                          & {22.57} & {0.29} \\
			&          &          &         &         & {22.75} \end{bmatrix}$ &
			
			$\begin{bmatrix}  {174.61} & {116.92} & {119.13} & {-0.34} & {1.11} & {-1.54} \\
			& {177.10} & {117.18} & {-1.80} & {1.09} & {1.17} \\
			&          & {174.88} & {0.18}  & {-0.27}& {0.08} \\
			&          &          & {32.96} & {0.17} & {-1.63} \\
			\multicolumn{4}{c}{\textbf{Sym.}}                         & {28.10}& {1.05} \\
			&          &          &         &        & {29.19} \end{bmatrix}$\\		
			
			\hline {50$^3$ unit cells, 16 grains in BCC system, 500020 atoms} & {50$^3$ unit cells, 54 grains in BCC system, 500058 atoms} \\
			{(50$^3$-16-BCC)} & {(50$^3$-54-BCC)} \\ \hline
			$\begin{bmatrix}  {182.57} & {115.48} & {115.55} & {-0.65} & {0.03} & {-6.23} \\
			& {182.04} & {115.56} & {-1.44} & {0.09} & {3.82} \\
			&          & {180.12} & {1.32}  & {-0.69}& {2.28} \\
			&          &          & {38.88} & {3.44} & {0.88} \\
			\multicolumn{4}{c}{\textbf{Sym.}}                          & {38.67}& {-1.52} \\
			&          &          &         &        & {40.46} \end{bmatrix}$ &
			
			$\begin{bmatrix}  {177.33} & {116.96} & {114.38} & {-1.49} & {-0.49} & {-2.32} \\
			& {179.06} & {116.59} & {-1.01} & {-1.65} & {2.03} \\
			&          & {175.49} & {3.30}  & {3.30}  & {0.30} \\
			&          &          & {36.85} & {1.42}  & {-0.36} \\
			\multicolumn{4}{c}{\textbf{Sym.}}                          & {36.20} &  {-1.05}\\
			&          &          &         &         & {35.09} \end{bmatrix}$\\
			
			\hline {50$^3$ unit cells, 250 grains in BCC system, 500008 atoms} & {50$^3$ unit cells, 125 grains in random system, 499836 atoms} \\
			{(50$^3$-250-BCC)} & {(50$^3$-125-Random)} \\ \hline
			$\begin{bmatrix}  {172.29} & {118.51} & {120.44} & {-0.14} & {0.45} & {-1.32} \\
			& {173.39} & {119.08} & {-0.48} & {0.46} & {0.35} \\
			&          & {173.41} & {1.24}  & {-0.55}& {0.75} \\
			&          &          & {28.53} & {-2.40}& {-1.77} \\
			\multicolumn{4}{c}{\textbf{Sym.}}                         & {27.57}& {-1.19} \\
			&          &          &         &        & {29.23} \end{bmatrix}$ &
			
			$\begin{bmatrix} {176.96} & {115.67} & {118.06} & {3.00}  & {2.65}  & {-2.05} \\
			& {174.31} & {118.15} & {-0.71} & {-1.01} & {-0.10} \\
			&          &  {175.02}& {3.97}  & {1.02}  & {0.04} \\
			&          &          & {31.71} & {2.56}  & {-2.30} \\
			\multicolumn{4}{c}{\textbf{Sym.}}                         & {30.66} & {0.05} \\
			&          &          &         &         & {29.92} \end{bmatrix}$\\
			
			\hline \multicolumn{2}{ c }{100$^3$ unit cells, 16 grains in BCC system, 4000010 atoms} \\
			\multicolumn{2}{ c }{(100$^3$-16-BCC)} \\ \hline
			\multicolumn{2}{ c }{$\begin{bmatrix}  {186.97} & {114.79}& {112.47} & {-0.88}& {-0.07} & {-6.21} \\
				& {186.11}& {113.14} & {-1.20}& {1.42}  & {2.40} \\
				&         & {187.21} & {2.26} & {-1.08} & {3.56} \\
				&         &          & {42.93}& {2.71}  & {1.77} \\
				\multicolumn{4}{c}{\textbf{Sym.}}                        & {42.50} & {-1.00} \\
				&         &          &        &         & {44.38} \end{bmatrix}$ }\\
			\hline
		\end{tabular}
	\end{threeparttable}
\end{table}

\subsection{Nanocrystalline aluminum}

\begin{table}[H] 
	\caption{Aluminum: Volume (\AA$^3$), box lengths (\AA), number of atoms, average grain diameter $d$ (\AA), fraction of transient shell atoms $f_{0}$ (\ref{def:fsa}), average cohesive energy $E_{c}$\,(eV/atom) of analysed computational samples.}
	\label{tab:SamplesAl}
	\centering
	\renewcommand{\arraystretch}{1.5}
	\tiny 
	\begin{tabular}{|c c c c c c c|}
		\hline Sample & V & L & No.of atoms & $d$ & $f_{0}$ &  E$_{c}$\\ 
		\hline Monocrystal & 66.43 & 4.05 & 4 & &  & -3.36 \\
		10$^3$-128-BCC & 68949.29 & 40.97 & 3992 & 10.1 & 1.00  &-3.288 \\
		15$^3$-128-BCC & 231371.86 & 61.52 & 13405 & 15.1 & 0.99 &3.289 \\
		50$^3$-128-BCC & 8449883 & 203.70 & 499847 & 50.1 & 0.46  &-3.325 \\
		50$^3$-16-BCC & 8382312.4 & 203.11 & 499935 &  100.0 & 0.33 & -3.341 \\
		50$^3$-54-BCC & 8419421.3 & 203.43 & 500038 & 66.9 & 0.58 &-3.333 \\
		50$^3$-250-BCC & 8480939.5 & 203.93 & 499991 & 40.2 & 0.67  &-3.318 \\
		50$^3$-125-Random & 8457020.3 & 203.76 & 499836 & 50.6 & 0.58 &-3.323 \\
		100$^3$-16-BCC & 66762145 & 405.67 & 4000010 & 199.7 & 0.18  & -3.350 \\
		\hline 
	\end{tabular}
\end{table} 

\begin{table}[H] 
	\caption{{{Aluminum: Elasticity tensors $\bar{\mathbb{C}}$\,[GPa]  of analysed samples (for notation used see Eq. (\ref{eqn:CuCij}))}}.}
	\label{tab:Cij-cAl}
	\begin{threeparttable}[b]
		\centering
		\scriptsize 
		\begin{tabular}{ c c }
			\hline
			{Monocrystal} & {10$^3$ unit cells, 128 grains in BCC system, 3992 atoms} \\
			& {(10$^3$-128-BCC)}	  \\ \hline
			$\begin{bmatrix}  {113.80} & {61.56} & {61.56} & 0       &       0 & 0 \\
			& {113.80} & {61.56} & 0       &       0 & 0 \\
			&          & {113.80} & 0       & 0       & 0 \\
			&          &          & {31.59} &       0 & 0 \\
			\multicolumn{4}{c}{\textbf{Sym.}}                          & {31.59} & 0 \\
			&          &          &         &         & {31.59} \end{bmatrix}$ &	
			
			$\begin{bmatrix}  {96.37} & {64.80} & {63.67} & {0.21} & {-0.20} & {0.51} \\
			& {93.89} & {62.94} & {-0.77} & {0.95}  & {-0.09} \\
			&          & {95.05} & {-0.92}  & {0.99} & {0.48} \\
			&          &          & {15.83} & {0.76} & {0.36} \\
			\multicolumn{4}{c}{\textbf{Sym.}}                          & {15.27} & {0.01} \\
			&          &          &         &         & {15.32} \end{bmatrix}$\\  
			
			\hline {15$^3$ unit cells, 128 grains in BCC system, 13405 atoms} & {50$^3$ unit cells, 128 grains in BCC system,  499847 atoms} \\  
			{(15$^3$-128-BCC)} & {(50$^3$-128-BCC)} \\  \hline
			$\begin{bmatrix}  {92.59} & {63.38} & {63.50} & {-0.47}  & {0.38} & {-0.82} \\
			& {93.90} & {62.95} & {-0.16}  & {0.09}  & {0.56} \\
			&          & {93.64} & {0.76} & {-0.01}  & {-0.07} \\
			&          &          & {15.38} & {-0.25} & {0.14} \\
			\multicolumn{4}{c}{\textbf{Sym.}}                          & {15.15} & {-0.07} \\
			&          &          &         &         & {15.34} \end{bmatrix}$ &
			
			$\begin{bmatrix}  {103.59} & {59.47} & {61.68} & {0.23} & {-0.62} & {0.03} \\
			& {105.25} & {61.58} & {0.09} & {-1.45} & {0.36} \\
			&          & {102.76} & {0.17}  & {0.03}& {0.22} \\
			&          &          & {21.26} & {0.69} & {0.08} \\
			\multicolumn{4}{c}{\textbf{Sym.}}                         & {22.16}& {-0.23} \\
			&          &          &         &        & {22.01} \end{bmatrix}$\\		
			
			\hline {50$^3$ unit cells, 16 grains in BCC system, 499935 atoms} & {50$^3$ unit cells, 54 grains in BCC system, 500038 atoms} \\
			{(50$^3$-16-BCC)} & {(50$^3$-54-BCC)} \\ \hline
			$\begin{bmatrix}  {108.60} & {60.66} & {61.81} & {-1.94} & {-0.12} & {0.40} \\
			& {107.68} & {62.68} & {-0.95} & {1.07} & {0.30} \\
			&          & {109.01} & {2.78}  & {-0.54}& {-0.84} \\
			&          &          & {24.35} & {-0.33} & {-0.56} \\
			\multicolumn{4}{c}{\textbf{Sym.}}                          & {24.46}& {-0.59} \\
			&          &          &         &        & {26.05} \end{bmatrix}$ &
			
			$\begin{bmatrix}  {103.17} & {61.26} & {63.36} & {3.03} & {1.01} & {0.59} \\
			& {104.90} & {62.18} & {-0.59} & {-1.08} & {-0.41} \\
			&          & {102.55} & {-1.10}  & {1.24}  & {-0.64} \\
			&          &          & {19.52} & {0.64}  & {0.46} \\
			\multicolumn{4}{c}{\textbf{Sym.}}                          & {21.13} &  {-1.42}\\
			&          &          &         &         & {23.01} \end{bmatrix}$\\
			
			\hline {50$^3$ unit cells, 250 grains in BCC system, 499991 atoms} & {50$^3$ unit cells, 125 grains in random system, 499836 atoms} \\
			{(50$^3$-250-BCC)} & {(50$^3$-125-Random)} \\ \hline
			$\begin{bmatrix}  {98.69} & {63.30} & {61.83} & {0.18} & {-2.44} & {-1.32} \\
			& {97.22} & {63.77} & {0.97} & {0.75} & {-0.74} \\
			&          & {99.85} & {-0.66}  & {0.53}& {1.30} \\
			&          &          & {20.00} & {-0.08}& {1.02} \\
			\multicolumn{4}{c}{\textbf{Sym.}}                         & {19.18}& {-2.04} \\
			&          &          &         &        & {17.86} \end{bmatrix}$ &
			
			$\begin{bmatrix} {102.04} & {60.08} & {62.42} & {-0.31}  & {-0.48}  & {-0.94} \\
			& {101.79} & {62.67} & {0.31} & {0.78} & {1.27} \\
			&          &  {102.71}& {1.81}  & {0.15}  & {-0.58} \\
			&          &          & {20.16} & {-1.57}  & {-2.17} \\
			\multicolumn{4}{c}{\textbf{Sym.}}                         & {19.29} & {-1.38} \\
			&          &          &         &         & {18.60} \end{bmatrix}$\\
			
			\hline \multicolumn{2}{ c }{100$^3$ unit cells, 16 grains in BCC system, 4000010 atoms} \\
			\multicolumn{2}{ c }{(100$^3$-16-BCC)} \\ \hline
			\multicolumn{2}{ c }{$\begin{bmatrix}  {111.82} & {60.96}& {60.27} & {-0.24}& {-0.02} & {-0.93} \\
				& {111.27}& {60.98} & {-0.01}& {0.16}  & {0.65} \\
				&         & {111.69} & {0.44} & {0.07} & {0.48} \\
				&         &          & {27.25}& {0.36}  & {-0.08} \\
				\multicolumn{4}{c}{\textbf{Sym.}}                        & {26.45} & {-0.28} \\
				&         &          &        &         & {26.61} \end{bmatrix}$ }\\
			\hline
		\end{tabular}
	\end{threeparttable}
\end{table}

\subsection{Nanocrystalline nickel}

\begin{table}[H] 
	\caption{Nickel: Volume (\AA$^3$), box lengths (\AA), number of atoms, average grain diameter $d$ (\AA), fraction of transient shell atoms $f_{0}$ (\ref{def:fsa}), average cohesive energy $E_{c}$\,(eV/atom) of analysed computational samples.}
	\label{tab:SamplesNi}
	\centering
	\renewcommand{\arraystretch}{1.5}
	\tiny 
	\begin{tabular}{|c c c c c c c|}
		\hline Sample & V & L & No.of atoms & $d$ & $f_{0}$ &  E$_{c}$\\ 
		\hline Monocrystal & 43.54 & 3.52 & 4 & &  & -4.39 \\
		14$^3$-128-BCC & 123418.2 & 49.79 & 11052 & 12.1 & 1.00  &-4.261 \\
		15$^3$-128-BCC & 182329.0 & 56.75 & 16335 & 13.0 & 0.99 &-4.264 \\
		50$^3$-128-BCC & 5527341.4 & 176.80 & 499984 & 44.7	 & 0.61  &-4.320 \\
		50$^3$-16-BCC & 5491908.5 & 176.39 & 500020 &  89.2	 & 0.35	 & -4.350 \\
		50$^3$-54-BCC & 5511531.4 & 176.64 & 500058 & 59.5	 & 0.49	 &-4.335 \\
		50$^3$-250-BCC & 5542256.3  & 176.97 & 500008 & 35.8 & 0.71  &-4.309 \\
		50$^3$-125-Random & 5528944.5 & 176.81 & 499836 & 45.1 & 0.61 &-4.318 \\
		100$^3$-16-BCC & 43749012 & 352.36 & 4000010 & 178.2 & 0.19	  & -4.367 \\
		\hline 
	\end{tabular}
\end{table} 

\begin{table}[H] 
	\caption{{{Nickel: Elasticity tensors $\bar{\mathbb{C}}$\,[GPa]  of analysed samples (for notation used see Eq. (\ref{eqn:CuCij}))}}.}
	\label{tab:Cij-cNi}
	\begin{threeparttable}[b]
		\centering
		\scriptsize 
		\begin{tabular}{ c c }
			\hline
			{Monocrystal} & {14$^3$ unit cells, 128 grains in BCC system, 11052 atoms} \\
			& {(14$^3$-128-BCC)}	  \\ \hline
			$\begin{bmatrix}  {247.00} & {147.30} & {147.30} & 0       &       0 & 0 \\
			& {247.00} & {147.30} & 0       &       0 & 0 \\
			&          & {247.00} & 0       & 0       & 0 \\
			&          &          & {122.77} &       0 & 0 \\
			\multicolumn{4}{c}{\textbf{Sym.}}                          & {122.77} & 0 \\
			&          &          &         &         & {122.77} \end{bmatrix}$ &	
			
			$\begin{bmatrix}  {131.33} & {68.93} & {71.44} & {-1.95} & {0.12} & {1.19} \\
			& {135.70} & {68.75} & {0.84} & {0.27}  & {0.24} \\
			&          & {135.85} & {0.81}  & {-0.75} & {-1.79} \\
			&          &          & {31.09} & {-1.34} & {-0.66} \\
			\multicolumn{4}{c}{\textbf{Sym.}}                          & {31.40} & {-0.53} \\
			&          &          &         &         & {30.47} \end{bmatrix}$\\  
			
			\hline {16$^3$ unit cells, 128 grains in BCC system, 16335 atoms} & {50$^3$ unit cells, 128 grains in BCC system,  499984 atoms} \\  
			{(16$^3$-128-BCC)} & {(50$^3$-128-BCC)} \\  \hline
			$\begin{bmatrix}  {136.64} & {71.04} & {70.69} & {-0.91}  & {0.63} & {0.76} \\
			& {134.17} & {71.92} & {0.88}  & {1.04}  & {0.46} \\
			&          & {136.52} & {0.10} & {-1.10}  & {-0.46} \\
			&          &          & {30.70} & {-0.66} & {0.60} \\
			\multicolumn{4}{c}{\textbf{Sym.}}                          & {34.12} & {-0.03} \\
			&          &          &         &         & {31.03} \end{bmatrix}$ &
			
			$\begin{bmatrix}  {195.44} & {98.34} & {103.61} & {-2.35} & {-2.03} & {-0.99} \\
			& {203.37} & {101.21} & {-2.48} & {2.58} & {2.43} \\
			&          & {200.19} & {2.86}  & {1.35}& {-3.22} \\
			&          &          & {53.21} & {2.76} & {-2.67} \\
			\multicolumn{4}{c}{\textbf{Sym.}}                         & {52.58}& {-4.71} \\
			&          &          &         &        & {50.49} \end{bmatrix}$\\		
			
			\hline {50$^3$ unit cells, 16 grains in BCC system, 500020 atoms} & {50$^3$ unit cells, 54 grains in BCC system, 500058 atoms} \\
			{(50$^3$-16-BCC)} & {(50$^3$-54-BCC)} \\ \hline
			$\begin{bmatrix}  {225.72} & {117.81} & {117.25} & {0.11} & {-4.94} & {-13.38} \\
			& {226.23} & {112.46} & {-2.22} & {5.98} & {10.43} \\
			&          & {234.17} & {3.80}  & {-0.42}& {8.47} \\
			&          &          & {69.93} & {5.45} & {5.80} \\
			\multicolumn{4}{c}{\textbf{Sym.}}                          & {60.36}& {-8.23} \\
			&          &          &         &        & {62.76} \end{bmatrix}$ &
			
			$\begin{bmatrix}  {215.37} & {107.09} & {109.49} & {-9.02} & {2.11} & {-0.67} \\
			& {204.12} & {112.09} & {-2.24} & {-0.25} & {-6.54} \\
			&          & {206.22} & {11.47}  & {1.01}& {5.18} \\
			&          &          & {50.53} & {-3.42} & {-2.24} \\
			\multicolumn{4}{c}{\textbf{Sym.}}                          & {52.64}& {2.20} \\
			&          &          &         &        & {55.21} \end{bmatrix}$\\
			
			\hline {50$^3$ unit cells, 250 grains in BCC system, 500008 atoms} & {50$^3$ unit cells, 125 grains in random system, 499836 atoms} \\
			{(50$^3$-250-BCC)} & {(50$^3$-125-Random)} \\ \hline
			$\begin{bmatrix}  {185.81} & {95.36} & {96.31} & {-4.77} & {2.21} & {-0.69} \\
			& {185.41} & {96.87} & {-3.02} & {0.55} & {1.13} \\
			&          & {183.42} & {1.69}  & {0.15}& {0.18} \\
			&          &          & {53.10} & {0.12}& {-1.76} \\
			\multicolumn{4}{c}{\textbf{Sym.}}                         & {46.81}& {-0.56} \\
			&          &          &         &        & {47.30} \end{bmatrix}$ &
			
			$\begin{bmatrix} {194.81} & {97.82} & {96.98} & {-7.23}  & {-4.55}  & {2.80} \\
			& {196.70} & {102.81} & {0.25} & {2.80} & {-3.07} \\
			&          &  {196.78}& {2.24}  & {-0.94}  & {0.70} \\
			&          &          & {51.82} & {0.16}  & {-3.44} \\
			\multicolumn{4}{c}{\textbf{Sym.}}                         & {53.48} & {-2.10} \\
			&          &          &         &         & {47.65} \end{bmatrix}$\\
			
			\hline \multicolumn{2}{ c }{100$^3$ unit cells, 16 grains in BCC system, 4000010 atoms} \\
			\multicolumn{2}{ c }{(100$^3$-16-BCC)} \\ \hline
			\multicolumn{2}{ c }{$\begin{bmatrix}  {250.16} & {124.20}& {124.80} & {-0.23}& {-0.41} & {-8.81} \\
				& {253.06}& {121.09} & {-3.20}& {0.00}  & {3.36} \\
				&         & {252.50} & {3.97} & {0.11} & {4.10} \\
				&         &          & {74.18}& {4.85}  & {0.31} \\
				\multicolumn{4}{c}{\textbf{Sym.}}                        & {74.58} & {-2.05} \\
				&         &          &        &         & {75.39} \end{bmatrix}$ }\\
			\hline
		\end{tabular}
	\end{threeparttable}
\end{table}

\subsection{Nanocrystalline tungsten}

\begin{table}[H] 
	\caption{Tungsten: Volume (\AA$^3$), box lengths (\AA), number of atoms, average grain diameter $d$ (\AA), fraction of transient shell atoms $f_{0}$ (\ref{def:fsa}), average cohesive energy $E_{c}$\,(eV/atom) of analysed computational samples.}
	\label{tab:SamplesT}
	\centering
	\renewcommand{\arraystretch}{1.5}
	\tiny 
	\begin{tabular}{|c c c c c c c|}
		\hline Sample & V & L & No.of atoms & $d$ & $f_{0}$ &  E$_{c}$\\ 
		\hline Monocrystal & 31.06 & 3.14 & 2 & &  & -8.90 \\
		14$^3$-128-BCC & 87959.41 & 44.50 &  5552 & 10.9 & 1.00  &-8.487 \\
		15$^3$-128-BCC & 107627.75 & 47.54 & 6789 & 11.7 & 0.99 &-8.487 \\
		63$^3$-128-BCC & 7929186 & 199.41 & 500269 & 49.1 & 0.53 &-8.699 \\
		63$^3$-16-BCC & 7861853.7 & 198.84 & 500183 &  98.2	 & 0.30 & -8.791 \\
		63$^3$-54-BCC & 7895657.6  & 199.13 & 500121 & 65.46 & 0.42	 &-8.743 \\
		63$^3$-250-BCC & 7944642.6 & 199.54 & 500207 & 39.27 & 0.63	  &-8.662 \\
		63$^3$-125-Random & 7929410.1 & 199.41 & 500118 & 50.6 & 0.58 &-8.689 \\
		126$^3$-16-BCC & 62509290 & 396.87 & 4000817 & 195.4 & 0.16	  & -8.845 \\
		\hline 
	\end{tabular}
\end{table}

\begin{table}[H] 
	\caption{{{Tungsten: Elasticity tensors $\bar{\mathbb{C}}$\,[GPa]  of analysed samples (for notation used see Eq. (\ref{eqn:CuCij}))}}.}
	\label{tab:Cij-cW}
	\begin{threeparttable}[b]
		\centering
		\scriptsize 
		\begin{tabular}{ c c }
			\hline
			{Monocrystal} & {14$^3$ unit cells, 128 grains in BCC system, 5552 atoms} \\
			& {(14$^3$-128-BCC)}	  \\ \hline
			$\begin{bmatrix}  {523.04} & {202.19} & {202.19} & 0       &       0 & 0 \\
			& {523.04} & {202.19} & 0       &       0 & 0 \\
			&          & {523.04} & 0       & 0       & 0 \\
			&          &          & {160.88} &       0 & 0 \\
			\multicolumn{4}{c}{\textbf{Sym.}}                          & {160.88} & 0 \\
			&          &          &         &         & {160.88} \end{bmatrix}$ &	
			
			$\begin{bmatrix}  {431.54} & {271.43} & {257.76} & {-1.95} & {1.90} & {23.55} \\
			& {449.18} & {275.95} & {0.78} & {-0.65}  & {-3.31} \\
			&          & {443.80} & {-1.62}  & {3.22} & {-2.75} \\
			&          &          & {87.17} & {-2.78} & {-1.45} \\
			\multicolumn{4}{c}{\textbf{Sym.}}                          & {82.91} & {-1.46} \\
			&          &          &         &         & {83.47} \end{bmatrix}$\\  
			
			\hline {15$^3$ unit cells, 128 grains in BCC system, 6789 atoms} & {63$^3$ unit cells, 128 grains in BCC system, 500269 atoms} \\  
			{(15$^3$-128-BCC)} & {(63$^3$-128-BCC)} \\  \hline
			$\begin{bmatrix}  {437.60} & {272.96} & {270.05} & {4.79}  & {2.28} & {0.18} \\
			& {448.75} & {273.76} & {-1.07}  & {3.72}  & {-2.47} \\
			&          & {429.72} & {-3.54} & {-1.30}  & {-0.25} \\
			&          &          & {84.64} & {2.23} & {-0.90} \\
			\multicolumn{4}{c}{\textbf{Sym.}}                          & {81.94} & {1.24} \\
			&          &          &         &         & {81.76} \end{bmatrix}$ &
			
			$\begin{bmatrix}  {464.42} & {238.75} & {232.20} & {0.66} & {-0.11} & {-4.26} \\
			& {463.07} & {229.66} & {-4.44} & {8.18} & {12.10} \\
			&          & {462.69} & {3.92}  & {0.72}& {10.65} \\
			&          &          & {124.57} & {2.16} & {-0.63} \\
			\multicolumn{4}{c}{\textbf{Sym.}}                         & {124.30}& {3.52} \\
			&          &          &         &        & {111.47} \end{bmatrix}$\\		
			
			\hline {63$^3$ unit cells, 16 grains in BCC system, 500183  atoms} & {63$^3$ unit cells, 54 grains in BCC system, 500121 atoms} \\
			{(63$^3$-16-BCC)} & {(63$^3$-54-BCC)} \\ \hline
			$\begin{bmatrix}  {478.88} & {220.10} & {225.16} & {-8.55} & {-5.88} & {-11.31} \\
			& {463.41} & {243.55} & {-0.28} & {-15.90} & {9.03} \\
			&          & {462.02} & {0.55}  & {24.64}& {-3.03} \\
			&          &          & {109.57} & {16.38} & {6.06} \\
			\multicolumn{4}{c}{\textbf{Sym.}}                          & {126.27}& {0.09} \\
			&          &          &         &        & {128.30} \end{bmatrix}$ &
			
			$\begin{bmatrix}  {474.50} & {220.26} & {238.31} & {-19.69} & {23.14} & {-32.94} \\
			& {440.61} & {231.82} & {-14.25} & {12.60} & {-4.48} \\
			&          & {478.11} & {11.17}  & {-3.64}  & {0.82} \\
			&          &          & {107.21} & {13.27}  & {1.73} \\
			\multicolumn{4}{c}{\textbf{Sym.}}                          & {109.28} &  {-4.59}\\
			&          &          &         &         & {141.24} \end{bmatrix}$\\
			
			\hline {63$^3$ unit cells, 250 grains in BCC system, 500207 atoms} & {63$^3$ unit cells, 125 grains in random system,  500118 atoms} \\
			{(63$^3$-250-BCC)} & {(63$^3$-125-Random)} \\ \hline
			$\begin{bmatrix}  {452.56} & {234.89} & {242.06} & {-2.32} & {-4.01} & {3.08} \\
			& {447.59} & {240.94} & {7.36} & {6.14} & {9.61} \\
			&          & {453.30} & {-2.46}  & {17.51}& {7.22} \\
			&          &          & {109.92} & {-3.21}& {-4.82} \\
			\multicolumn{4}{c}{\textbf{Sym.}}                         & {119.58}& {7.81} \\
			&          &          &         &        & {107.57} \end{bmatrix}$ &
			
			$\begin{bmatrix} {465.75} & {235.10} & {234.81} & {-1.04}  & {-1.32}  & {1.56} \\
			& {469.61} & {232.24} & {-2.52} & {-1.63} & {0.82} \\
			&          &  {469.58}& {5.07}  & {0.20}  & {0.63} \\
			&          &          & {118.90} & {-0.44}  & {0.38} \\
			\multicolumn{4}{c}{\textbf{Sym.}}                         & {117.17} & {-1.80} \\
			&          &          &         &         & {118.33} \end{bmatrix}$\\
			
			\hline \multicolumn{2}{ c }{126$^3$ unit cells, 16 grains in BCC system, 4000817 atoms} \\
			\multicolumn{2}{ c }{(126$^3$-16-BCC)} \\ \hline
			\multicolumn{2}{ c }{$\begin{bmatrix}  {493.32} & {217.42}& {216.53} & {-1.06}& {-1.68} & {-1.50} \\
				& {494.54}& {216.01} & {0.96}& {1.95}  & {0.77} \\
				&         & {494.96} & {0.67} & {-1.14} & {0.04} \\
				&         &          & {147.42}& {0.58}  & {0.19} \\
				\multicolumn{4}{c}{\textbf{Sym.}}                        & {147.28} & {0.79} \\
				&         &          &        &         & {146.85} \end{bmatrix}$ }\\
			\hline
		\end{tabular}
	\end{threeparttable}
\end{table}

\subsection{Nanocrystalline iron}

\begin{table}[H] 
	\caption{Iron: Volume (\AA$^3$), box lengths (\AA), number of atoms, average grain diameter $d$ (\AA), fraction of transient shell atoms $f_{0}$ (\ref{def:fsa}), average cohesive energy $E_{c}$\,(eV/atom) of analysed computational samples.}
	\label{tab:SamplesFe}
	\centering
	\renewcommand{\arraystretch}{1.5}
	\tiny 
	\begin{tabular}{|c c c c c c c|}
		\hline Sample & V & L & No.of atoms & $d$ & $f_{0}$ &  E$_{c}$\\ 
		\hline Monocrystal & 23.54 & 2.87 & 2 & &  & -4.29 \\
		14$^3$-128-BCC &67251.15 & 40.63 &  5552 & 10.6	 & 1.00  &-4.143 \\
		17$^3$-128-BCC & 119035.24 & 49.18 & 9841 & 12.00 & 0.99 &-4.143 \\
		63$^3$-128-BCC & 5973535.1 & 181.43 & 500130 & 44.46 & 0.59	 &-4.217 \\
		63$^3$-16-BCC & 5934590.9 & 181.04 & 500095 &  88.92 & 0.34	 & -4.251 \\
		63$^3$-54-BCC & 5954905.6  & 181.26 & 500121 & 59.28 & 0.47	 &-4.233 \\
		63$^3$-250-BCC & 5986928.7 & 181.54 & 500152 & 35.57 & 0.68	  &-4.203 \\
		63$^3$-125-Random & 5975973.4 & 181.49 & 500118 & 44.81	 & 0.58	 &-4.214 \\
		126$^3$-16-BCC & 47294569 & 361.62 & 4000811 & 177.83	 & 0.18  & -4.269\\
		\hline 
	\end{tabular}
\end{table}

\begin{table}[H] 
	\caption{{{Iron: Elasticity tensors $\bar{\mathbb{C}}$\,[GPa]  of analysed samples (for notation used see Eq. (\ref{eqn:CuCij}))}}.}
	\label{tab:Cij-cFe}
	\begin{threeparttable}[b]
		\centering
		\scriptsize 
		\begin{tabular}{ c c }
			\hline
			{Monocrystal} & {14$^3$ unit cells, 128 grains in BCC system, 5552 atoms} \\
			& {(14$^3$-128-BCC)}	  \\ \hline
			$\begin{bmatrix}  {229.65} & {135.50} & {135.50} & 0       &       0 & 0 \\
			& {229.65} & {135.50} & 0       &       0 & 0 \\
			&          & {229.65} & 0       & 0       & 0 \\
			&          &          & {116.76} &       0 & 0 \\
			\multicolumn{4}{c}{\textbf{Sym.}}                          & {116.76} & 0 \\
			&          &          &         &         & {116.76} \end{bmatrix}$ &	
			
			$\begin{bmatrix}  {188.58} & {107.73} & {105.40} & {-0.35} & {1.02} & {-0.42} \\
			& {187.06} & {106.98} & {1.15} & {-0.30}  & {2.04} \\
			&          & {192.44} & {0.78}  & {0.21} & {0.11} \\
			&          &          & {44.56} & {0.78} & {-0.59} \\
			\multicolumn{4}{c}{\textbf{Sym.}}                          & {42.36} & {0.66} \\
			&          &          &         &         & {44.55} \end{bmatrix}$\\  
			
			\hline {17$^3$ unit cells, 128 grains in BCC system, 9841 atoms} & {63$^3$ unit cells, 128 grains in BCC system, 500130 atoms} \\  
			{(17$^3$-128-BCC)} & {(63$^3$-128-BCC)} \\  \hline
			$\begin{bmatrix}  {186.58} & {106.35} & {107.00} & {-0.05}  & {0.34} & {1.24} \\
			& {187.66} & {109.23} & {1.49}  & {-0.72}  & {1.99} \\
			&          & {186.80} & {-1.54} & {0.77}  & {-2.43} \\
			&          &          & {43.17} & {0.02} & {-0.05} \\
			\multicolumn{4}{c}{\textbf{Sym.}}                          & {41.73} & {-1.78} \\
			&          &          &         &         & {40.76} \end{bmatrix}$ &
			
			$\begin{bmatrix}  {223.27} & {112.99} & {114.35} & {-3.22} & {-0.75} & {-0.39} \\
			& {223.38} & {114.89} & {-1.65} & {-0.06} & {2.64} \\
			&          & {219.49} & {2.36}  & {-0.59}& {-2.04} \\
			&          &          & {60.79} & {-0.04} & {-0.03} \\
			\multicolumn{4}{c}{\textbf{Sym.}}                         & {59.82}& {-0.08} \\
			&          &          &         &        & {57.48} \end{bmatrix}$\\		
			
			\hline {63$^3$ unit cells, 16 grains in BCC system, 500095  atoms} & {63$^3$ unit cells, 54 grains in BCC system, 500121 atoms} \\
			{(63$^3$-16-BCC)} & {(63$^3$-54-BCC)} \\ \hline
			$\begin{bmatrix}  {233.60} & {119.91} & {115.65} & {-4.11} & {0.39} & {-9.20} \\
			& {237.17} & {119.22} & {-3.89} & {-1.02} & {4.84} \\
			&          & {243.01} & {5.32}  & {1.35}& {2.16} \\
			&          &          & {69.99} & {5.69} & {0.69} \\
			\multicolumn{4}{c}{\textbf{Sym.}}                          & {67.76}& {-4.03} \\
			&          &          &         &        & {73.41} \end{bmatrix}$ &
			
			$\begin{bmatrix}  {235.69} & {112.66} & {114.37} & {-1.63} & {-1.79} & {-4.26} \\
			& {235.54} & {114.83} & {-1.27} & {-0.61} & {3.39} \\
			&          & {233.87} & {3.09}  & {2.20}  & {0.83} \\
			&          &          & {67.86} & {1.26}  & {-0.49} \\
			\multicolumn{4}{c}{\textbf{Sym.}}                          & {66.61} &  {-1.87}\\
			&          &          &         &         & {65.87} \end{bmatrix}$\\
			
			\hline {63$^3$ unit cells, 250 grains in BCC system, 500152 atoms} & {63$^3$ unit cells, 125 grains in random system,  500118 atoms} \\
			{(63$^3$-250-BCC)} & {(63$^3$-125-Random)} \\ \hline
			$\begin{bmatrix}  {217.72} & {113.46} & {109.90} & {-0.96} & {-1.15} & {-2.63} \\
			& {219.30} & {112.86} & {0.79} & {-0.29} & {0.25} \\
			&          & {220.87} & {0.91}  & {1.52}& {0.90} \\
			&          &          & {57.10} & {-1.87}& {0.05} \\
			\multicolumn{4}{c}{\textbf{Sym.}}                         & {57.56}& {-0.05} \\
			&          &          &         &        & {51.84} \end{bmatrix}$ &
			
			$\begin{bmatrix} {230.85} & {126.02} & {112.78} & {5.53}  & {-3.55}  & {2.37} \\
			& {230.04} & {113.79} & {7.13} & {0.84} & {8.48} \\
			&          &  {214.39}& {3.07}  & {4.98}  & {-1.92} \\
			&          &          & {57.73} & {-10.16}  & {2.20} \\
			\multicolumn{4}{c}{\textbf{Sym.}}                         & {54.91} & {-3.31} \\
			&          &          &         &         & {59.32} \end{bmatrix}$\\
			
			\hline \multicolumn{2}{ c }{126$^3$ unit cells, 16 grains in BCC system, 4000811 atoms} \\
			\multicolumn{2}{ c }{(126$^3$-16-BCC)} \\ \hline
			\multicolumn{2}{ c }{$\begin{bmatrix}  {253.03} & {116.99}& {115.26} & {-0.61}& {0.77} & {-10.71} \\
				& {252.56}& {116.04} & {-3.98}& {1.09}  & {5.68} \\
				&         & {254.30} & {4.72} & {-1.56} & {4.28} \\
				&         &          & {77.19}& {5.09}  & {1.45} \\
				\multicolumn{4}{c}{\textbf{Sym.}}                        & {77.93} & {2.74} \\
				&         &          &        &         & {79.34} \end{bmatrix}$ }\\
			\hline
		\end{tabular}
	\end{threeparttable}
\end{table}

\subsection{Nanocrystalline sodium}

\begin{table}[H] 
	\caption{Sodium: Volume (\AA$^3$), box lengths (\AA), number of atoms, average grain diameter $d$ (\AA), fraction of transient shell atoms $f_{0}$ (\ref{def:fsa}), average cohesive energy $E_{c}$\,(eV/atom) of analysed computational samples.}
	\label{tab:SamplesNa}
	\centering
	\renewcommand{\arraystretch}{1.5}
	\tiny 
	\begin{tabular}{|c c c c c c c|}
		\hline Sample & V & L & No.of atoms & $d$ & $f_{0}$ &  E$_{c}$\\ 
		\hline Monocrystal & 75.57 & 4.23 & 2 & &  & -1.111 \\
		17$^3$-128-BCC &375856.55 & 72.10 &  9841 & 17.77 & 1.00  &-1.092 \\
		19$^3$-128-BCC &520818.75 & 80.46 & 13638 & 19.81 & 0.99 &-1.092 \\
		63$^3$-128-BCC & 19037348 & 267.02 & 500074 & 65.57	 & 0.63	 &-1.101 \\
		63$^3$-16-BCC & 18979403 & 266.73 & 500100 &  131.15 & 0.36	 & -1.106 \\
		63$^3$-54-BCC & 19013873  & 266.95 & 500173 & 87.43	 & 0.51	 &-1.104 \\
		63$^3$-250-BCC & 19059847  & 267.15 & 500237 & 52.46 & 0.73  &-1.10 \\
		63$^3$-125-Random & 19038515 & 267.02 & 500118 & 66.09 & 0.62 &-1.101 \\
		126$^3$-16-BCC & 151533591 & 533.13 & 4000898 & 262.29 & 0.19  & -1.108\\
		\hline 
	\end{tabular}
\end{table}

\begin{table}[H] 
	\caption{{{Sodium: Elasticity tensors $\bar{\mathbb{C}}$\,[GPa]  of analysed samples (for notation used see Eq. (\ref{eqn:CuCij}))}}.}
	\label{tab:Cij-cNa}
	\begin{threeparttable}[b]
		\centering
		\scriptsize 
		\begin{tabular}{ c c }
			\hline
			{Monocrystal} & {17$^3$ unit cells, 128 grains in BCC system, 9841 atoms} \\
			& {(17$^3$-128-BCC)}	  \\ \hline
			$\begin{bmatrix}  {8.26} & {6.83} & {6.83} & 0       &       0 & 0 \\
			& {8.26} & {6.83} & 0       &       0 & 0 \\
			&          & {8.26} & 0       & 0       & 0 \\
			&          &          & {5.84} &       0 & 0 \\
			\multicolumn{4}{c}{\textbf{Sym.}}                          & {5.84} & 0 \\
			&          &          &         &         & {5.84} \end{bmatrix}$ &	
			
			$\begin{bmatrix}  {8.86} & {5.68} & {5.60} & {0.00} & {-0.04} & {-0.01} \\
			& {8.73} & {5.72} & {0.09} & {0.03}  & {0.05} \\
			&          & {8.87} & {-0.07}  & {0.02} & {0.00} \\
			&          &          & {1.74} & {-0.04} & {-0.02} \\
			\multicolumn{4}{c}{\textbf{Sym.}}                          & {1.72} & {0.02} \\
			&          &          &         &         & {1.75} \end{bmatrix}$\\  
			
			\hline {19$^3$ unit cells, 128 grains in BCC system, 13638 atoms} & {63$^3$ unit cells, 128 grains in BCC system, 500074 atoms} \\  
			{(19$^3$-128-BCC)} & {(63$^3$-128-BCC)} \\  \hline
			$\begin{bmatrix}  {9.08} & {5.58} & {5.50} & {0.04}  & {-0.01} & {-0.01} \\
			& {9.06} & {5.58} & {0.01}  & {-0.04}  & {0.02} \\
			&          & {9.09} & {-0.03} & {0.04}  & {0.00} \\
			&          &          & {1.78} & {0.03} & {0.04} \\
			\multicolumn{4}{c}{\textbf{Sym.}}                          & {1.64} & {0.03} \\
			&          &          &         &         & {1.75} \end{bmatrix}$ &
			
			$\begin{bmatrix}  {9.74} & {5.71} & {5.91} & {0.13} & {-0.04} & {0.01} \\
			& {9.95} & {5.85} & {-0.06} & {-0.04} & {0.10} \\
			&          & {9.85} & {0.13}  & {0.07}& {-0.07} \\
			&          &          & {2.24} & {-0.04} & {-0.05} \\
			\multicolumn{4}{c}{\textbf{Sym.}}                         & {2.14}& {-0.02} \\
			&          &          &         &        & {2.08} \end{bmatrix}$\\		
			
			\hline {63$^3$ unit cells, 16 grains in BCC system, 500100  atoms} & {63$^3$ unit cells, 54 grains in BCC system, 500173 atoms} \\
			{(63$^3$-16-BCC)} & {(63$^3$-54-BCC)} \\ \hline
			$\begin{bmatrix}  {9.93} & {5.92} & {5.95} & {-0.05} & {0.08} & {-0.57} \\
			& {10.12} & {5.75} & {-0.17} & {0.11} & {0.31} \\
			&          & {10.10} & {0.22}  & {-0.19}& {0.25} \\
			&          &          & {2.56} & {0.24} & {0.10} \\
			\multicolumn{4}{c}{\textbf{Sym.}}                          & {2.59}& {-0.17} \\
			&          &          &         &        & {2.73} \end{bmatrix}$ &
			
			$\begin{bmatrix}  {9.84} & {5.74} & {6.02} & {-0.17} & {-0.07} & {-0.37} \\
			& {10.01} & {6.01} & {-0.09} & {-0.02} & {0.15} \\
			&          & {9.72} & {0.25}  & {0.15}  & {0.22} \\
			&          &          & {2.57} & {0.12}  & {-0.04} \\
			\multicolumn{4}{c}{\textbf{Sym.}}                          & {2.44} &  {-0.08}\\
			&          &          &         &         & {2.30} \end{bmatrix}$\\
			
			\hline {63$^3$ unit cells, 250 grains in BCC system, 500237 atoms} & {63$^3$ unit cells, 125 grains in random system,  500118 atoms} \\
			{(63$^3$-250-BCC)} & {(63$^3$-125-Random)} \\ \hline
			$\begin{bmatrix}  {9.67} & {5.78} & {5.83} & {-0.01} & {0.00} & {0.04} \\
			& {9.61} & {5.89} & {-0.06} & {-0.08} & {0.01} \\
			&          & {9.54} & {0.06}  & {0.24}& {-0.01} \\
			&          &          & {2.11} & {-0.14}& {0.03} \\
			\multicolumn{4}{c}{\textbf{Sym.}}                         & {2.16}& {-0.04} \\
			&          &          &         &        & {1.78} \end{bmatrix}$ &
			
			$\begin{bmatrix} {9.86} & {5.77} & {5.87} & {-0.16}  & {-0.19}  & {-0.02} \\
			& {9.84} & {5.85} & {0.06} & {0.15} & {0.03} \\
			&          &  {9.72}& {0.05}  & {0.05}  & {-0.04} \\
			&          &          & {2.10} & {-0.24}  & {0.13} \\
			\multicolumn{4}{c}{\textbf{Sym.}}                         & {1.69} & {0.12} \\
			&          &          &         &         & {2.08} \end{bmatrix}$\\
			
			\hline \multicolumn{2}{ c }{126$^3$ unit cells, 16 grains in BCC system, 4000898 atoms} \\
			\multicolumn{2}{ c }{(126$^3$-16-BCC)} \\ \hline
			\multicolumn{2}{ c }{$\begin{bmatrix}  {10.22} & {5.85}& {5.84} & {-0.14}& {0.14} & {-0.64} \\
				& {10.31}& {5.69} & {-0.18}& {0.08}  & {0.33} \\
				&         & {10.34} & {0.33} & {-0.22} & {0.27} \\
				&         &          & {2.78}& {0.33}  & {0.15} \\
				\multicolumn{4}{c}{\textbf{Sym.}}                        & {2.78} & {-0.22} \\
				&         &          &        &         & {2.91} \end{bmatrix}$ }\\
			\hline
		\end{tabular}
	\end{threeparttable}
\end{table}

\subsection{Nanocrystalline niobium}

\begin{table}[H] 
	\caption{Niobium: Volume (\AA$^3$), box lengths (\AA), number of atoms, average grain diameter $d$ (\AA), fraction of transient shell atoms $f_{0}$ (\ref{def:fsa}), average cohesive energy $E_{c}$\,(eV/atom) of analysed computational samples.}
	\label{tab:SamplesNb}
	\centering
	\renewcommand{\arraystretch}{1.5}
	\tiny 
	\begin{tabular}{|c c c c c c c|}
		\hline Sample & V & L & No.of atoms & $d$ & $f_{0}$ &  E$_{c}$\\ 
		\hline Monocrystal & 36.19 & 3.31 & 2 & &  & -7.091 \\
		10$^3$-128-BCC &37081.65 & 33.33 &  1989 & 8.21	 & 1.00  &-6.897 \\
		13$^3$-128-BCC &81017.13 & 43.16 & 4343 & 10.65	 & 0.99 &-6.895 \\
		63$^3$-128-BCC & 9172528.1 & 209.32 & 500256 & 51.30 & 0.46 &-7.004 \\
		63$^3$-16-BCC & 9115524.1 & 208.88 & 500157 &  102.61 & 0.25 & -7.045 \\
		63$^3$-54-BCC & 9143533.5  & 209.11 & 500121 & 68.41 & 0.36	 &-7.024 \\
		63$^3$-250-BCC & 9193627.1  & 209.49 & 500192 & 41.04 & 0.55  &-6.988 \\
		63$^3$-125-Random & 9176933.2 & 209.34 & 500118 & 51.71	 & 0.46	 &-6.999 \\
		126$^3$-16-BCC & 72671923 & 417.29 & 4000874 & 205.22	 & 0.13	  & -7.067\\
		\hline 
	\end{tabular}
\end{table}

\begin{table}[H] 
	\caption{{{Niobium: Elasticity tensors $\bar{\mathbb{C}}$\,[GPa]  of analysed samples (for notation used see Eq. (\ref{eqn:CuCij}))}}.}
	\label{tab:Cij-cNb}
	\begin{threeparttable}[b]
		\centering
		\scriptsize 
		\begin{tabular}{ c c }
			\hline
			{Monocrystal} & {10$^3$ unit cells, 128 grains in BCC system, 1989 atoms} \\
			& {(10$^3$-128-BCC)}	  \\ \hline
			$\begin{bmatrix}  {233.08} & {123.89} & {123.89} & 0       &       0 & 0 \\
			& {233.08} & {123.89} & 0       &       0 & 0 \\
			&          & {233.08} & 0       & 0       & 0 \\
			&          &          & {32.13} &       0 & 0 \\
			\multicolumn{4}{c}{\textbf{Sym.}}                          & {32.13} & 0 \\
			&          &          &         &         & {32.13} \end{bmatrix}$ &	
			
			$\begin{bmatrix}  {236.02} & {148.34} & {148.06} & {-1.41} & {0.23} & {-0.31} \\
			& {235.85} & {147.82} & {-1.47} & {0.90}  & {0.25} \\
			&          & {237.56} & {0.60}  & {1.06} & {-0.04} \\
			&          &          & {44.45} & {-1.06} & {-0.12} \\
			\multicolumn{4}{c}{\textbf{Sym.}}                          & {44.69} & {0.17} \\
			&          &          &         &         & {45.45} \end{bmatrix}$\\  
			
			\hline {13$^3$ unit cells, 128 grains in BCC system, 4343 atoms} & {63$^3$ unit cells, 128 grains in BCC system, 500256 atoms} \\  
			{(13$^3$-128-BCC)} & {(63$^3$-128-BCC)} \\  \hline
			$\begin{bmatrix}  {239.34} & {146.48} & {148.07} & {0.65}  & {-0.87} & {0.32} \\
			& {237.38} & {146.20} & {-0.42}  & {0.96}  & {0.57} \\
			&          & {238.63} & {0.68} & {-0.44}  & {0.22} \\
			&          &          & {45.48} & {-0.29} & {-0.24} \\
			\multicolumn{4}{c}{\textbf{Sym.}}                          & {45.39} & {-0.03} \\
			&          &          &         &         & {46.49} \end{bmatrix}$ &
			
			$\begin{bmatrix}  {221.13} & {140.54} & {139.15} & {0.98} & {-0.46} & {1.45} \\
			& {220.79} & {140.95} & {-0.41} & {-0.17} & {0.87} \\
			&          & {221.93} & {-1.24}  & {-0.21}& {0.87} \\
			&          &          & {40.81} & {0.14} & {-0.39} \\
			\multicolumn{4}{c}{\textbf{Sym.}}                         & {40.29}& {1.07} \\
			&          &          &         &        & {42.28} \end{bmatrix}$\\		
			
			\hline {63$^3$ unit cells, 16 grains in BCC system, 500157 atoms} & {63$^3$ unit cells, 54 grains in BCC system, 500121 atoms} \\
			{(63$^3$-16-BCC)} & {(63$^3$-54-BCC)} \\ \hline
			$\begin{bmatrix}  {219.76} & {136.19} & {137.22} & {0.13} & {0.58} & {1.79} \\
			& {220.32} & {135.74} & {0.75} & {-0.10} & {-0.70} \\
			&          & {219.96} & {-1.49}  & {0.31}& {-0.99} \\
			&          &          & {40.69} & {-1.12} & {-0.32} \\
			\multicolumn{4}{c}{\textbf{Sym.}}                          & {41.36}& {0.26} \\
			&          &          &         &        & {39.47} \end{bmatrix}$ &
			
			$\begin{bmatrix}  {222.33} & {138.38} & {138.01} & {0.05} & {-0.39} & {0.20} \\
			& {222.32} & {138.10} & {0.75} & {0.42} & {-0.63} \\
			&          & {222.32} & {-0.34}  & {-0.80}  & {-0.76} \\
			&          &          & {40.81} & {-0.48}  & {-0.15} \\
			\multicolumn{4}{c}{\textbf{Sym.}}                          & {41.16} &  {0.45}\\
			&          &          &         &         & {42.38} \end{bmatrix}$\\
			
			\hline {63$^3$ unit cells, 250 grains in BCC system, 500192 atoms} & {63$^3$ unit cells, 125 grains in random system,  500118 atoms} \\
			{(63$^3$-250-BCC)} & {(63$^3$-125-Random)} \\ \hline
			$\begin{bmatrix}  {219.87} & {139.55} & {141.16} & {1.09} & {2.55} & {-0.77} \\
			& {223.63} & {140.01} & {0.10} & {0.31} & {-2.04} \\
			&          & {222.48} & {-0.10}  & {1.88}& {0.67} \\
			&          &          & {40.87} & {0.32}& {1.20} \\
			\multicolumn{4}{c}{\textbf{Sym.}}                         & {41.31}& {0.15} \\
			&          &          &         &        & {40.53} \end{bmatrix}$ &
			
			$\begin{bmatrix} {222.57} & {139.80} & {140.60} & {1.37}  & {0.84}  & {-0.52} \\
			& {222.99} & {139.96} & {1.95} & {0.59} & {-0.48} \\
			&          &  {223.18}& {0.04}  & {0.13}  & {0.40} \\
			&          &          & {40.30} & {1.77}  & {0.50} \\
			\multicolumn{4}{c}{\textbf{Sym.}}                         & {40.94} & {0.44} \\
			&          &          &         &         & {40.11} \end{bmatrix}$\\
			
			\hline \multicolumn{2}{ c }{126$^3$ unit cells, 16 grains in BCC system, 4000874 atoms} \\
			\multicolumn{2}{ c }{(126$^3$-16-BCC)} \\ \hline
			\multicolumn{2}{ c }{$\begin{bmatrix}  {217.40} & {133.88}& {135.30} & {0.30}& {0.46} & {2.62} \\
				& {217.82}& {134.64} & {1.25}& {-0.13}  & {-1.42} \\
				&         & {216.78} & {-1.57} & {-0.22} & {-1.20} \\
				&         &          & {39.57}& {-1.24}  & {-0.28} \\
				\multicolumn{4}{c}{\textbf{Sym.}}                        & {40.14} & {0.25} \\
				&         &          &        &         & {38.93} \end{bmatrix}$ }\\
			\hline
		\end{tabular}
	\end{threeparttable}
\end{table}

\subsection{Nanocrystalline vanadium}

\begin{table}[H] 
	\caption{Vanadium: Volume (\AA$^3$), box lengths (\AA), number of atoms, average grain diameter $d$ (\AA), fraction of transient shell atoms $f_{0}$ (\ref{def:fsa}), average cohesive energy $E_{c}$\,(eV/atom) of analysed computational samples.}
	\label{tab:SamplesV}
	\centering
	\renewcommand{\arraystretch}{1.5}
	\tiny 
	\begin{tabular}{|c c c c c c c|}
		\hline Sample & V & L & No.of atoms & $d$ & $f_{0}$ &  E$_{c}$\\ 
		\hline Monocrystal & 27.82 & 3.03 & 2 & &  & -5.300 \\
		10$^3$-128-BCC &28264.68 & 30.45 &  1990 & 7.50	 & 1.00  &-5.103 \\
		13$^3$-128-BCC &61984.58 & 39.57 & 4357 & 9.74	 & 0.99 &-5.110 \\
		63$^3$-128-BCC & 7042771.7 & 191.67 & 500130 & 47.19 & 0.42	 &-5.215 \\
		63$^3$-16-BCC & 7006050.2 & 191.35 & 500095 &  94.22 & 0.23 & -5.255 \\
		63$^3$-54-BCC & 7026628.8  & 191.54 & 500121 & 62.87 & 0.33	 &-5.234 \\
		63$^3$-250-BCC & 7057669.1  & 191.82 & 500152 & 37.78 & 0.50  &-5.198 \\
		63$^3$-125-Random & 7047184.1 & 191.71 & 500118 & 47.57	 & 0.42	 &-5.210 \\
		126$^3$-16-BCC & 55861236 & 382.26 & 4000817 & 188.22 & 0.23 & -5.277\\
		\hline 
	\end{tabular}
\end{table}

\begin{table}[H] 
	\caption{{{Vanadium: Elasticity tensors $\bar{\mathbb{C}}$\,[GPa]  of analysed samples (for notation used see Eq. (\ref{eqn:CuCij}))}}.}
	\label{tab:Cij-cV}
	\begin{threeparttable}[b]
		\centering
		\scriptsize 
		\begin{tabular}{ c c }
			\hline
			{Monocrystal} & {10$^3$ unit cells, 128 grains in BCC system, 1990 atoms} \\
			& {(10$^3$-128-BCC)}	  \\ \hline
			$\begin{bmatrix}  {227.57} & {119.10} & {119.10} & 0       &       0 & 0 \\
			& {227.57} & {119.10} & 0       &       0 & 0 \\
			&          & {227.57} & 0       & 0       & 0 \\
			&          &          & {43.16} &       0 & 0 \\
			\multicolumn{4}{c}{\textbf{Sym.}}                          & {43.16} & 0 \\
			&          &          &         &         & {43.16} \end{bmatrix}$ &	
			
			$\begin{bmatrix}  {283.79} & {137.46} & {148.78} & {-0.73} & {-0.16} & {0.28} \\
			& {224.10} & {135.56} & {-17.57} & {-3.76}  & {-9.65} \\
			&          & {287.03} & {-0.67}  & {-0.55} & {1.40} \\
			&          &          & {69.01} & {0.47} & {-1.12} \\
			\multicolumn{4}{c}{\textbf{Sym.}}                          & {70.04} & {0.12} \\
			&          &          &         &         & {69.82} \end{bmatrix}$\\  
			
			\hline {13$^3$ unit cells, 128 grains in BCC system, 4357 atoms} & {63$^3$ unit cells, 128 grains in BCC system, 500130 atoms} \\  
			{(13$^3$-128-BCC)} & {(63$^3$-128-BCC)} \\  \hline
			$\begin{bmatrix}  {259.63} & {138.40} & {139.74} & {-8.38}  & {11.75} & {10.62} \\
			& {288.83} & {142.92} & {-4.38}  & {5.77}  & {8.32} \\
			&          & {285.48} & {-3.57} & {2.73}  & {5.46} \\
			&          &          & {65.82} & {2.27} & {5.63} \\
			\multicolumn{4}{c}{\textbf{Sym.}}                          & {66.06} & {-4.61} \\
			&          &          &         &         & {59.31} \end{bmatrix}$ &
			
			$\begin{bmatrix}  {235.38} & {128.77} & {129.22} & {-3.04} & {-0.75} & {-0.87} \\
			& {234.94} & {134.76} & {-2.67} & {-1.61} & {-0.41} \\
			&          & {234.90} & {0.49}  & {-0.33}& {-2.34} \\
			&          &          & {52.10} & {-1.00} & {-1.73} \\
			\multicolumn{4}{c}{\textbf{Sym.}}                         & {50.63}& {-0.84} \\
			&          &          &         &        & {49.63} \end{bmatrix}$\\		
			
			\hline {63$^3$ unit cells, 16 grains in BCC system, 500095 atoms} & {63$^3$ unit cells, 54 grains in BCC system, 500121 atoms} \\
			{(63$^3$-16-BCC)} & {(63$^3$-54-BCC)} \\ \hline
			$\begin{bmatrix}  {228.63} & {131.16} & {129.15} & {-1.43} & {0.32} & {-2.57} \\
			& {228.70} & {128.55} & {-0.16} & {1.19} & {1.26} \\
			&          & {230.89} & {1.24}  & {-0.71}& {-1.32} \\
			&          &          & {48.12} & {1.36} & {-0.04} \\
			\multicolumn{4}{c}{\textbf{Sym.}}                          & {50.15}& {-2.00} \\
			&          &          &         &        & {51.27} \end{bmatrix}$ &
			
			$\begin{bmatrix}  {224.73} & {128.38} & {131.94} & {2.23} & {-4.74} & {-3.80} \\
			& {228.62} & {128.88} & {1.80} & {-0.54} & {0.70} \\
			&          & {232.11} & {-2.69}  & {1.82}  & {0.96} \\
			&          &          & {52.36} & {1.71}  & {0.53} \\
			\multicolumn{4}{c}{\textbf{Sym.}}                          & {54.20} &  {-1.39}\\
			&          &          &         &         & {52.24} \end{bmatrix}$\\
			
			\hline {63$^3$ unit cells, 250 grains in BCC system, 500152 atoms} & {63$^3$ unit cells, 125 grains in random system,  500118 atoms} \\
			{(63$^3$-250-BCC)} & {(63$^3$-125-Random)} \\ \hline
			$\begin{bmatrix}  {240.59} & {133.04} & {130.85} & {-4.09} & {-1.55} & {0.91} \\
			& {232.44} & {133.58} & {-4.64} & {0.00} & {-0.57} \\
			&          & {227.86} & {-1.03}  & {4.71}& {-2.22} \\
			&          &          & {48.48} & {-2.38}& {-2.77} \\
			\multicolumn{4}{c}{\textbf{Sym.}}                         & {53.98}& {0.16} \\
			&          &          &         &        & {53.22} \end{bmatrix}$ &
			
			$\begin{bmatrix} {230.56} & {125.92} & {127.74} & {-1.53}  & {-4.12}  & {-8.51} \\
			& {240.11} & {125.42} & {-1.89} & {4.91} & {3.68} \\
			&          &  {224.47}& {-1.92}  & {1.68}  & {-3.04} \\
			&          &          & {54.09} & {-0.67}  & {2.67} \\
			\multicolumn{4}{c}{\textbf{Sym.}}                         & {50.29} & {-2.90} \\
			&          &          &         &         & {53.27} \end{bmatrix}$\\
			
			\hline \multicolumn{2}{ c }{126$^3$ unit cells, 16 grains in BCC system, 4000817 atoms} \\
			\multicolumn{2}{ c }{(126$^3$-16-BCC)} \\ \hline
			\multicolumn{2}{ c }{$\begin{bmatrix}  {225.54} & {126.12}& {125.77} & {-0.09}& {0.20} & {-0.08} \\
				& {225.56}& {125.69} & {0.21}& {0.07}  & {-0.01} \\
				&         & {224.84} & {-0.43} & {0.07} & {-0.02} \\
				&         &          & {49.60}& {0.13}  & {3.99} \\
				\multicolumn{4}{c}{\textbf{Sym.}}                        & {48.69} & {-0.08} \\
				&         &          &        &         & {49.24} \end{bmatrix}$ }\\
			\hline
		\end{tabular}
	\end{threeparttable}
\end{table}

\clearpage

\section{A core-shell model with an inhomogeneous shell properties}

{In this appendix an alternative variant of a core-shell model for nanocrystalline material, based on the concept introduced in \cite{Sevostianov2006}, originally proposed for metal-matrix composites with nanosized inclusions, is shortly outlined. Within this variant, first, effective properties of an anisotropic composite grain are established and next the overall properties of polycrystal composed of composite grains are found applying one of the available averaging schemes for the coarse-grained one-phase polycrystalline materials.
	
	In the first step anisotropic grain core is embedded in the interphase (shell) matrix and the effective anisotropic properties of a composite grain are found in the spirit of a differential scheme. Thus, it is assumed that the respective effective properties are known for the shell layer of thickness $h$ and then the layer is increased by ${\rm{d}}h$. Such an approach enables to consider \emph{inhomogeneous} interphase. When the interphase layer is assumed as locally isotropic then the composite inclusion will be of cubic symmetry, inheriting the anisotropy of a grain core.  Using the result of \cite{Shen03} the following differential equations are to be solved to find three effective Kelvin moduli (\ref{Eq:moduli}):
	\begin{equation}\label{Eq:DiffCS}
	\frac{{\rm{d}}\hat{C}(r)}{{\rm{d}}r}
	=-\frac{3}{r}
	\left((\hat{C}(r)-C_s(r))+\frac{\alpha_s^C(r)}{C_s(r)}(\hat{C}(r)-C_s(r))^2\right)
	\end{equation}  
	where $\hat{C}(r)$ stands subsequently for $\hat{K}(r)$, $\hat{G}_1(r)$, $\hat{G}_2(r)$, function $C_s(r)$  for  ${K}_s(r)$, ${G}_{1s}(r)=G_{2s}(r)=G_s(r)$, respectively, and
	\begin{displaymath}
	\alpha_s^{K}(r)=\frac{1 + \nu_s(r)}{3 (1 - \nu_s(r))}\,,\quad \alpha_s^{G_1}(r)=\alpha_s^{G_2}(r)=\alpha_s^{G}(r)=\frac{8 - 10 \nu_s(r)}{15 (1 - \nu_s(r))}
	\end{displaymath}
	where $\nu_s(r)$ is local Poisson's ratio of the shell, which is the function of the shell bulk and shear moduli $K_s(r)$ and $G_s(r)$ according to the formula (\ref{Eq:E-nu})$_2$. Differential equations (\ref{Eq:DiffCS}) are solved with initial conditions $\hat{K}(r_0)=K$, $\hat{G}_1(r_0)=G_1$ and $\hat{G}_2(r_0)=G_2$ ($r_0$ is a radius of the grain core). The effective Kelvin moduli of a composite grain are then calculated as $\hat{K}_{\Delta}=\hat{K}(r_0+\Delta)$, $\hat{G}_{1\Delta}=\hat{G}_1(r_0+\Delta)$ and  $\hat{G}_{2\Delta}=\hat{G}_2(r_0+\Delta)$. Finally, in the second step, in order to obtain the effective elastic stiffness of a nanocrystalline material, the obtained values are used in place of local properties of the grain within one of the averaging schemes developed for a one phase coarse-grained polycrystals (see Table A.5 in \cite{Kowalczyk18}). In this appendix the Hashin-Shtrikman lower bound is used.
	
	\begin{figure}[bh!]
		\centering
		\centering
		\begin{tabular}{cc}
			(a)&(b)\\
			\includegraphics[angle=0,width=0.47\textwidth]{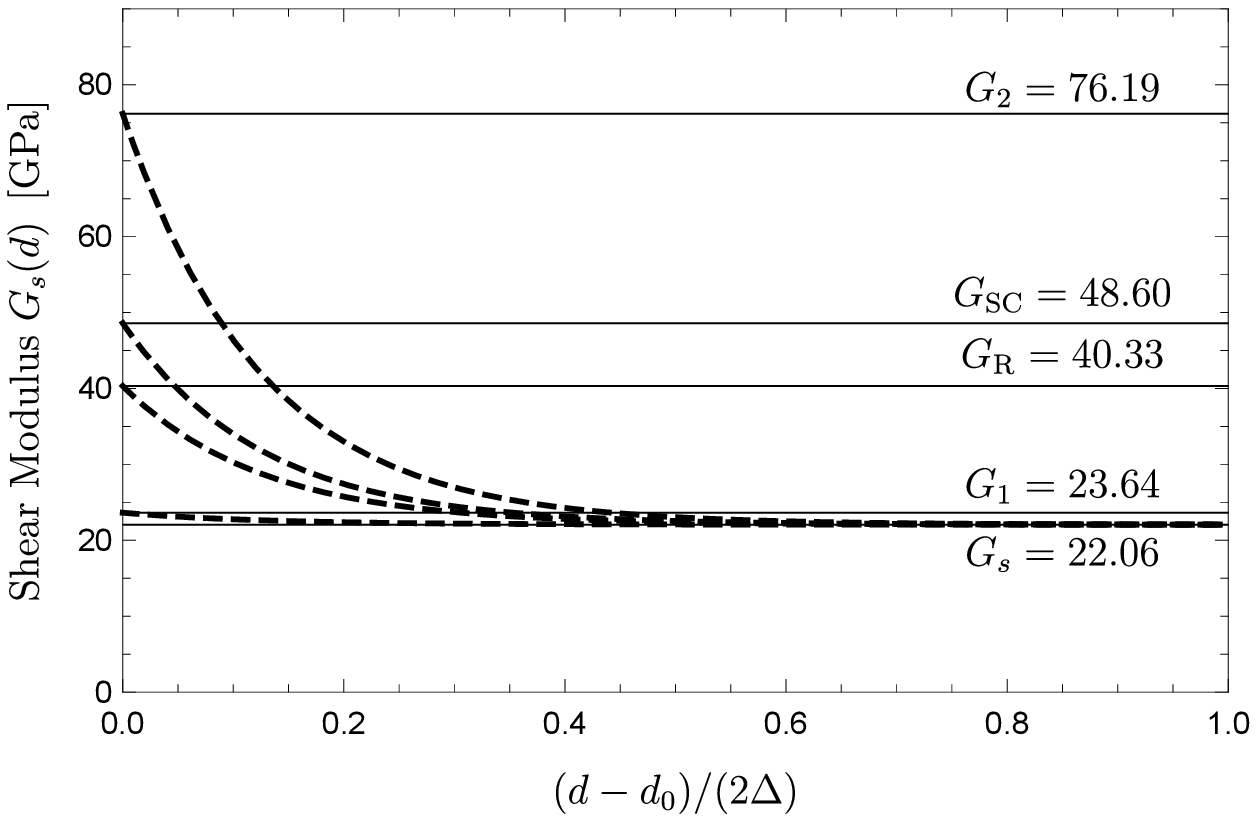}&
			\includegraphics[angle=0,width=0.47\textwidth]{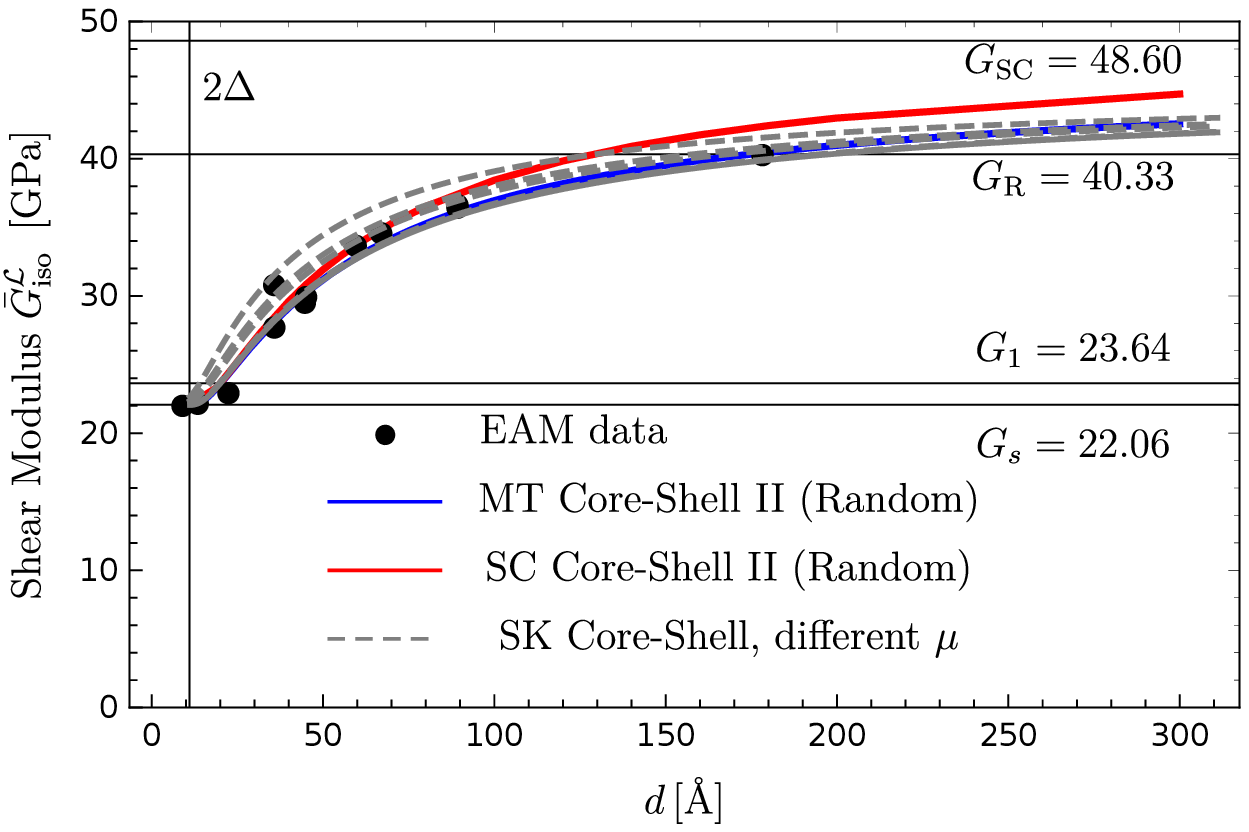}\\
		\end{tabular}
		\caption{({a) Assumed variation of the shear modulus $G_s(r)$ along the coating thickness (see Eq. \ref{Eq:Csvar}), (b) the isotropic shear modulus of nanocrystalline copper (Cu) as a function of the grain diameter by the differential variant (SK) of core-shell model with shell properties varying according to the formula (\ref{Eq:Csvar}) ($b=5$ and $\mu=G_2/G_s,G_{\rm{SC}}/G_s,G_{\rm{R}}/G_s,G_1/G_s,1$) - comparison with results of atomistic simulations and estimates of MT and SC variant of core-shell model with a uniform shell properties. }}
		\label{fig:CoreShellSK}
	\end{figure}
	
	Let us now shortly assess the predictive capabilities of this alternative approach on the example of nanocrystalline copper. As already discussed identification of functions $K_s(r)$ and $G_s(r)$ by means of atomistic simulations is not a trivial task. Here, in this preliminary study, following the work by \cite{Sevostianov2006} the ad-hoc relation, ensuring smooth variation in accordance with results of \cite{Kluge1990}, is proposed 
	\begin{equation}\label{Eq:Csvar}
	C_s(r)=C_s\left(1 + (\mu - 1) \exp\left(-\frac{b}{\Delta}(r - r_0)\right)\right)\,,
	\end{equation}
	in which $C_s$ are $K_s$ or $G_s$ from Table \ref{tab:Shell-CutOff}. In Fig. \ref{fig:CoreShellSK}a the variation the shell shear modulus $G_s(r)$  is presented for varying value of $\mu$ and $b=5$. Note that the assumed form ensures constant Poisson's ratio $\nu_s$ along the shell. Using this relation three differential equations specified by Eq. (\ref{Eq:DiffCS}) have been solved numerically using Wolfram Mathematica. The resulting effective shear modulus of nanocrystalline copper with perfectly random distribution of composite grain orientations, estimated by means of the Hashin-Shtrikman lower bound, is shown in Fig. \ref{fig:CoreShellSK}b.    
	It is worth mentioning that equations (\ref{Eq:DiffCS}) can be solved analytically for $C_s(r)=C_s$ (equivalent to $\mu=1$ in Eq. \ref{Eq:Csvar}). Effective Kelvin moduli of a composite grain of a diameter $d$ (see Fig. \ref{fig:CoreShell}) are then given by a closed-form formula,
	\begin{displaymath}
	\hat{C}_{\Delta}=C_s \left(1 + \frac{(C - 
		C_s) (1 - 
		2 \Delta/d)^3}{C_s + \alpha_s^C (C - 
		C_s) (1 - (1 - 2 \Delta/d)^3)}\right)
	\end{displaymath} 
	
	Fig. \ref{fig:CoreShellSK}b demonstrates that for constant shell moduli the shear modulus predicted by the alternative variant of core-shell model (SK) are very close to the MT variant. When an inhomogeneous shell is assumed ($\mu\neq 1$), the highest differences between the variants are obviously seen for small grains. As compared to the data obtained by atomistic simulations, good predictions of the SK variant are seen for smaller values of $\mu$ in relation (\ref{Eq:Csvar}). Nevertheless the improvement of predictions as compared to MT and SC variant is not readily seen. It should be stressed that MT and SC variants, as compared to the alternative model with an inhomogeneous interphase, have the advantage of being simpler (e.g. they do not require numerical integration). The are specified by closed-formed formulas with less uncertainty stemming from the applied variation of shell properties.}

\section*{References}

\bibliography{References}

\end{document}